\documentclass[reprint, onecolumn, 10pt]{revtex4-2}
\usepackage{CJK}
\usepackage{graphicx,amsmath,amssymb,bm}
\usepackage{color}
\usepackage{hyperref}
\usepackage{natbib}
\usepackage[utf8]{inputenc}
\usepackage{titlesec}
\usepackage{float}
\usepackage{caption}
\usepackage{ulem} 

\usepackage{ragged2e}
\newcommand{\medium}{\fontsize{12pt}{14pt}\selectfont}
\titleformat{\section}[hang]{\normalfont\medium\bfseries}{\thesection}{1em}{}
\titleformat{\subsection}[hang]{\normalfont\small\bfseries}{\thesubsection}{1em}{}
\begin{document}
\title{Machine-learning Guided Search for Phonon-mediated Superconductivity in Boron and Carbon Compounds}
\author{Niraj K. Nepal$^1$}
\email[]{nepalneeraz@gmail.com}
\author{Lin-Lin Wang$^{1,2}$}
\email[]{llw@ameslab.gov}
\affiliation{[1] Ames National Laboratory, Ames, Iowa 50011, USA}
\affiliation{[2] Department of Physics and Astronomy, Iowa State University, Ames, Iowa 50011, USA}
\date{\today}

\begin{abstract}
\normalsize
\justifying
\section*{Abstract}
 We present a workflow that iteratively combines \textit{ab-initio} calculations with a machine-learning (ML) guided search for superconducting compounds with both dynamical stability and instability from imaginary phonon modes, the latter of which have been largely overlooked in previous studies. Electron-phonon coupling (EPC) properties and critical temperature (T$_c$) of 417 boron, carbon, and borocarbide compounds have been calculated with density functional perturbation theory (DFPT) and isotropic Eliashberg approximation. Our study addresses T$_c$ convergence of Brillouin zone sampling with an ansatz test, stabilizing imaginary phonon modes for significant EPC contributions and comparing performance of two ML models especially when including compounds of dynamical instability.  We predict a few promising superconducting compounds with formation energy just above the ground state convex hull, such as Ca$_5$B$_3$N$_6$ (35 K), TaNbC$_2$ (28.4 K), Nb$_3$B$_3$C (16.4 K), Y$_2$B$_3$C$_2$ (4.0 K), Pd$_3$CaB (7.0 K), MoRuB$_2$ (15.6 K), RuVB$_2$ (15.0 K), RuSc$_3$C$_4$ (6.6 K) among others.
\end{abstract}

\maketitle
\textbf{$^*$Corresponding author:} Niraj K. Nepal, Email: \texttt{nepalneeraz@gmail.com}\\

$^\dagger${llw@ameslab.gov}
\section{\textbf{Introduction}}
 The pursuit of high-temperature superconductivity (SC) is a challenging and active research area. The recent discovery of superconducting temperature (T$_c$) near 200 K in H$_3$S at the high pressure of 150 GPa\cite{DETKS15} has re-energized the field to focus on phonon-mediated SC. { For phonon-mediated SC, high phonon frequency from light atomic mass with strong electron-phonon coupling (EPC) is beneficial, as shown by the recent discovery of near room temperature T$_c$ for metal hydrides under high pressure\cite{MWXYWZLY22, LHZWZJF22, KMKDBM21, TSKSSPIP21, STIKKHSS21, LHLLM14, AKTNA15, ECPNNL15, PKMP15}. But making metallic bonds with hydrogen requires extremely high pressure. B and C compounds are known for their diverse structures and due to their light atomic mass, sizable Tc has been found in compounds \cite{WYG21, ZLSMGSKCPY23, DQBB22, ZLYWYL22, GHZWSZ23} ranging from MgB$_2$ to metal intercalated graphite and fullerenes. Therefore, metallic B and C containing compounds with diverse structures are promising pools to search for new phonon-mediated superconductors.} However, at ambient pressure, increasing T$_c$ even just higher than the 40 K of MgB$_2$ \cite{NNMZA01, BLPCAC01, BHR01, HCJ01, WPJ01} has been difficult \cite{KAKFP01, TCSRBC02, CVM04, XLYHZ04, BAC04, MN01, MSVBLLC03, TBNOTH04, HH54, KXVBA10, KXKBJSARRB07, AAMZA04, HBMD73}. While the recent report of a T$_c$=32 K for MoB$_2$ under the pressure of 110 GPa is a promising development \cite{PZWZGG23}, the large temperature and pressure gaps between MgB$_2$ and H$_3$S still remain, which motivates intensive search for new phonon-mediated SC compounds. Despite the challenges faced in experiments, theoretical studies continue to provide valuable insights into potential SC materials and their properties, {  such as SC in FeB$_4$ was first predicted\cite{KSMBHD10} and then verified\cite{GDBTKSRMHA13}}. Density functional perturbation theory (DFPT) \cite{BDDG01,D01} is one of the most robust \textit{ab-initio} methods \cite{SCDFT07} to compute the EPC matrices over the full Brillouin zone (BZ). One can then employ either isotropic Eliashberg approximation or Green function-based anisotropic Migdal-Eliashberg equations to compute T$_c$ \cite{M58,E60,A72,AD72,MG13,F17_review,WMH15,LHHPSZPADE22}. Therefore, computational exploration of compounds containing atoms slightly heavier than H, such as B and C\cite{WYG21, ZLSMGSKCPY23, DQBB22, ZLYWYL22, GHZWSZ23, KMTMMK23}, with strong  EPC is a promising venue to search for high-temperature SC at ambient pressure and also fill the large materials gap between MgB$_2$ and H$_3$S.

 Machine learning (ML) and Artificial intelligence (AI) are increasingly taking important roles for predicting materials properties including SC. {ML models usually offer a computationally inexpensive alternative to DFPT calculations for evaluating superconducting properties, completing predictions in minutes compared to the months, typically required for high throughput DFPT calculations of hundreds of compounds.} Previous studies performing ML predictions based on random forest \cite{SOKRPCT18}, regression \cite{RD20_reg1,XQHDDSSFL22,XSHHH19,ZZXZHZQHL22}, classification \cite{RD20_reg1}, natural language processing (NLP) \cite{CC20_NLP}, and deep learning \cite{KKNSOHM21} models have trained on experimental data, mostly from SuperCon database \cite{supercon}. Comparing to the more expensive and time consuming experiments to explore many new compounds to generate SC data for ML, the high throughput (HT) \textit{ab-initio} first-principles approaches are valuable tools to obtain the data that can be trained to predict potential SC compounds. Several recent studies have utilized \textit{ab-initio} computed data for training ML model and predicting SC. One approach involves performing BCS-inspired screening of materials to identify potential candidates based on certain key properties such as the Debye temperature and the density of states at the Fermi level (N(E$_F$)) \cite{CG22}. A study similar to that described in Ref.~\cite{CG22} has been conducted on a vast range of materials, but restricting the size of the compounds to eight or fewer atoms, as reported in Ref.~\cite{CSM23}. { More recently, a new ML model suitable for the limited size of EPC dataset has been developed to predict Eliashberg spectral functions $\alpha^2 F(\omega)$ rather than the overall EPC strength, which represents an important step to account for the correlation between phonon density of state and EPC spectral function \cite{GHDBGHH25}.} In a recent study \cite{HSN20}, a ML approach was used to predict the maximum T$_c$ and corresponding pressure of binary metal hydrides. The input layer consisted of atomic properties of the heavier metallic atom, while the output layer had two nodes representing T$_c$ and pressure. The \textit{ab-initio} data utilized in this study has been collected from literature. Recently, ML-driven search with experimental feedback was also performed to discover a novel superconductor in Zr-In-Ni systems \cite{PMGDHMT22}.

 Notably in these previous HT and ML studies, compounds of dynamical instability with imaginary phonon modes have been largely discarded. However, as shown by our recent EPC study \cite{NCW24} on Y$_2$C$_3$ with experimentally known T$_c$=18K\cite{AAMZA04}, imaginary phonon of C dimer wobbly motion once stabilized can carry significant EPC contributions, which explains well the observed sizable T$_c$. In recent model Hamiltonian studies, phonon softening and anharmonicity have also been found to enhance T$_c$ \cite{JBBSZ23,SBZ20}. Here we present a workflow that iteratively combines \textit{ab-initio} calculations with an ML-guided search across the dataset of compounds with both dynamical stability and instability from imaginary phonon modes by focusing on boron/carbon/borocarbide (B/C/B+C) compounds. \textit{Ab-initio} calculations were performed to compute the EPC strength ($\lambda$), the logarithmic average phonon frequency ($\omega_{log}$) and T$_c$ of 417 compounds employing DFPT and isotropic Eliashberg approximation. Two major issues arise during DFPT calculations: choosing appropriate BZ sampling (\textbf{k} and \textbf{q}-mesh) for convergence and the problem of calculated dynamic instability. To address the convergence problem, we developed an ansatz test to check the convergence of EPC properties, particularly the T$_c$. For dynamically unstable compounds, we employed large electronic smearing, lattice distortion, and pressure to stabilize them. We then calculated their EPC properties, which were included in building the ML models. We evaluated ML models, specifically the crystal graph convolutional neural network (CGCNN) and the atomistic line graph neural network (ALIGNN), trained utilizing \textit{ab initio} computed data to predict SC properties. { The key new findings of this study are two-fold. First is the inclusion of the previously overlooked dynamical unstable compounds in the EPC training dataset, accounting for a sizable 20\% of all compounds being studied, which would otherwise be omitted. Second, by including compounds with imaginary phonon modes, we reveal a clear performance gap between the two ML architectures, reinforcing the widely accepted principle that the completeness of a training dataset is as crucial as the ML model itself.} Among the two models, especially when including the dynamically unstable compounds, ALIGNN consistently outperforms the CGCNN in predicting EPC properties. We predict a few promising SC compounds with formation energy just above the ground state convex hull. For dynamically stable systems, we predict TaNbC$_2$ (28.4 K), Nb$_3$B$_3$C (16.4 K), Y$_2$B$_3$C$_2$ (4.0 K) among others. For systems with dynamical instability and imaginary phonon modes, we predict Ca$_5$B$_3$N$_6$ with a T$_c$ as high as 35-42.4 K, besides Pd$_3$CaB (7.0 K), and a few Ru compounds of MoRuB$_2$ (15.6 K), RuVB$_2$ (15.0 K), and RuSc$_3$C$_4$ (6.6 K).

\section{Results}
\subsection{Machine learning guided workflow}
 A ML-guided search workflow is employed in this study. It can be divided into three parts: data extraction, DFPT calculations, and training ML models. We will discuss each part in the following three sections (A, B and C) before presenting the main results in the last three sections (D, E and F). As illustrated in Fig.~\ref{fig:3} for obtaining the crystal structures of all the known B/C/B+C compounds, we utilized the Materials Project (MP) database \cite{JOHCRD13_MP,OCJBGCP15,Pymatgen13}, which offers a diverse range of compounds, including experimentally synthesized and theoretically predicted ones. We selected those B and C compounds in the MP database that meet certain criteria: being metallic with negative formation energy, excluding oxides, C$_{60}$ and Lanthanides except for La. Approximately 1500 compounds fall within this category, out of which 400 exhibit magnetic moments, and these magnetic cases are not considered in the present study. Consequently, our focus narrows down to around 1100 nonmagnetic compounds, forming the pool for investigating phonon-mediated superconductivity. To manage computational cost, we set a further criterion that considers only systems with primitive cells containing 40 atoms or less and composed of up to four different elements ($N_{type} \leq 4$). This refinement narrows our selection to approximately 700 compounds. These 700 include 121 compounds with known T$_c$ from experimental measurement as in SuperCon database (113 being dynamically stable and 8 dynamically unstable) and the other 579 compounds of unknown T$_c$. We will first discuss the 113 compounds with known T$_c$ and also dynamical stability (no imaginary phonon modes), while the remaining 8 compounds with dynamical instability will be discussed later.

{Figure~\ref{fig:1} provides a statistical description of the 113 B/C/B+C compounds with known T$_c$ and dynamic stability, which are 53 superconductors (SC) and 60 non-superconductors (NSC). We reviewed and corrected any inaccuracies or discrepancies found in the SuperCon database through an extensive literature review \cite{MH52,HH54,GSSB63,WBM67,MJWPDD71,J77,R78,SR79,KMAJ80,LCEH81,SAS82,KL87,KPKV89,BH92,SOMVBEKA92,CBSKPCFTV94,CSBTECKP94,CTBZKPVFSM94,SRMCGC94,ZJCKP94,KTHM95,MHDH95,HSKS96,S97,YFJWCCTP97,AMS98,OOOYY00,BLLS01,BY01,GSZIK01,NNMZA01,SWNOK01,RWJPT01,KMTMA03,CSPKPO05,SWHDRSE15,THYTEMRCO05,BSDDPTM06,LWTAMP06,AGH07,BY01,KTSEA07,MPDK07,MC07,SNVPJ07,TKEMHT07,BFSDM09,HDFFE10,SMBEPJ10,MCNDGF11,MKKH11,TE11,IMH12,KHFSHAA12,GDBTKSRMHA13,TETRCFTDG12,WO12,BJKSOS12,XWLCGKMCR15,CDDCDCFM16,ECCMHV16,Takada16,BSDGMFOM17,KSRSKVMKS17,SGDAABVGMK17,CXKLGSOC18,SJSGTS18,BRSMPBLDBH14,SMBNPMHSL21,PZGWGZTCLL21,DHSS22,CSCDTS23,KM23,PZWZGG23,ACKMWWFHMV75,SWRG82,SDSZKIC01,JP04,PGPLLC03,HKFYSTK15,IZSF19,MBTLHHZXZC19,SXZCG21,NGLLEHMLL08,HBMD73,CP11,BGHBSYG04,BSKMDR96,HSKS96,KEISSON85,BBFFPSW95,RDVHSSS20,KBHP22}. 

Figure~\ref{fig:1} (a)-(c) illustrate the distribution of different elements in these systems, allocation of B/C/B+C compounds and thermodynamic stability, respectively. The thermodynamic stability of these compounds, indicated by the energy above the ground state convex hull ($\Delta E_h$), are obtained from the MP database\cite{JOHCRD13_MP}. Approximately 71\% of the compounds are on the ground state convex hull with $\Delta E_h \sim 0$. Around 18\% of the cases have an $\Delta E_h$ within 0.05 eV/atom, while the remaining 11\% have $\Delta E_h$ larger than 0.05 eV/atom. Figure~\ref{fig:1}(d) shows the distributions of these compounds based on their space groups (SGs). In Fig.~\ref{fig:1}(a) and (d), each bar is partitioned into two segments with the red segment representing the number of SC, while blue segment for NSC. From Fig.~\ref{fig:1}(a), many known SC compounds are associated with transition metals (TM) such as Y, La, Ni, Rh, Mo, Nb and others. Figure~\ref{fig:1}(e)-(m) depict the crystal structures of the representative SC compounds in the top 6 SGs among the experimentally known SC. In these structures, B and C atoms form various structural motifs: honeycomb lattice of B in MgB$_2$ (Fig.~\ref{fig:1}(e)); monomers in NbC (Fig.~\ref{fig:1}(f)), MgCNi$_3$ (Fig.~\ref{fig:1}(h)) and Mo$_2$GaC (Fig.~\ref{fig:1}(i)); dimers in YC$_2$ (Fig.~\ref{fig:1}(k)); graphene sheets in SrC$_6$ (Fig.~\ref{fig:1}(j)); chains of lighter elements in Mo$_2$BC (Fig.~\ref{fig:1}(g)) and LaPt$_2$B$_2$C (Fig.~\ref{fig:1}(m)); octahedral cage structures in YB$_6$ (Fig.~\ref{fig:1}(l)). In terms of the lattice types of these known SC compounds, highly symmetric structures with hexagonal, tetragonal and cubic SGs have the most compounds, as shown in Fig.~\ref{fig:1}(d). Similarly, statistical description for the SC compounds with predicted T$_c$ that have not been measured, akin to Fig.~\ref{fig:1}, is illustrated in Supplementary Materials (SM) Fig.S1.

\subsection{Overview of DFPT calculations}
 After the crystal structures of B and C compounds have been collected, the next step as shown in Fig.1 is to do HT calculations with DFPT on EPC properties for compounds with both known and unknown T$_c$ using our recently developed high-throughput electronic structure package (HTESP) \cite{htesp}. We performed DFPT calculations and computed EPC properties using the isotropic Eliashberg approximation. The accuracy of the EPC data is crucial for building reliable ML models. We have encountered two major challenges in the HT calculations with DFPT. One is the convergence with respect to BZ sampling and the other is dynamic instability. The first obstacle involved determining appropriate BZ samplings (\textbf{k-} and \textbf{q-} meshes) to compute the EPC properties, as these calculations become computationally expensive with dense meshes. To reduce the computational cost, we initiated an efficient screening by using a \textbf{k-}point mesh that accurately describes the ground-state structures and energetics. EPC quantities are then interpolated to fine \textbf{k}-mesh only twice the size of the coarse \textbf{k}-mesh. But we noticed that calculations with such grid combination can lead to inaccurate predictions, with discrepancies as high as approximately 10\% of the total compounds with known T$_c$, giving NSC for known SC and vice versa. To address this problem, we developed an ansatz test to assess the convergence of T$_c$ with respect to the \textbf{k-}point mesh. This approach leverages the decaying behavior of T$_c$ with respect to Gaussian broadening width ($\sigma$), which is used in the double-delta integration. {Initially, we acquired results using the DFPT method with the \textbf{k}-point mesh size from the MP database. Subsequently, we assessed the convergence of these results based on the convergence ansatz test. To ensure convergence, we repeated calculations with denser k-mesh for the cases where results did not pass this test. We applied this technique to the 113 dynamically stable compounds, and obtained reasonable accuracy for the calculated T$_c$ with a mean absolute error (MAE) of 2.21 K compared to experimental data, as presented in Supplementary Figure S2 (b). { Moreover, incorporating dynamically unstable compounds further improves the dataset quality, as evidenced in supplementary Fig.~S2(c)–(d) by the enhanced $R^2_{log}$ score in the $\log(T_c)$ comparison with experiment.} Further details regarding the convergence with respect to the \textbf{k-} and \textbf{q-}point meshes are discussed in the Method section and SM. In addition to the 113 dynamically stable compounds with known T$_c$, we also computed the EPC properties of 268 compounds of unknown T$_c$ with dynamical stability, resulting from the ML-guided search as summarized in Fig.1. 

The second obstacle encountered in EPC calculations was the presence of imaginary phonon modes and dynamical instability in almost 146 compounds. Among them, the imaginary phonon modes in 36 compounds show large EPC contribution. Stabilizing these imaginary modes is crucial for calculating EPC properties in such systems. These imaginary modes can be stabilized through lattice distortion, pressure, and electronic smearing, with the later method being particularly effective in HT screening\cite{NCW24}. A comprehensive analysis of dynamical instability and its implications for superconductivity will be presented in the ``Imaginary phonon modes and superconductivity" section. In addition to these instances, there are cases up to 173 compounds, where EPC calculations are not complete due to either numerical problems or too large size of unit cells. For now, we will set these cases aside and will revisit them in the future.

\subsection{Training and testing ML models}
 Besides the data extraction and DFPT calculations, training and testing ML models also play crucial roles in the ML-guided search workflow in Fig.1. We utilized two different ML models: CGCNN)\cite{XG_CGCNN18} and ALIGNN)\cite{CD21_ALIGNN}. CGCNN maps 3D crystal structures to 2D graphs by using chemical element information and neighbor bonding distance to encode them into local chemical environments through convolution operations and updating the node features based on these descriptors. The updated node features are then aggregated to represent the entire crystal, which is connected to the output via a neural network. ALIGNN includes extra local chemical information, such as bonding angles, in addition to the crystal graphs in CGCNN with another auxiliary graph of bonding distances and angles. The parameters, including weights and biases, of the neural network connections are learned through training with available DFPT data.

 As shown in Fig.~\ref{fig:3}, we trained the initial ML models in Run 1 using a dataset of 250 dynamically stable compounds with 109 SC (calculated T$_c$ $>$ 1 K) and 141 NSC (calculated T$_c$ $<$ 1 K) including those 113 from SuperCon database that were already measured in experiments. This dataset encompasses 45 distinct SGs. For the purpose of evaluation, a separate collection of 58 stable compounds (18 SC and 40 NSC), belonging to 27 unique SGs (6 of which are not part of the training 45), was reserved exclusively for independent testing and was not involved in the ML training process. { Moving from Run 1 to Run 2, we use the Run 1 ML model (trained on 250 dynamically stable compounds) to predict Tc for the remaining B–C compounds, which is fewer than 450, because some compounds were already found to be dynamically unstable but without EPC data during Run 1 ML training.} We then sort the predictions and select the top candidates of both SC and NSC for more DFPT calculations to close the ML-guided loop. In Run2, the original 250 stable systems were augmented with an additional 73 dynamically stable systems to refine the ML models. In the concluding stage of Run 3, we incorporated the results derived from the compounds with dynamic instability. Notably, this stage included results from 2 new SGs. Among the 36 results in this category, we incorporated 28 into the training set and added 8 compounds into the independent test set. In the overall count of 417 compounds with converged EPC properties, 181 were classified as SCs, whereas 236 were categorized as NSCs. The progress from Run 1 to Run 3 constitutes a loop, as illustrated in Fig.~\ref{fig:3}. In summary, our methodology has a series of iterations involving training and testing ML models with increasingly comprehensive datasets from dynamically stable to unstable compounds.

\subsection{Comparison of ML models}
 Next, we will present the main results of the ML models and guided search, highlight the notable compounds, including cases of dynamical stability and instability. Figures~\ref{fig:4}(a-d) depict the ML-predicted vs. DFPT-calculated $\lambda$, $\omega_{log}$, T$_c$, and T$^{\prime}_c$ using the dynamically stable systems in Run 1, respectively. Here, T$_c$ represents the critical temperature computed using DFPT-computed $\lambda$ and $\omega_{log}$ or predicted directly from ML models, while T$^{\prime}_c$ is calculated from the ML-predicted $\lambda^{ML}$ and $\omega^{ML}_{log}$ using Eq.~\ref{eq6} in a postprocessing manner.
\begin{equation}\label{eq6}
    T^{\prime}_c = \frac{\omega^{ML}_{log}}{1.2} exp \left[- \frac{1.04 (1 + \lambda^{ML})}{\lambda^{ML} - \mu^*_c(1 + 0.62\lambda^{ML})}  \right],
\end{equation}
Here, It is important to note that the data first used do not include dynamically unstable cases. For Run 1, we trained ML models using 250 compounds, divided into training, validation, and testing sets in a ratio of 0.8:0.1:0.1. The mean absolute errors (MAEs) for the 10\% test set in each training iteration are documented in Supplementary Table S1. An additional independent test set of 58 compounds was used to assess the predictability of the ML models and their MAE and predictions are plotted in Fig.~\ref{fig:4}. {Apart from MAE, we also computed the R$^2$-score for the ML parity plot, which is presented in Supplementary Table S2. As expected, accurately predicting the exact EPC properties is challenging. However, the performance of the ML models improves when superconductors with stabilized phonons are included.} The training process consisted of 3000 epochs using default settings provided by the ML packages, which have been thoroughly investigated in the original work \cite{XG_CGCNN18,CD21_ALIGNN}. In both CGCNN and ALIGNN, training and validation procedures are employed. Checkpoints are established at regular intervals to store crucial parameters such as model weights and architecture. The ML packages operate automatically to retain and update the model exhibiting the best performance, determined by the lowest validation error. This iterative process also acts to mitigate overfitting concerns. {Our Supporting Figure S3 and Supporting Table S2 demonstrate that training up to 1500 epochs is sufficient to achieve the lowest validation error for both ALIGNN and CGCNN model.} The performance of the models was evaluated by computing the MAE between the predicted and target quantities for the independent test set. The MAEs for CGCNN-predicted $\lambda$ and $\omega_{log}$ stand at 0.23 and 105 K, respectively, while the corresponding numbers for ALIGNN are slightly higher at 0.28 and 113 K (Figs.~\ref{fig:4}(a) and (b)). In terms of predicting T$_c$, the CGCNN and ALIGNN models yield MAEs of 2.4 K and 3.8 K, respectively. Despite the relatively small magnitude of MAEs, the ML outcomes exhibit distinct clustering patterns, as shown in Fig.~\ref{fig:4}(c). Specifically, for the CGCNN model, the results tend to cluster closely along the ``DFPT" axis (red arrow in Fig.~\ref{fig:4}(c)), whereas for ALIGNN, the clustering is pronounced along the ``ML prediction" axis (blue arrow in Fig.~\ref{fig:4}(c)). An alternative approach, rather than directly training and predicting T$_c$, is to utilize ML-predicted values, specifically $\lambda^{ML}$ and $\omega^{ML}_{log}$, to estimate T$^{\prime}_c$ using Eq.~\ref{eq6}. This modification not only enhances predictive performance for both models but also slightly ameliorates the issue of clustering [Fig.~\ref{fig:4}(d)]. A comparable approach has been employed in a recent study~\cite{TV22}, wherein $\lambda$ and $\omega_{log}$ can be directly acquired from first principles calculations. 

Then we use the predicted T$^{\prime}_c$ to rank the remaining compounds to pick the ones with high and low T$^{\prime}_c$ for additional DFPT calculations. In this next stage, we utilized ML-guided search to expand the dataset to 323 systems characterized by 54 distinct SGs, the ML models underwent further training. Subsequently, the improved ML models were subjected to the same independent testing, employing the same set of 58 system test cases. The outcomes of the Run 2 are presented in Figs.~\ref{fig:4}(e)-(h). The results demonstrate a notable improvement in addressing the issue of prediction clustering with the expansion of the training dataset, all while maintaining reasonable accuracy. ALIGNN improves the prediction of $\lambda$ and $\omega_{log}$ with MAEs of 0.24 and 93K, respectively, compared to CGCNN's MAEs of 0.26 and 97K. However, T$^\prime_c$ calculated from ML-predicted $\lambda^{ML}$ and $\omega^{ML}_{log}$ shows a significant improvement for ALIGNN, with an MAE of 2.7K. Notably, $\omega_{log}$ is more accurately predicted than $\lambda$ and then T$_c$, because the former is directly related to the overall bonding strength and cohesive energy, while the later depends on the details of the Fermi surface and the EPC matrix elements. 

 As mentioned above, from the ML-guided search, we find quite some number of compounds with imaginary phonon and dynamical instability. In all the previous high throughput phonon-mediated SC studies, these dynamically unstable compounds are simply discarded, but as shown by our recent study \cite{NCW24} on Y$_2$C$_3$, some imaginary phonon modes once stabilized can carry a large EPC and give rise to a large T$_c$. Thus, specifically here we also include dynamically unstable compounds in ML. As far as we know, this is the first ML trained with both dynamically stable and unstable compounds for phonon-mediated SC. We carried out a distinct training and testing phase for ML models, incorporating dynamically unstable cases after stabilization, denoted as Run 3. These results predominantly included nonzero T$_c$. Among these, 28 results were added to the training dataset, while the remaining 8 were added for independent testing. In total, the independent test set now consists of the original 58 dynamically stable and additional 8 stabilized compounds with imaginary phonon modes. Both the training and test datasets were balanced across various SGs. The outcomes of Run 3 are illustrated in Fig. ~\ref{fig:5}(a)-(d). In Run 3, the ALIGNN model consistently outperforms the CGCNN model across different superconducting properties. The ALIGNN model achieves MAEs of 0.27 for $\lambda$, 86 K for $\omega_{log}$, 4.3 K for T$_c$, and 3.5 K for T$^{\prime}_c$. In contrast, the CGCNN model records MAEs of 0.34, 104 K, 4.5 K, and 5.3 K for the respective properties. The reason is that ALIGNN includes the bonding angles as part of training parameters, which better describe the dynamically unstable compounds, because imaginary phonon modes often involves lower-energy bond rotation with changing angle than higher-energy bond stretching vibration. Based on the information presented in Figs.~\ref{fig:4} and ~\ref{fig:5}, it is apparent that the ML model performs better in learning the parameters $\omega_{\text{log}}$ than $\lambda$ and then T$_c$. As a result, it is justifiable to utilize Eq.~\ref{eq6} to estimate the ML-predicted critical temperature T$^{\prime}_c$ rather than relying solely on the directly predicted T$_c$ values. { Furthermore, we separately computed the MAE for SC and NSC, as shown in Supplementary Table S3. ALIGNN demonstrates better accuracy in predicting $\omega_{log}$ compared to CGCNN for both categories. However, it predicts $\lambda$ more accurately than CGCNN for NSC, while performing slightly worse for SC. Additionally, a classification model was employed, with the results presented in the Supplementary Table S4.}

\subsection{Dynamically stable systems}
\begin{table}[h!]
    \centering
        \caption{DFPT calculated EPC results for experimentally unknown systems with T$_c$ $>$ 10 K, { whose superconductivity has not yet been measured}; SG is spacegroup, and $\Delta E_h$ is  taken from materials project. \cite{JOHCRD13_MP}}
         \scalebox{1.2}{%
    \begin{tabular}{|c|c|c|c|c|c|}
    \hline
Compound&SG&$\Delta E_h$ (eV/atom) &$\lambda$&$\omega_{log}$ (K)&T$_c$ (K) \\
\hline
B$_2$CN&R3m&0.34&1.66&578&60.7\\
B$_2$CN&P3m1&0.35&1.11&567&34.9\\
Mo$_7$B$_{24}$&P-6m2&0.15&1.28&386&29.6\\
TaNbC$_2$&R-3m&0.02&1.41&326&28.4\\
TcB&P6$_3$/mmc&0.26&1.32&333&26.5\\
TaC&P-6m2&0.41&1.54&235&22.8 \\
ZrBC&P6$_3$/mmc&0.45&1.12&321&20.0\\
Ta$_2$CN&I4$_1$/amd&0.13&1.94&162&19.8\\
Ta$_2$CN&P4/mmm&0.15&2.70&127&19.5\\
B$_2$CN&P-4m2&0.32&0.80&644&18.8\\
TaB$_2$&P6/mmm&0.0&1.22&254&18.1\\
NbFeB&P-6m2&0.42&1.64&173&18.0\\
Nb$_3$B$_3$C&Cmcm&0.02&1.25&229&16.4\\
V$_2$CN&R-3m&0.12&1.00&323&16.2\\
NbC & P6$_3$/mmc&0.15& 0.84 & 455 & 15.3 \\
ZrMoB$_4$&P6/mmm&0.09&1.30&193&15.1\\
NbVCN&R3m&0.16&0.99&274&13.4\\
Ta$_4$C$_3$&Pm-3m&0.13&1.46&138&12.6\\
NbVC$_2$&R-3m&0.11&0.83&365&11.9\\
Nb$_2$CN&R-3m&0.08&0.88&310&11.7\\
Nb$_4$B$_3$C$_2$&Cmcm&0.05&1.01&219&11.3\\
TaVC$_2$&R-3m&0.08&0.80&343&10.3\\
\hline
\end{tabular}
}
\label{tab:10K}
\end{table}

 In Table~\ref{tab:10K}, we present the results of our ML-guided search for dynamically stable compounds with DFPT-calculated T$_c$ exceeding 10 K, while the complete list of EPC properties of the 381 dynamically stable compounds are presented in SM from Table S10 to S17. It is worth noting that the compounds with high T$_c$ tend to be thermodynamically metastable, as indicated by their formation energy significantly above the convex hull. B$_2$CN in different phases, with $\Delta E_h$ up to 0.35 eV/atom, has been predicted to exhibit SC, consistent with earlier theoretical findings \cite{B2CN}. TaC and NbC, both have sizable T$_c$ in the metastable hexagonal structure, while their more stable cubic structures in $Fm-3m$ have already been observed with SC \cite{YYWSYS21, JA12}. In Table~\ref{tab:10K}, the first compound close to the ground state convex hull ($\Delta E_h$ $\sim$ 0.02 eV/atom) is TaNbC$_2$, whose crystal structure and EPC properties are plotted in Fig.~\ref{fig:6}(a)-(c). It crystallizes in the trigonal structure of $R-3m$. The calculated EPC properties are \(\lambda = 1.4\), \(\omega_{\log} = 326\) K, and \(T_c = 28.4\) K, with the majority of the contribution coming from phonons within the 3-6 THz range (Fig.~\ref{fig:6}(b) and (c)), as well as a significant contribution from the 16-20 THz range of C-dominated modes. Other compounds close to the ground state convex hull with $\Delta E_h$ $\leq$ 0.02 is TaB$_2$. The presence of SC in TaB$_2$ remains a subject of debate as discussed in the previous studies \cite{BY01}.

 Moreover, we predict other ternary superconductors, such as Nb$_3$B$_3$C ($\Delta E_h$ of 0.018 eV/atom \cite{JOHCRD13_MP}), with T$_c$ of 16.4 K (Figs.~\ref{fig:6} (d), (f), (h) and (j)). Comparison between the calculated and experimental T$_c$ are also plotted (blue circles) for the Nb-B-C superconductors with known T$_c$. These ternary metallic borocarbides were all experimentally synthesized \cite{HG01_NbBC}, however, some of their SC (red circles) have not been reported yet, with Nb$_3$B$_3$C has a predicted T$_c$ as high as 16.4 K. { To differentiate them from experimentally measured systems, we color-coded them in red and presented them along the y = x line.} The crystal structure of Nb$_3$B$_3$C exhibits { an interesting} layer-like arrangement as in Fig.~\ref{fig:6}(f), where B atoms form strips of honeycomb lattice, while C being monomers. The $\lambda$-projected phonon dispersion (Fig.~\ref{fig:6}(h)) analysis indicates that the SC is attributed to the presence of low-frequency soft phonon modes in the vicinity of the $Z$ and $T$ points of the BZ. For ternary Nb-B-C systems, EPC calculations with isotropic approximation tend to overestimate T$_c$, while for Y-B-C systems, EPC calculations have also notable underestimations as plotted in Fig.~\ref{fig:6}(e). Therefore, we also show the EPC properties of Y$_2$B$_3$C$_2$ with a predicted T$_c$ of 4 K, but having $\Delta E_h$ $\sim$ 0\cite{JOHCRD13_MP}, along with other experimentally measured Y-B-C systems in Fig.~\ref{fig:6}(e). Unlike Nb$_3$B$_3$C, the crystal structure of Y$_2$B$_3$C$_2$ shows a layer of mixed B and C network sandwiching the Y layer (Fig.~\ref{fig:6}(g)). The $\lambda$-projected phonon dispersion shows that the EPC properties are mostly contributed by phonon within the 4-10 THz energy range around the $\Gamma$ and $Z$ points (Figs.~\ref{fig:6} (i) and (k)). Other compounds with calculated T$_c$ $<$ 10 K and their EPC properties are listed in SM.

\subsection{Imaginary phonon modes and superconductivity}
 In this section, we address the issue of dynamical instability observed in the DFPT-computed phonon dispersion (146 out of 700 compounds). Out of these 146 dynamical unstable compounds, 34 exhibit large EPC in their imaginary phonon modes. Figure \ref{fig:7} (a) and (b) provide the breakdown of these 34 compounds based on their constituent elements, while (c) and (d) show the distribution of systems in terms of formation energy and SG, respectively. As expected, a considerable portion of these compounds with imaginary phonon have formation energies above the convex hull. Superconductivity in Sc$_2$C$_3$, shown in Fig.~\ref{fig:7} (d), are very likely because SC has been observed in the same structure with larger cations of the same group as in Y$_2$C$_3$ \cite{Y2C3} and La$_2$C$_3$ \cite{La2C3}, which are closer to the convex hull. We have categorized these compounds into two groups based on whether the imaginary phonon modes occur at the $\Gamma$-point or elsewhere. The instabilities at $\Gamma$-point are represented by Sc$_2$C$_3$, Ta$_2$B, and La$_3$InB (Fig.~\ref{fig:7}(i-l)) which constitutes 11 cases, whereas the other 23 compounds including MoB$_2$ represents the case of dynamic instability outside of $\Gamma$-point, as illustrated in Figs.~\ref{fig:7}(l) for MoB$_2$. The unstable phonon modes possess a large mode-resolved $\lambda$ in the vicinity of instability, $\lambda_{\textbf{q}\nu} = $ $\frac{\gamma_{\textbf{q}\nu}}{\pi N(E_F) \omega^2_{\textbf{q}\nu}}$, where $\gamma_{\textbf{q}\nu}$ is the change in phonon linewidth due to EPC (See Fig.~\ref{fig:7}). The systems with dynamical instability at $\Gamma$-point can be stabilized by obtaining a low-symmetry ground state structure through lattice distortion along the direction of the imaginary eigenmodes at $\Gamma$ and performing a full ionic relaxation. The analysis presented here is expanded from our earlier work on Y$_2$C$_3$ \cite{NCW24}. The application of smearing to stabilize imaginary phonon modes has been explored in other systems, including $\beta$-phase Ni$_x$Al$_{(1-x)}$ \cite{ZH92_NiAl1-x}, NiTi \cite{ZH93_NiTi} and EuAl$_4$\cite{WNC24}. In these cases, instead of optical modes, the instability arises from acoustic modes. It was proposed that the dynamic instability observed in these materials is the result of strong EPC between nested electronic states near the Fermi level \cite{ZH92_NiAl1-x,ZH93_NiTi,WNC24}. Previously as exmplified with Y$_2$C$_3$ \cite{NCW24} these imaginary phonon modes can significantly contribute to $\lambda$ after stabilization. Although these cases represent only a small fraction of the compounds, disregarding them would exclude potential SC compounds with a sizable T$_c$.

Performing EPC calculations on low-symmetry structures can be computationally expensive, especially when the system contains a large number of atoms like Sc$_2$C$_3$. Hence, it is recommended to first try stabilization using pressure and smearing. The former option of stabilizing through pressure can give interesting pressure-dependent properties, whereas the latter with increasing electronic smearing can be beneficial for HT computations. To stabilize the dynamically unstable systems, we applied pressure (ranging from 5 to 60 GPa) or used a larger electronic smearing (0.05, 0.06, 0.08, 0.1 Ry). In each case, we fully relaxed the structures. We computed the EPC properties for the stabilized systems around $\Gamma$-point and presented the results in Table~\ref{tab:distort} together with their SG and \(\Delta E_h\). Most of these compounds crystallize in high-symmetry structures such as cubic, hexagonal and tetragonal lattices. The relative ground state energy of relaxed low-symmetry structures compared to the original high-symmetry ones are listed as Distorted GS. We utilize pressure and the electronic smearing to stabilize the imaginary phonons in Y$_2$C$_3$, La$_2$C$_3$ and Sc$_2$C$_3$. However, for Ta$_2$B and La$_3$InB systems, pressure and smearing were insufficient for stabilization, so lattice distortion was employed. The calculated T$_c$ using DFPT for the stabilized systems agrees well with experimental data. For example, Al\(_2\)Mo\(_3\)C exhibits an instability at \(\Gamma\) akin to that observed in Sc\(_2\)C\(_3\). After stabilization with electronic smearing, the calculated \(T_c\) is 12.05 K, which is comparable to the experimental \(T_c\) of 9.2 K \cite{KXVBAYTCKP10}. We predict the EPC properties for Sc$_2$C$_3$, YBC, and  MoB$_4$ with a sizable T$_c$ of 27.9 K, 10.15 K and 7.58 K under ambient or moderate pressure. These compounds have formation energy higher (0.11, 0.42, and 0.26 eV/atom respectively) than the ground state convex hull, indicating they are metastable.
\begin{table}[htbp!]
    \centering
    
        \caption{Stabilization of imaginary phonon modes (around $\Gamma$) utilizing Distortion (D), Pressure (P), and Smearing (S). The values of pressure and smearing are presented as P-value (GPa) and S-value (Ry) respectively. Crystal is distorted along the direction represented by eigenmode at $\Gamma$ and relaxed to obtain low-symmetry ground-state structures. Distorted GS is the difference between ground-state total energy of an undistorted and distorted structures at equilibrium. SG is spacegroup and $\Delta E_h$ represents the energy above the ground state hull.}
         \scalebox{0.95}{%
    \begin{tabular}{|c|c|c|c|c|c|c|c|}
    \hline
       Compound  & SG & $\Delta E_h$ (eV/atom)\cite{JOHCRD13_MP}  & Distorted GS (meV/atom) & EPC ($\lambda$) & $\omega_{log}$ (K) & $T_c$ (K) &T$^{Expt}_c$ (K)\\
       \hline
       Y$_2$C$_3$ & I-43d & 0.04   & 0.9 & 1.94 (P-10) & 175.23 (P-10) & 21.27 (P-10) & 18 \cite{Y2C3} \\
        & &  &  & 1.14 (S-0.1) & 227.97 (S-0.1) & 14.50 (S-0.1)  \cite{NCW24} &   \\
       \hline
       La$_2$C$_3$ & I-43d & 0.0   & 2.2 & 1.15 (S-1) & 219.4 (S-1) & 14.3 (S-0.1) & 13.4 \cite{La2C3} \\
       \hline
       Sc$_2$C$_3$ &I-43d & 0.11  & 3.6 & 1.435 (P-30) & 303.18 (P-30) & 27.90 (P-30) & - \\
        & &   &  & 1.99 (S-0.1) & 205.6 (S-0.1) & 25.5 (S-0.1)  & - \\
       \hline
       Al$_2$Mo$_3$C & P4\_132& 0.05  & 0.6 & 1.19 (S-0.1) & 174.95 (S-0.1) & 12.05 (S-0.1) & 9.2 \cite{KXVBAYTCKP10}\\
       \hline
       YBC & Cmmm& 0.42  & 150 & 0.73 (P-20) & 454.42 (P-20) & 10.15 (P-20) &  -  \\
        \hline
       W$_2$B & I4/mcm & 0.0   & 0.6 & 0.81 (D) & 215.68 (D) & 6.61 (D) & 3.10 \cite{HH54} \\
        \hline
       Mo$_2$B &I4/mcm &0.03  & 3.6 & 0.79 (D) & 284.65 (D) & 8.12 (D) & 4.74 \cite{HH54}\\
        \hline
       Ta$_2$B &I4/mcm  & 0.03 & 8.9 & 0.44 (D) & 236.84 (D) & 0.34 (D) & 3.12 \cite{HH54} \\
        \hline
       MoB$_4$ & P6/mmm & 0.26  & 0.7 & 0.69 (D) & 418.42 (D) & 7.58 (D) & - \\
       \hline
       La$_3$InC & Pm-3m& 0.0 & 0.6 & 1.04 (D) & 108.63 (D) & 5.844 (D)  & 2.6 \cite{ZDVOC95}  \\
              \hline
        La$_3$InB & Pm-3m &0.20   & 2.3 & 1.18 (D) & 83.03 (D) & 5.60 (D) & 10 \cite{ZDVOC95}\\
         \hline
    \end{tabular}
    }
    \label{tab:distort}
\end{table}

 Figure~\ref{fig:7}(m)-(p) displays the $\lambda$-projected phonon dispersion for four different stabilized compounds: Sc$_2$C$_3$ (P-30), Ta$_2$B (D), La$_3$InB (D), and MoB$_2$ (S-0.1), utilizing pressure of 30 GPa, distortion (D), distortion (D), and smearing of 0.1 Ry, respectively. By comparing these plots with Figs.~\ref{fig:7}(i)-(l), we can see that the soft optical phonon modes are stabilized and contribute significantly to $\lambda$ near the $\Gamma$-point, represented by green open circles. Despite the slight lifting of phonon band degeneracy caused by distortion, the large contribution to EPC remains, which give rise to SC. The discovery of such systems is interesting as it presents opportunities for stabilization through pressure and alloying, leading to potentially high T$_c$ metastable compounds that may be synthesizable in experiment.

 For the compounds with imaginary phonon modes away from the $\Gamma$ point, i.e. MoB$_2$-type, we also stabilized these dynamically unstable compounds with larger electronic smearing of 0.1 Ry and tabulated the results in Table.~\ref{tab:Neg_MoB2}. For example, compounds like MoB$_2$ in the MgB$_2$ structure have shown experimentally measured T$_c$ of 32 K under high pressure \cite{PZWZGG23}. { Moreover, we demonstrate that large electronic smearing can play a role similar to pressure in stabilizing imaginary phonon modes. The pressure-induced stabilization of MoB$_2$ is presented in Supplementary Fig. S4. Our calculated structural phase transition from R-3m to P6/mmm phase at 75 GPa agrees well with the experimentally reported critical pressure of 70 GPa \cite{PZWZGG23}. The calculated electron–phonon coupling (EPC) at 75 and 90 GPa yields T$_c$ values of 28.4 K and 28.6 K, respectively, which are consistent with the experimental value of 32 K, as well as with results obtained from the electronic smearing approach used for high-throughput calculations. Notably, both approaches give comparable EPC parameters for the same stabilized low-lying phonon modes: at 75 GPa we obtain $\lambda$ = 1.85, $\omega_{log}$ = 244 K, and T$_c$ = 28.4 K, while the smearing-based approach (Table III) gives $\lambda$ = 2.12, $\omega_{log}$ = 209 K, and T$_c$ = 27.3 K.} Another notable example in Table~\ref{tab:Neg_MoB2} is Ca$_5$B$_3$N$_6$ with $\Delta E_h$ of 0.03 eV/atom, which exhibits dynamical instability at the $H$ point and displayed a significant mode-resolved $\lambda$ near its unstable phonon modes, as depicted in Fig.~\ref{fig:9} (d)-(e). After stabilizing the system with electronic smearing of 0.06 Ry, Ca$_5$B$_3$N$_6$ exhibits a $\lambda$ of 1.5. The $\omega_{log}$ is found to be 372 K. Moreover, when computing $\lambda$ (with broadening parameter $\sigma$ = 0.01 Ry), the T$_c$ is determined to be 35 K with Coulomb potential $\mu^*_c$=0.16 and 42.4 K with $\mu^*_c$=0.10, much larger than that of MgB$_2$ (16-20 K) computed from isotropic approximation. It should be noted that Ca$_5$B$_3$N$_6$ has been synthesized in cubic structure and Im3$^\prime$m (229) SG with partial occupancy in the 8c site of Ca \cite{WZHN98}. The crystal structure of this boronitride (Fig.~\ref{fig:9}(a)-(c)) has a cage-like structure, similar to XB$_3$C$_3$ borocarbides (X $=$ Ca,Ba,Sr,Y,La). For the stoichiometric Ca$_5$B$_3$N$_6$, Fig.~\ref{fig:9}(f) and (g) present the isotropic Eliashberg spectral function and electronic band structure, respectively. It shows an electron-doped band structure that connects to strong EPC as also found in other predicted SC compounds and studies. As listed in Table III, interestingly some of the trigonal compounds of NbMoC$_2$ are in the same structure as the stable TaNbC$_2$ in Table~\ref{tab:10K}. This shows that in this particular structure, with the substitution of neighboring group of early TMs, although bringing dynamical instability, the phonons once stablized can still provide a large EPC contribution for a sizable T$_c$. As also listed in Table III, other compounds with sizable predicted T$_c$ after stabilization that also near GS hull are some notable ternary Ru compounds,  MoRuB$_2$ at 15.6 K, RuVB$_2$ at 15.0 K, Pd$_3$CaB at 7.0 K and RuSc$_3$C$_4$ at 6.6 K.

\begin{table}[h!]\centering
        \caption{DFPT calculated EPC results for MoB$_2$-type instability (outside of $\Gamma$), stabilized with electronic smearing of 0.10 Ry; SG is spacegroup, and $\Delta E_h$ is $\Delta E_h$ taken from MP database. Experimental results for some high-pressure T$_c$ are also reported.}
         \scalebox{1.2}{%
    \begin{tabular}{|c|c|c|c|c|c|c|}
    \hline
Compound&SG&$\Delta E_h$ (eV/atom) \cite{JOHCRD13_MP} &$\lambda$&$\omega_{log}$ (K)&T$_c$ (K) & T$^{Expt}_c$ (K) \\
\hline
Ca$_5$B$_3$N$_6$ &Im3$^\prime$m&0.03&1.5&372&35.0 \footnote{Stabilized with electronic smearing of 0.06 Ry}& \\
MoB$_2$&P6/mmm&0.156&2.12&209&27.3&32\cite{PZWZGG23}\\
NbMoC$_2$&R-3m&0.172&1.72&215&23.5&\\
TaMo$_2$C$_3$&P-3m1&0.197&2.01&171&21.4&\\
TaMoC$_2$&R-3m&0.15&1.69&189&20.3&\\
WVC$_2$&R-3m&0.225&1.31&249&19.8&\\
TaTiWC$_3$&P3m1&0.119&1.42&213&18.9&\\
ScC&P6$_3$/mmc&0.6&0.95&410&18.5&\\
TaWC$_2$&R-3m&0.211&2.04&129&16.3&\\
MoRuB$_2$&Pmc2$_1$&0.09&1.29&202&15.6&\\
RuC&P-6m2&0.649&1.02&288&15.1&\\
RuVB$_2$&Pmc2$_1$&0.073&1.09&255&15.0&\\
Ni$_3$AlC&Pm-3m&0.166&1.86&116&13.6&\\
Nb$_4$C$_3$&Pm-3m&0.167&0.99&267&13.2&\\
Ta$_4$C$_3$&Pm-3m&0.134&0.99&203&10.1&\\
RhC&F-43m&0.561&0.91&228&9.3&\\
MoC&F-43m&0.585&0.97&191&9.0&\\
Pd$_3$CaB&Pm-3m&0.0&1.23&96&7.0&\\
RuSc$_3$C$_4$&C2/m&0.0&0.72&321&6.6&\\
Ta$_2$S$_2$C&P-3m1&0.007&0.69&268&5.1&\\
Mo$_3$ZrB$_2$C$_2$&Amm2&0.046&0.64&282&4.0&\\
HfC&P-6m2&0.763&1.06&56&3.2&\\
\hline

\end{tabular}
}
\label{tab:Neg_MoB2}
\end{table}

\section{Discussion}
 In this work, we have employed the machine learning (ML) models that utilize the data generated from \textit{ab-initio} calculations using the DFPT and the isotropic Eliashberg approximation to iteratively guide the search for new phonon-mediated superconductors among B and C compounds. Our study also focuses on addressing the challenges encountered during DFPT calculations, such as convergence of Brillouin zone (BZ) sampling and the problem of calculated dynamic instability. To address the convergence issue, we developed an ansatz test to verify the convergence of superconductivity (SC) critical temperature \( T_c \). This test uses the variation of \( T_c \) with respect to Gaussian broadening to compute the double delta summation. For dynamically unstable compounds, we applied large electronic smearing, lattice distortion, and pressure to stabilize imaginary phonon modes. We then calculated their EPC properties and incorporated these into the ML models. Between the two ML models, ALIGNN consistently outperforms CGCNN in predicting EPC properties especially after including the stabilized compounds with imaginary phonon and dynamical instability. Our ML-guided search demonstrates promising predictability for T$_c$ values. For example, we predict SC in compounds with calculated dynamically stability such as TaNbC$_2$ (28.4 K), Nb$_3$B$_3$C (16.4 K), Y$_2$B$_3$C$_2$ (4.0 K), among others. In addition to studying dynamically stable compounds, we also focus on compounds with calculated dynamic instability, an area that, to our knowledge, has not been systematically explored before. We predicted SC in compounds showing dynamic instability such as Ca$_5$B$_3$N$_6$ (35 K), Pd$_3$CaB (7.0 K), some ternary Ru compounds, MoRuB$_2$ (15.6 K), RuVB$_2$ (15.0 K), Pd$_3$CaB (7.0 K) and RuSc$_3$C$_4$ (6.6 K) with $\Delta E_h$ mostly below 0.1 eV/atom. With further refinement and larger dataset, our workflow can be improved in accurately predicting more SC compounds. In this regard, identifying metastable compounds with calculated dynamic instability, where soft phonon exhibit significant EPC contribution, plays a crucial role.

\section{METHODS}
\subsection{Convergence with respect to Brillouin zone sampling (k-point mesh)}
To identify cases of convergence failure (i.e., incorrect prediction of SC/NSC) related to Brillouin zone sampling, we analyzed the variation of T$_c$ with respect to Gaussian broadening ($\sigma$) in the double delta integration and developed a simple ansatz based on the converged results of MgB$_2$, as described in the ``Convergence tests for MgB$_2$ and AlB$_2$ (Supplementary Figures S5$-$S10, Supplementary Tables S5$-$S6)" and "Convergence ansatz (Supplementary Figures S11$-$S14, Supplementary Tables S7$-$S9)" sections of the Supplemental Material (SM). This ansatz involves extracting T$_c$, similar to MgB$_2$, and estimating the decay parameter (A) in the exponential variation of T$_c$ with $\sigma$, 
\begin{equation}
    T_c = \exp(-A\sigma^{1/3} + B)
\end{equation}
where A is the variable that quantifies the rate of exponential decay, and B is the constant associated with Debye temperature. Unconverged results show larger values of A, which decrease with denser \textbf{k}-meshes. With larger \textbf{k}- or \textbf{q}- grids, the T$_c$ vs $\sigma$ curve becomes less steep. Our convergence analysis of MgB$_2$ suggests that A$_{MgB_2}$ = 12$-$13 can be used as a threshold for this study: calculations are considered as unconverged for A $>$ A$_{MgB_2}$, requiring a denser \textbf{k}-mesh, while those with A $<$ A$_{MgB_2}$ can be regarded as converged for accurate prediction.

\subsection{Computational details}
 Data extraction from the MP database, as well as input preparation for ground state calculations, calculation submission, result extraction and input preparation for ML studies, and plotting, were performed using the high-throughput electronic structure package (HTESP) \cite{htesp}. A script, `fitting\_elph\_smearing.py`, is included in the HTESP package to compute the decay parameter from \(T_c\) vs. broadening \(\sigma\) data. Ground-state DFT and EPC calculations were performed using the QE code \cite{QE09,QE217}. Ultrasoft or norm-conserving pseudopotentials (PP) from the efficiency standard solid-state pseudopotentials (SSSP) dataset \cite{SSSP18} were employed, with the replacement of projector-augmented wave (PAW) PPs by GBRV ultrasoft norm-conserving high-throughput PPs \cite{GBRV14}. The exchange-correlation energy was approximated using the Perdew-Burke-Ehrnzerhof (PBE) generalized-gradient approximation (GGA) \cite{PBE96}. The Brillouin zone was sampled using a k-point mesh from the MP database to compute the ground-state charge density for EPC calculations. The q-point mesh required for EPC calculations was obtained by halving the k-point mesh ($\textbf{q} = \textbf{k}/2$), with odd k-points changed to even by adding 1 to it. The structures were fully relaxed using Broyden–Fletcher–Goldfarb–Shanno (BFGS) minimization \cite{bfgs} until the total energy and forces for ionic minimization converged within 10$^{-5}$ Ry and 10$^{-4}$ Ry/Bohr, respectively. The self-consistent electronic energy and charge density were minimized with a convergence threshold of 10$^{-12}$ Ry. Similarly, the SCF convergence for phonon calculations is achieved with an energy cutoff of \(10^{-14}\). Starting from the default \(\alpha_{\text{mix}}\) value of 0.7, it is reduced to 0.3 if the EPC calculation does not converge. A fine k-point grid, twice the size of the k-point mesh used for charge density convergence, was used for interpolating EPC matrix elements to compute double-delta integration and the $\lambda$. Gaussian smearing of 0.02 Ry was applied for charge-density optimization. To compute $\lambda$ for various $\sigma$ values, we computed double-delta integration using 10 broadening ($\sigma$) values ranging from 0.005 to 0.05 Ry for the \textbf{k}-mesh. For the \textbf{q}-mesh integration, a fixed smearing of 0.5 meV was employed. The reported results were obtained for $\sigma = 0.01$ Ry with a Coulomb potential $\mu^*_c$ of 0.16.

\subsection{Limitations}
{When evaluating the significance of machine-learning (ML) models trained on ab-initio data, it is essential to recognize both their capabilities and inherent limitations. First, the performance of any ML model is strongly dependent on the quality, diversity, and representativeness of its training dataset. Limited or biased datasets, which are common in computational materials science, can restrict a model’s predictive accuracy and transferability. Second, ML
models are susceptible to overfitting, particularly when trained on small or noisy datasets, although careful validation procedures during model development can mitigate this issue to some extent. Third, the fidelity of ML predictions is ultimately bounded by the accuracy of the underlying ab-initio methods used to generate the training data. For example, the widely used isotropic Eliashberg approximation in superconductivity calculations neglects the inherently anisotropic and multi-band character of electron–phonon coupling. Although fully solving the anisotropic Migdal-Eliashberg equations (e.g., using EPW) yield more accurate estimates of superconducting properties, they are not computationally practical for large scale data generation. Consequently, ML models trained on data obtained from the isotropic Eliashberg approximation can function as efficient screening tools to identify promising candidate materials, which may then be examined with more accurate, anisotropic, and computationally demanding methods.}

\section{Data availability}
{ Data computed and utilized in this study are tabulated in Supplementary Tables S10$-$S17, Tables.~\ref{tab:distort} and \ref{tab:Neg_MoB2}. The raw data can be downloaded from the Figshare platform \cite{rawdata}.}

\section{Acknowledgements}
We thank Dr. Paul C. Canfield for the funding support, initiating the idea of searching for new phonon-mediated superconductors among boron and carbon compounds, and the helpful discussion throughout the project.

\section{Funding}
This work was supported by Ames National Laboratory LDRD and U.S. Department of Energy, Office of Basic Energy Science, Division of Materials Sciences and Engineering. Ames National Laboratory is operated for the U.S. Department of Energy by Iowa State University under Contract No. DE-AC02-07CH11358.

\section{Author information}
\subsection{Authors and Affiliations}
\textbf{Ames National Laboratory, Ames, Iowa 50011, USA}\\

Niraj K. Nepal, Lin-Lin Wang\\

\textbf{Department of Physics and Astronomy, Iowa State University, Ames, Iowa 50011, USA}\\

Lin-Lin Wang
\subsection{Contributions}
L.-L.W. conceived and supervised the work. N.K.N. and L.-L.W. designed and performed the high throughput calculations with the machine learning guided approach. N.K.N. developed the ansatz to test the convergence of Tc calculation. All authors discussed the results and contributed to the final manuscript.

\subsection{Corresponding author}
Correspondence to Niraj K. Nepal (nepalneeraz@gmail.com)

\section{Competing interests}
The authors declare no competing interests.

\newpage
\begin{figure}[h]\centering
    \includegraphics[scale=0.3]{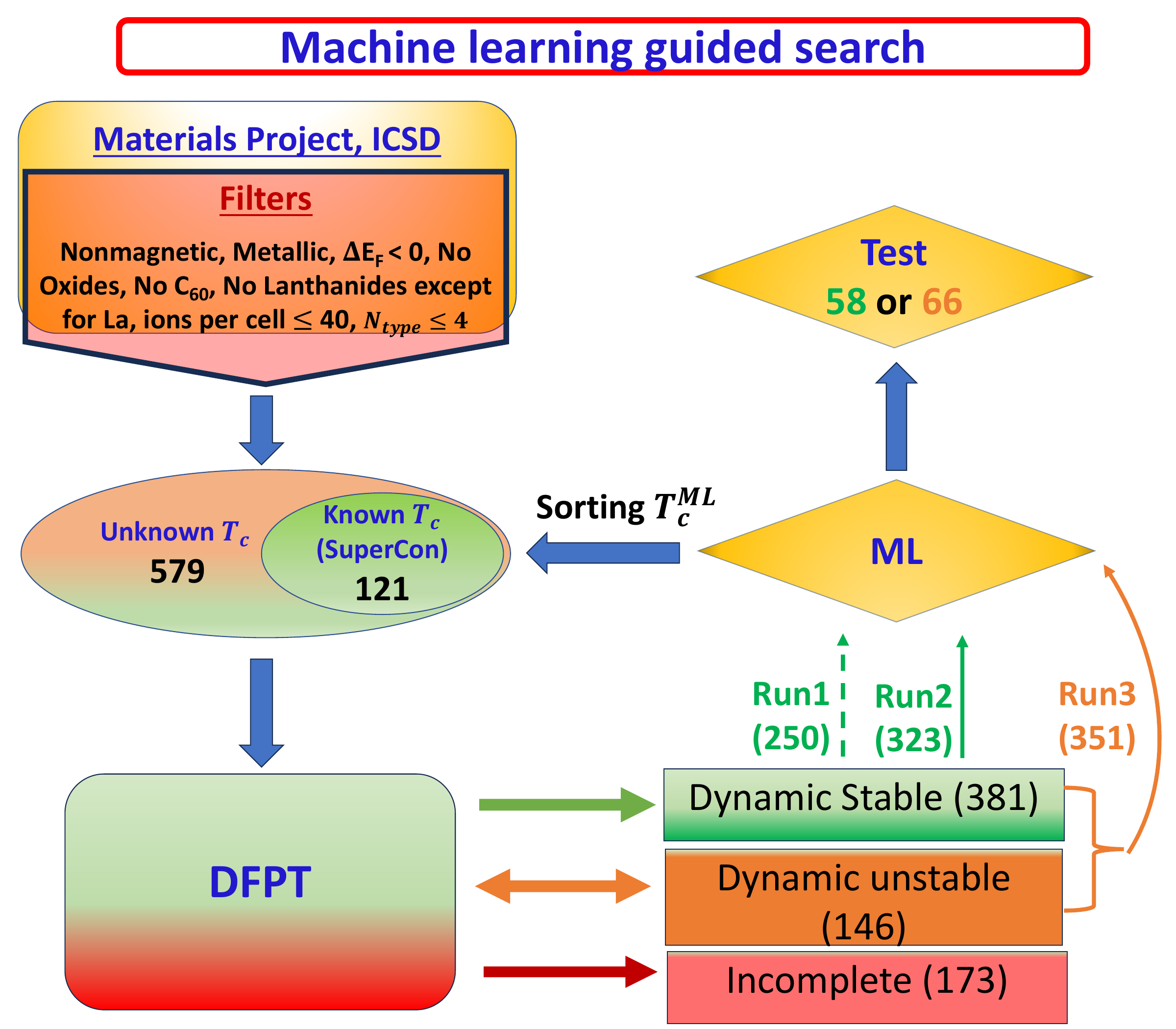}
    \caption{Machine learning workflow. The numbers are the counts of compounds involved in each step. See main text for the detailed summary.}
    \label{fig:3}
\end{figure}

\newpage
\begin{figure}[h!]\centering
    \centering
    \includegraphics[scale=0.3]{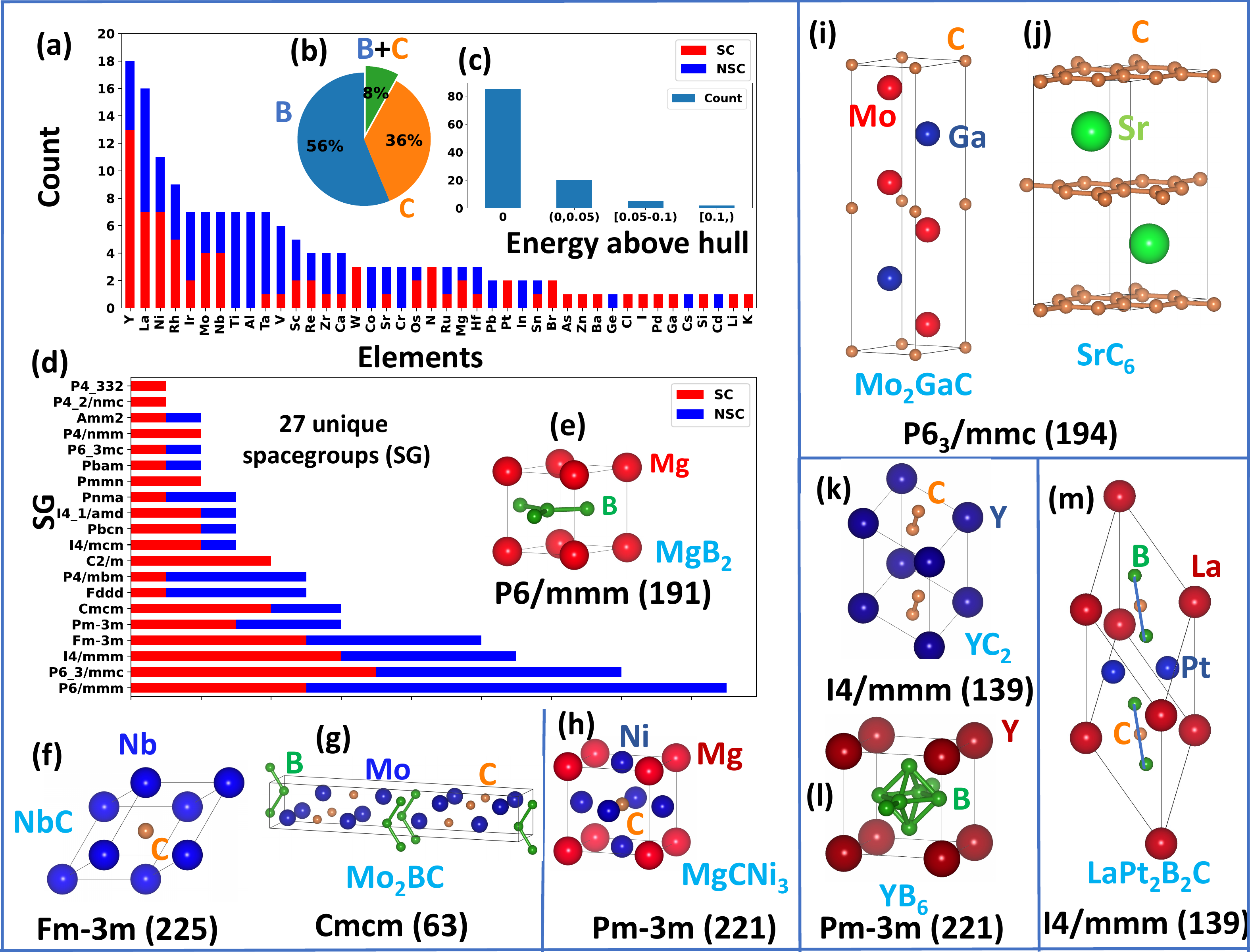}
    \caption{Statistical description of dynamically stable 113 compounds, whose (non-) superconductivity have been experimentally measured. (a) Number of compounds according to elements, each bar is partitioned into two color segments, with red representing the number of superconductors (SC), while blue denote non-superconductors (NSC).; (b) Proportion of boron and carbon compounds (c) Histogram for energy above the convex hull in (eV/atom), (d) Distribution according to space group, (e) Crystal structure of MgB$_2$. (f)-(m): Crystal structures of known superconductors. (f) NbC (10.03 K)\cite{WBM67}, (g) Mo$_2$BC (7.5 K) \cite{ECCMHV16}, (h) MgCNi$_3$ (8 K)\cite{HHSI01_MgNi3C}, (i) Mo$_2$GaC (3.9 K)\cite{T67_Mo2GaC}, (j) SrC$_6$ (1.65 K)\cite{KBORK07_SrC6}, (k) YC$_2$ (3.89K)\cite{BJKSOS12}, (l) YB$_6$ (7.2 K)\cite{LWTAMP06}, and (m) LaPt$_2$B$_2$C (10 K)\cite{CBSKPCFTV94}.}
    \label{fig:1}
\end{figure} 

\newpage
\begin{figure}[h!]\centering
    \includegraphics[scale=0.45]{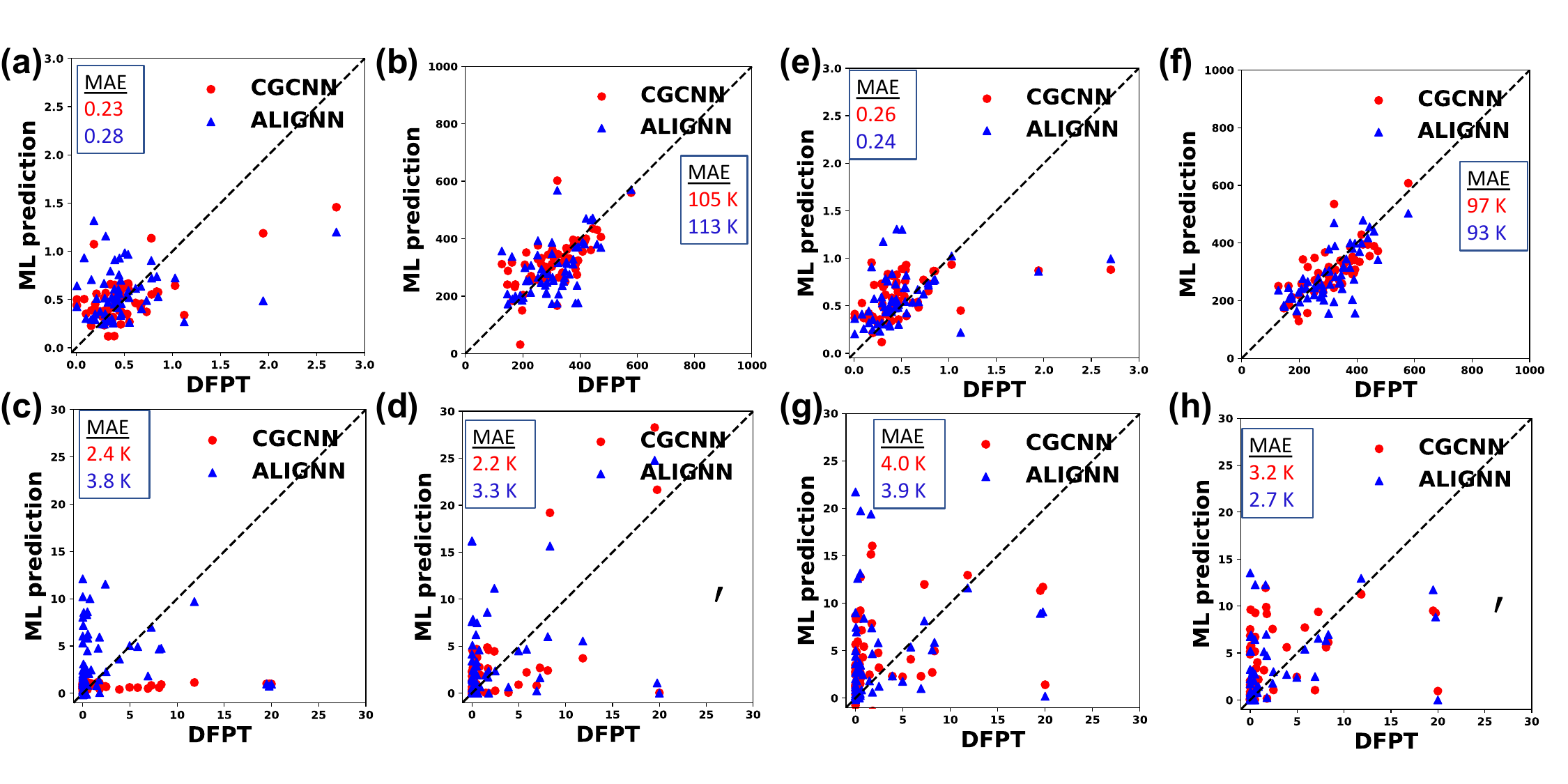}
    \caption{Graph neural network-based regression models for dynamically stable compounds. Predicting $\lambda$, $\omega_{log}$, T$_c$, and T$^{\prime}_c$ using CGCNN (red circles) and ALIGNN (blue upper triangle) models for the independent test set in comparison to the DFPT-calculated results in Run 1 (left panel, (a)-(d)) and Run 2 (right panel, (e)-(h)). Run 1 uses 250 training samples (45 space groups) and 58 test samples (27 space groups). Run 2 employs 323 training samples (54 space groups), with the test set unchanged. Similarly,  Mean absolute error (MAE) are presented in inset. Red and blue arrows show the clustering pattern, discussed in the main text.}
    \label{fig:4}
\end{figure}

\newpage

\begin{figure}[h!]\centering
    \includegraphics[scale=0.45]{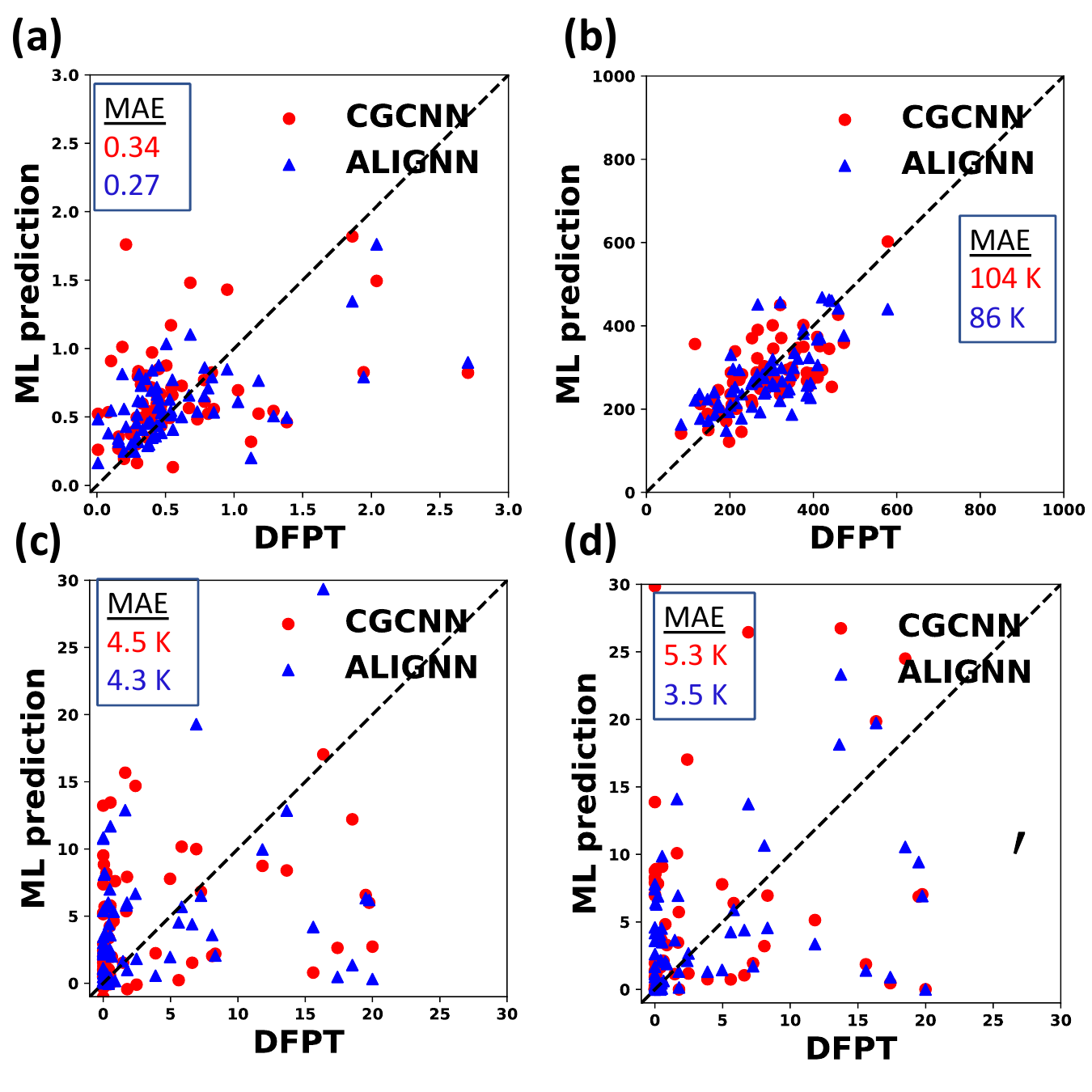}
    \caption{Graph neural network-based regression models similar to Fig. 3, but including both stable and stabilized unstable compounds. Predicting $\lambda$, $\omega_{log}$, T$_c$, and T$^{\prime}_c$ using CGCNN (red circles) and ALIGNN (blue upper triangle) models for the independent test set including both stable and dynamically unstable cases in comparison to the DFPT-calculated results in Run3. Run 3 employs 351 training samples (56 space groups) and 66 test samples (27 space groups). MAEs are presented in inset.}
    \label{fig:5}
\end{figure}

\newpage
\begin{figure}[h!]\centering
    \includegraphics[scale=0.75]{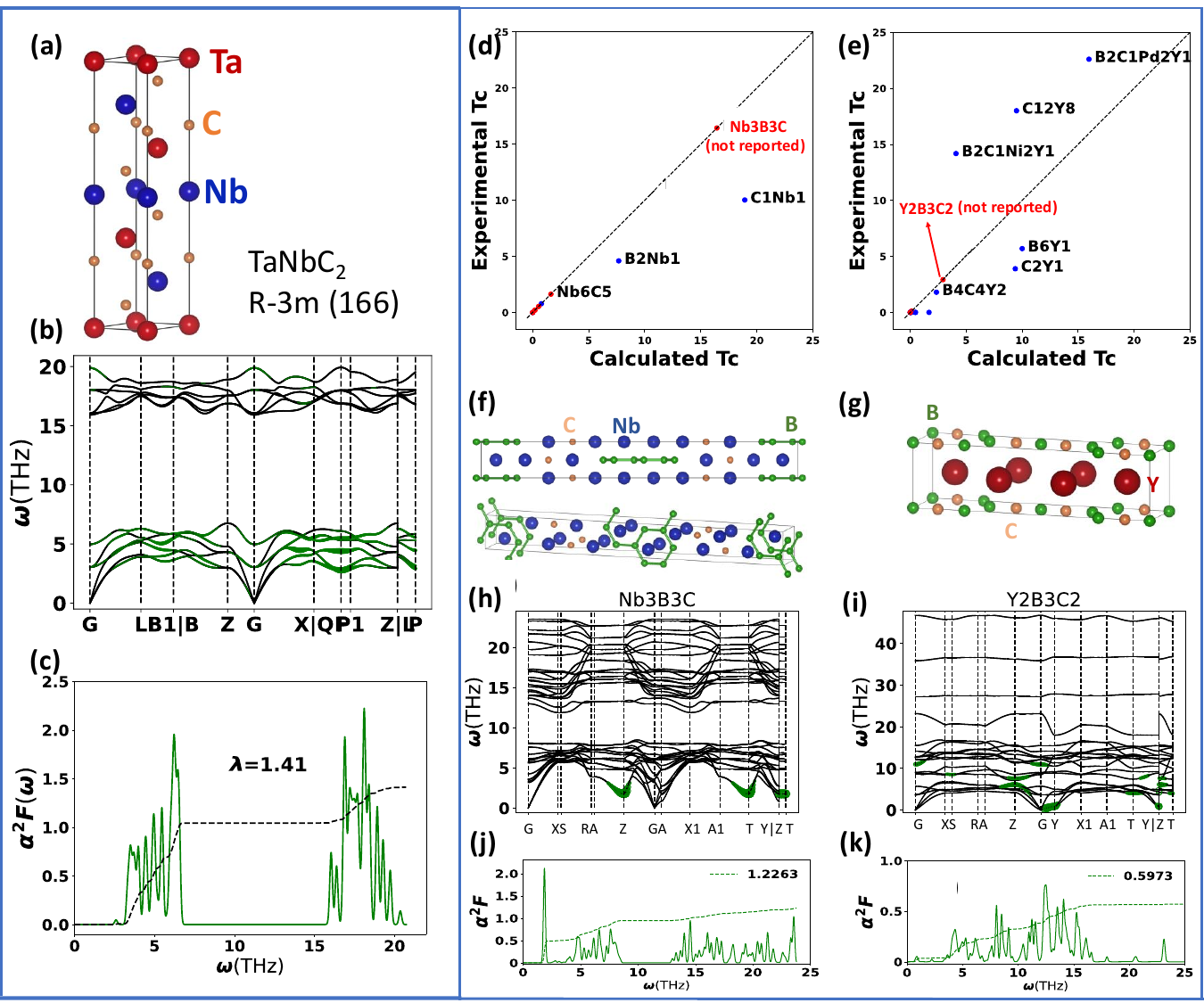}
    \caption{DFPT calculated EPC properties of a few dynamically stable compounds. (a) Crystal structure, (b) phonon dispersion, and (c) Eliashberg isotropic spectral function of TaNbC$_2$, respectively. (d)-(k) X-B-C compounds (X=Nb,Y): Comparison between DFPT computed T$_c$ values with experiments for (d) Nb-B-C and (e) Y-B-C systems. {Experimentally known results are represented by blue circle, while red circles are not reported ones shown along the y$=$x dashed line as labeled.} Crystal structures respectively for (f) Nb$_3$B$_3$C, and (g) Y$_2$B$_3$C$_2$. Atoms are highlighted by the colorized symbols. Phonon dispersion projected with mode-resolved $\lambda$ (green open circles) for (h) Nb$_3$B$_3$C, and (i) Y$_2$B$_3$C$_2$ respectively. Eliashberg spectral functions for (j) Nb$_3$B$_3$C, and (k) Y$_2$B$_3$C$_2$ respectively.}
    \label{fig:6}
\end{figure}

\newpage
\begin{figure}[h!]\centering
    \includegraphics[scale=0.8]{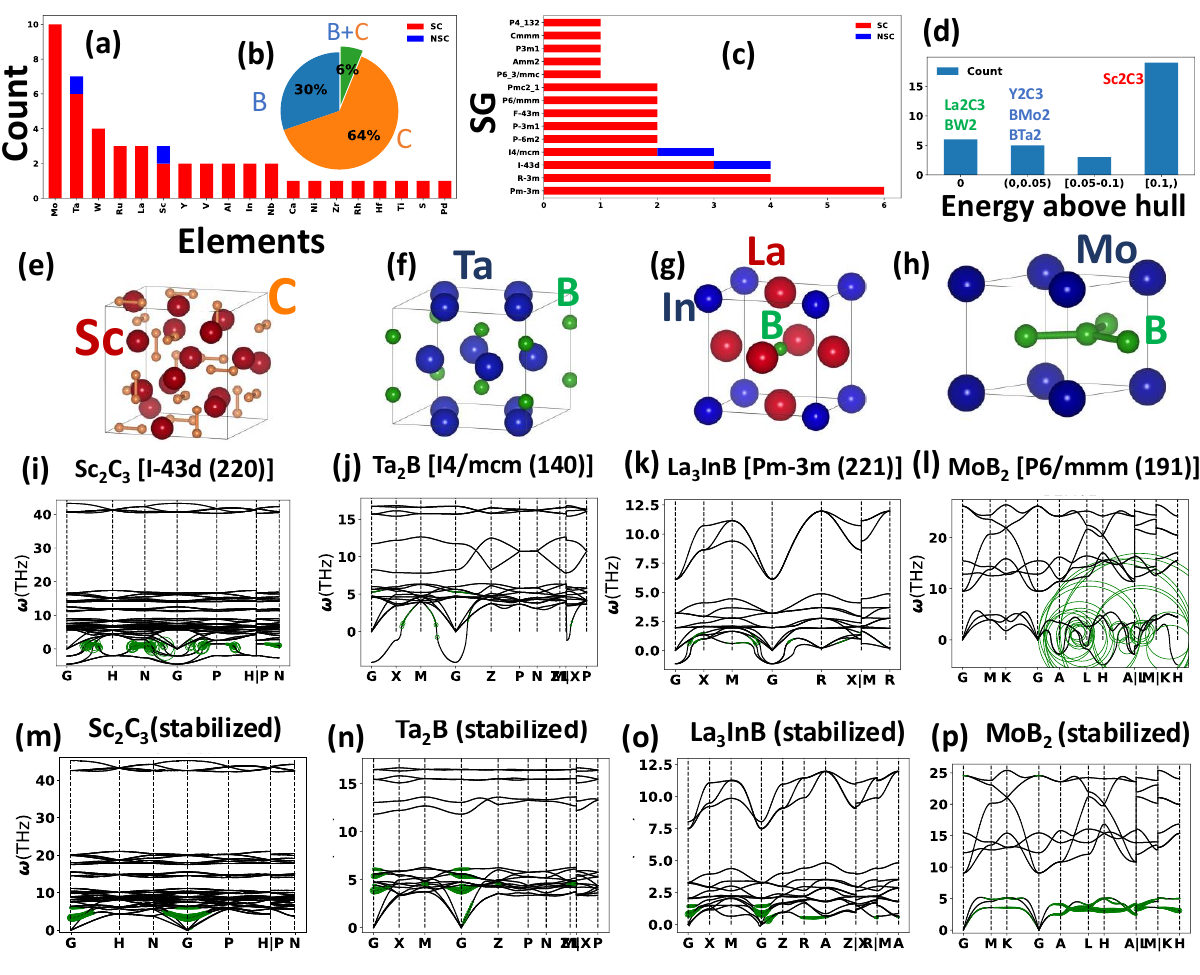}
    \caption{Imaginary frequency modes with large EPC on the soft modes. Fig. (a) and (b) represent the count of different elements and allocation of B/C/B+C compounds in dynamically unstable systems respectively. Statistics according to spacegroup(c) and (d) energy above the convex hull. (e)-(h) { Crystal structures of Sc$_2$C$_3$, Ta$_2$B, La$_3$InB, and MoB$_2$ are shown, respectively. (i)-(l) display the phonon dispersion plots of the dynamically unstable systems corresponding to Sc$_2$C$_3$, Ta$_2$B, La$_3$InB, and MoB$_2$. First 3 plots represent the instability of soft phonon modes at $\Gamma$-point, while the last one corresponds to the instability outside of $\Gamma$. (m)-(p): EPC results for compounds with stabilized imaginary phonon modes. EPC projected (highlighted by green circles) phonon band dispersions for Sc$_2$C$_3$ (P-30), Ta$_2$B (D), La$_3$InB (D), and MoB$_2$ (S-0.1) respectively.}}
    \label{fig:7}
\end{figure}

\newpage
\begin{figure}[h!]\centering
    \includegraphics[scale=0.7]{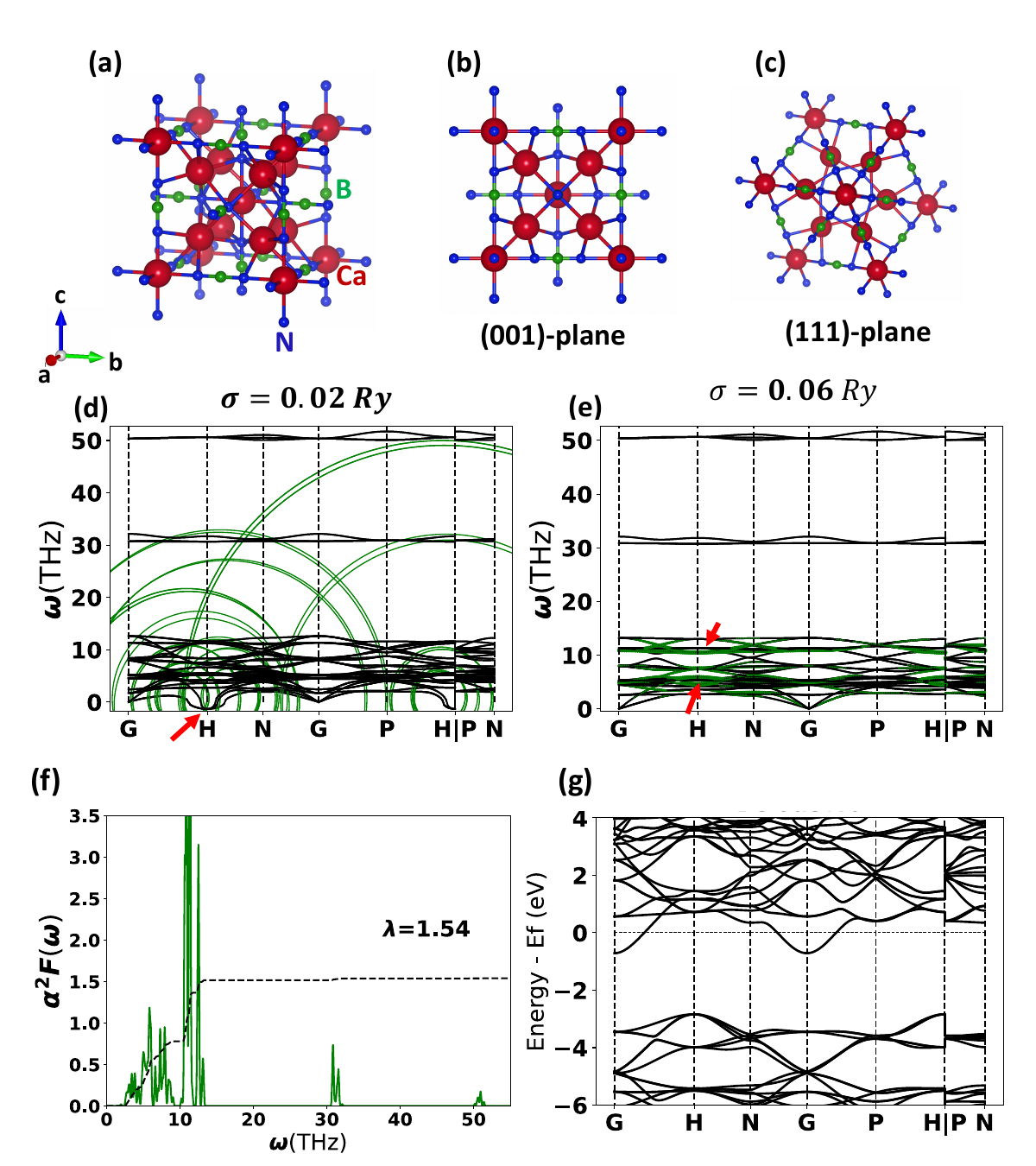}
    \caption{Structures and EPC properties of Ca$_5$B$_3$N$_6$. (a)-(c): Crystal structures of Ca$_5$B$_3$N$_6$, viewed from different directions (d)-(e): EPC results for compounds with stabilized imaginary phonon modes for Ca$_5$B$_3$N$_6$. (d): EPC projected (highlighted by green circles) phonon band dispersions for dynamically unstable systems represented by small well at H-point (shown by red arrow), (e): Similar results for systems, stabilized with large electronic smearing of 0.06 Ry. (f) Eliashberg spectral function (g) Electronic band structure}
    \label{fig:9}
\end{figure}

\clearpage
\newpage
\section*{Supplementary Materials for ``Machine-learning Guided Search for Phonon-mediated Superconductivity in Boron and Carbon Compounds"}

\author{Niraj K. Nepal$^1$}
\author{Lin-Lin Wang$^{1,2}$}
\affiliation{[1] Ames National Laboratory, Ames, Iowa 50011, USA}
\affiliation{[2] Department of Physics and Astronomy, Iowa State University, Ames, Iowa 50011, USA}

\date{\today}

\maketitle

\section{Statistics for compounds with unknown T$_c$}
In Fig.~\ref{fig:S1}, we present the descriptions of 268 dynamically stable compounds for which the experimental T$_c$ is not known. Figure~\ref{fig:S1} (a)-(c) illustrate the distribution of different elements in these systems, the allocation of B/C/B+C compounds, and the deviation of their formation energy ($\Delta E_h$) from stable structures, respectively. Most unknown compounds are associated with transition metals (TM) such as Nb, Ta, Mo, V, Y, and others. Approximately 80\% of the structures are close to the ground state convex hull with $\Delta E_h \leq 0.05$ eV/atom, while the remaining 20\% have $\Delta E_h$ greater than 0.05 eV/atom. Panel~\ref{fig:S1}(d) showcases the distributions of these compounds based on their space groups. In Fig.~\ref{fig:S1}(a) and (d), each bar is partitioned into segments, with the segments colored in red representing the number of SC, while the segments in blue denote the NSC. The overall description is similar to that of known compounds, as shown in Fig. 2 of the main text.
\begin{figure}[h!]\centering
\renewcommand{\thefigure}{S1}
    \includegraphics[scale=0.3]{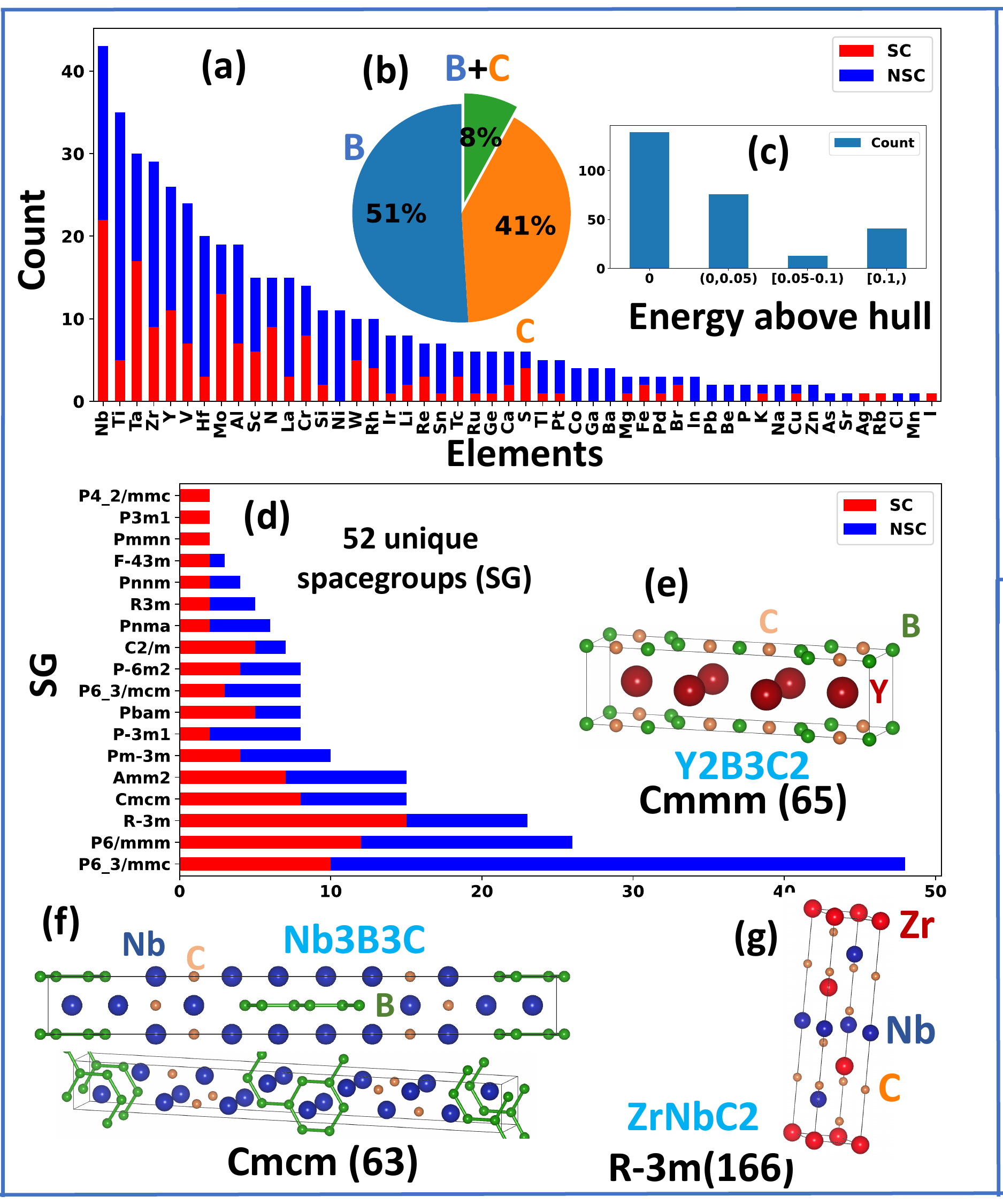}
    \caption{Statistical description of dynamically stable materials, whose (non-) superconductivity hasn't been experimentally measured. Each bar is partitioned into segments, with the segments colored in "red" representing the number of superconductors (SC), while segments in "blue" denote nonsuperconductors (NSC).; (a) Number of compounds according to elements, (b) Proportion of boron and carbon compounds (c) Statistics in terms of energy above the convex hull, (d) Distribution according to spacegroups, (e)-(g): Crystal structures of some potential superconductors with their respective spacegroups.}
    \label{fig:S1}
\end{figure} 

\begin{figure}[H]
    \centering
    \renewcommand{\thefigure}{S2}
    \includegraphics[scale=0.52]{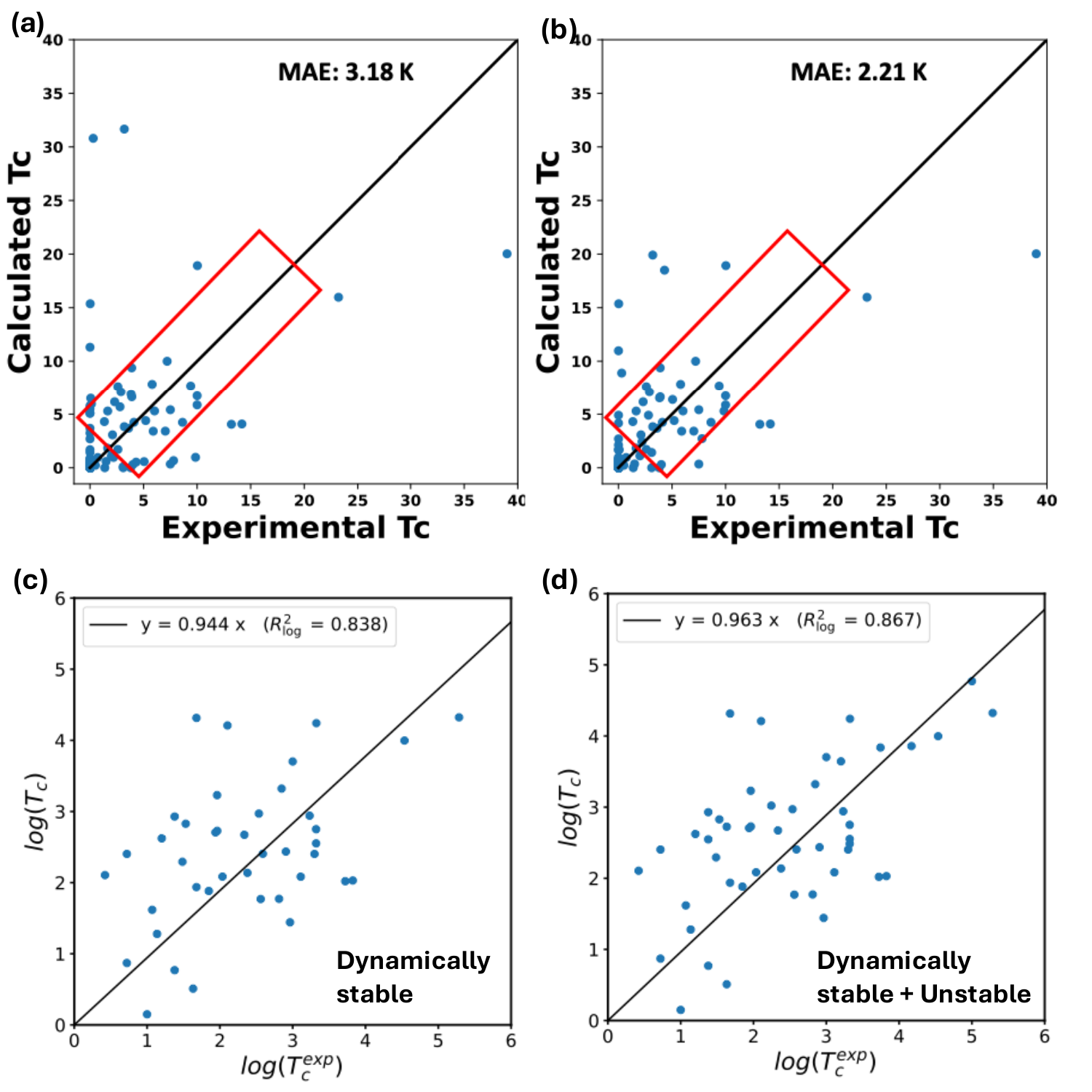}
    \caption{Comparison of Predicted and Experimental Critical Temperatures Under k-Grid Refinement and Dataset Expansion. (a) Comparing calculated and experimental critical temperature obtained from efficient calculations; Some largely deviated results are not enclosed within red rectangular box. (b) Same plot with improved results with denser \textbf{k}- grids for compounds with  A $>$ A$_{MgB_2}$. Red rectangular box on the right plot, enclosed more data points along the y = x axis (black solid line), compared to plot on the left. {(c) Predicted $\log(T_c)$ vs.\ measured $\log(T^{exp}_c)$ for dynamically stable superconductors.  
(d) Same comparison after incorporating additional data obtained by stabilizing dynamically unstable compounds. The inclusion of these superconductors improves the coefficient of determination $R^2_{log}$ score from 0.838 to 0.867.}
}
    \label{fig:s11}
\end{figure}

\begin{figure}[H]
    \centering
    \renewcommand{\thefigure}{S3}
    \includegraphics[scale=0.5]{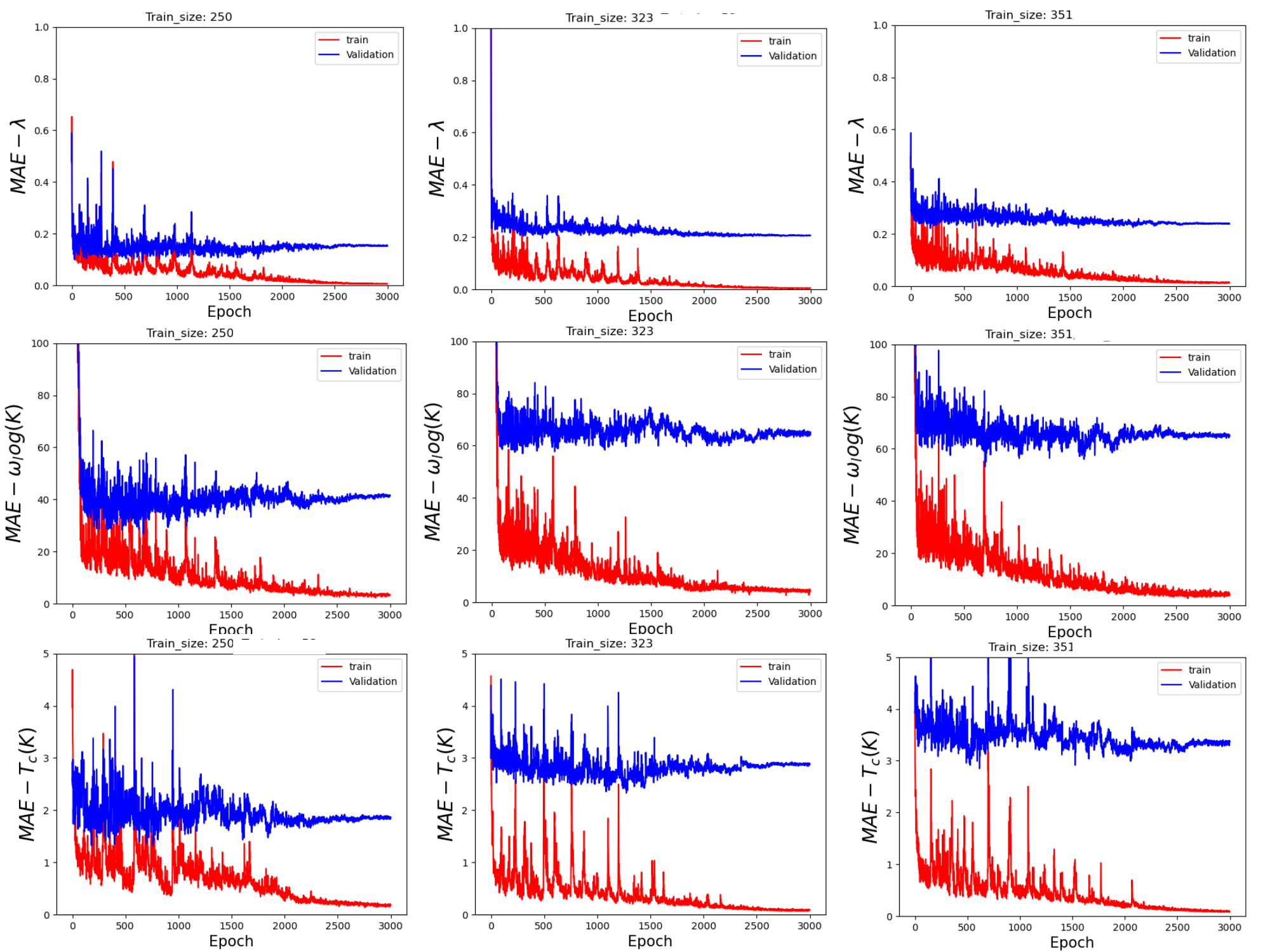}
    \caption{{Learning curves for various training and validation during the ALIGNN model training process.}}
    \label{fig:s12}
\end{figure}

\begin{figure}[H]
    \centering
    \renewcommand{\thefigure}{S4}
    \includegraphics[scale=0.5]{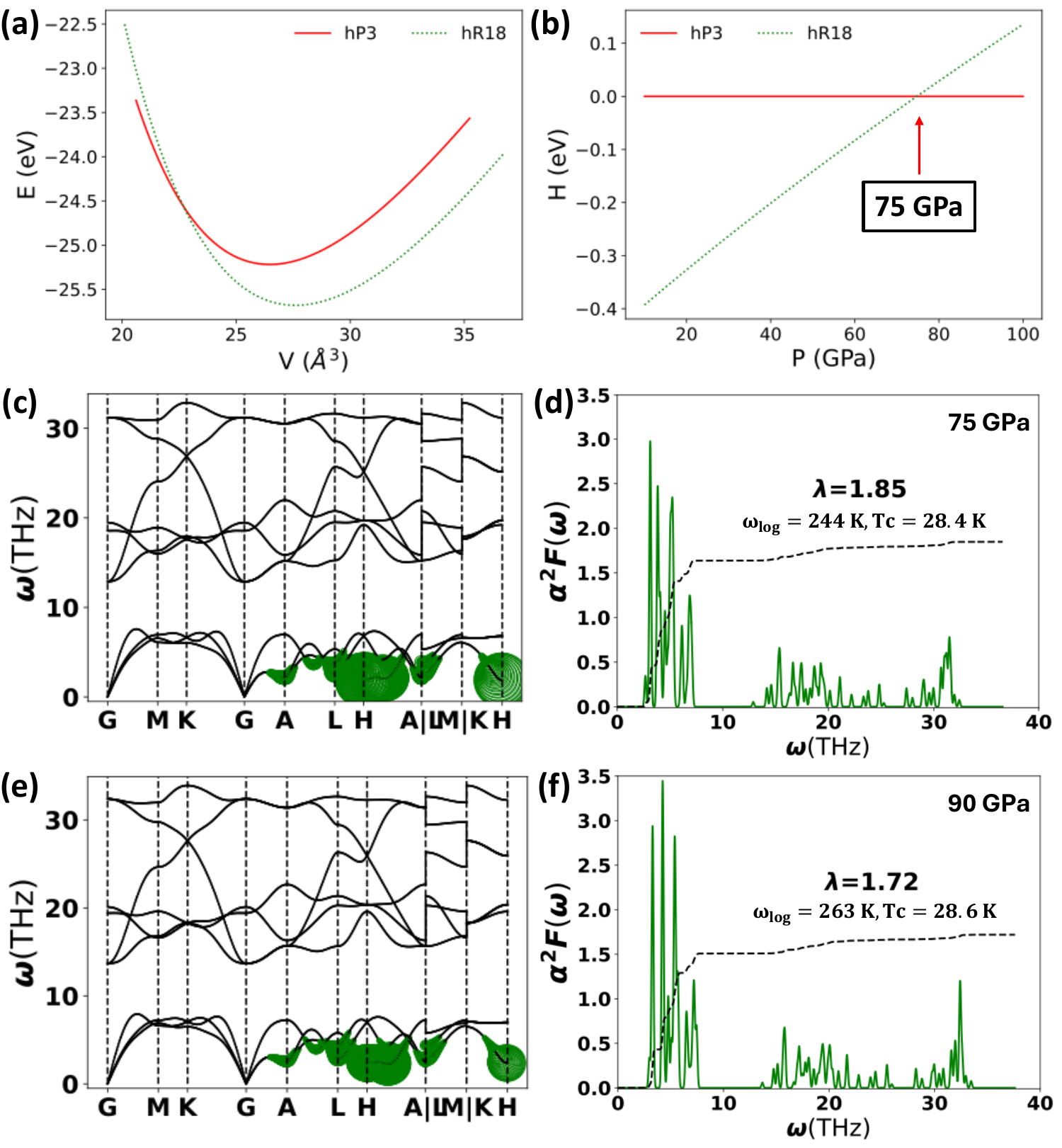}
    \caption{{Pressure-induced stabilization of phonons in MoB$_2$ and electron–phonon coupling (EPC) calculations. (a) Energy (E) versus volume (V) equation of state for the two phases of MoB$_2$: the high-pressure hP3 (P6/mmm) phase and the ambient pressure hR18 (R-3m) phase. (b) Enthalpy (H = E + PV) as a function of pressure (P, GPa) for two phases referenced to the hP3 phase. The crossover between the two phases occurs near 75 GPa (indicated by the arrow). (c) Phonon dispersion of hP3 MoB$_2$ at 75 GPa showing stabilization with EPC strength projected in green shading and (d) the corresponding Eliashberg spectral function. (e), (f) Same as (c), (d), but at 90 GPa.}}
    \label{fig:s12}
\end{figure}

\begin{table}[H]
 \renewcommand{\thetable}{S1}
    \centering
        \caption{Mean Absolute Errors (MAEs) for the 10\% test set in each training iteration with different sizes running for 3000 epochs.}
    \begin{tabular}{|c|c|c|c|c|c|c|c|c|}
    \hline
       Training size  & \multicolumn{4}{|c|}{CGCNN}  & \multicolumn{4}{|c|}{ALIGNN} \\
       \hline
       & $\lambda$ & $\omega_{log}$ (k) & T$_c$ (K) & T$^\prime_c$ (K) & $\lambda$ & $\omega_{log}$ (K) & T$_c$ (K) & T$^\prime_c$ (K)\\
       \hline
     250 (Run1) & 0.15  &  72 & 1.8 & 1.3 &0.15 & 56 & 1.7 & 2.2\\
      \hline
       323 (Run2) &  0.27 & 85 & 6.4 & 5.5 &0.16 & 83 & 3.0 & 3.3 \\
         \hline
      351 (Run3)  & 0.31  & 70 & 3.9 & 6.5 &0.24 & 60 & 3.0 & 4.4 \\
         \hline
    \end{tabular}
    \label{tab6}
\end{table}

\begin{table}[H]
\renewcommand{\thetable}{S2}
\centering
\caption{The best validation MAEs and their corresponding epochs during the training process are reported. CGCNN-250 represents the CGCNN model trained with 250 compounds and so on. Additionally, the coefficient of determination ($R^2$-score) is also provided { for a linear-fit $Y^{ML} = slope \times Y^{DFPT}, Y=(\lambda,\omega_{log},T_c)$}. A test set of 58 compounds is used for the model trained with 250 and 323 compounds, while 66 compounds are used for the model trained with 351 compounds. The R$^2$-score with higher values close to 1 indicating better accuracy, while a score of 0 suggests no better performance than predicting the mean.}
 \begin{tabular}{|c|c|c|c|c|}
\hline
Model & EPC & Epoch & Best MAE  & $R^2$-score \\
\hline
CGCNN-250 & $\lambda$ & 48 & 0.145  & 0.75 \\
CGCNN-250 & $\omega_{log}$ & 60 & 49 &  0.45\\
CGCNN-250 & T$_c$ & 13 & 2.80 &  0.56 \\
\hline
CGCNN-323 & $\lambda$ & 1190 & 0.203  & 0.66 \\
CGCNN-323 & $\omega_{log}$ & 590 & 57 &  0.5 \\
CGCNN-323 & T$_c$ & 736 & 2.73 &  0.31\\
\hline
CGCNN-351 & $\lambda$ & 64 & 0.204 & 0.62\\
CGCNN-351 & $\omega_{log}$ & 62 & 51 &  0.51\\
CGCNN-351 & T$_c$ & 46 & 3.34 &  0.263 \\
\hline
ALIGNN-250 & $\lambda$ & 671 & 0.09 &  0.58\\
ALIGNN-250 & $\omega_{log}$ & 672 & 27 &  0.47\\
ALIGNN-250 & T$_c$ & 760 & 1.32 &  0.23 \\
\hline
ALIGNN-323 & $\lambda$ & 483 & 0.196 &  0.64 \\
ALIGNN-323 & $\omega_{log}$ & 293 & 57 &  0.63 \\
ALIGNN-323 & T$_c$ & 1275 & 2.33  & 0.32 \\
\hline
ALIGNN-351 & $\lambda$ & 186 & 0.22 &  0.72 \\
ALIGNN-351 & $\omega_{log}$ & 701 & 53 &  0.77 \\
ALIGNN-351 & T$_c$ & 617 & 2.85 &  0.38 \\
\hline
\end{tabular}
\label{}
\end{table}

\begin{table}[H]
\renewcommand{\thetable}{S3}
\centering
\caption{MAE calculated separately for nonsuperconductors and superconductors. The test set of 58 includes 40 nonsuperconductors and 18 superconductors, while the test set of 66 consists of 41 nonsuperconductors and 25 superconductors, among which 7 are stabilized superconductors.}
\begin{tabular}{|c|c|c|c|c|c|}
\hline
Training/test size& Property & \multicolumn{2}{|c|}{Tc $<=$ 1}  &  \multicolumn{2}{|c|}{Tc $>$ 1}   \\
 \hline

 & & CGCNN & ALIGNN &  CGCNN & ALIGNN   \\
\hline
& $\lambda$ & 0.28 & 0.21  & 0.20 & 0.32 \\
323/58 & $\omega_{log}$ & 56 & 54  & 188 & 111  \\
& T$_c$ & 3.60 & 1.98  & 2.23 & 4.30 \\
\hline
& $\lambda$ & 0.39 & 0.20 & 0.27 & 0.39 \\
351/66 & $\omega_{log}$ & 125 & 99  & 69 & 63 \\
& T$_c$ & 4.67 & 1.97 & 6.23 & 5.95 \\
\hline
\end{tabular}
\label{}
\end{table}

{
\section{Classification models}
In addition to regression, we also employ machine learning models for binary classification of materials into two categories: SC or NSC. We utilized the similar settings (batch, epochs, etc in config.json) as that of regression model training. Instead of finding a threshold T$_c$ that maximizes the accuracy metrics, we simply perform the classification ML for a fix threshold T$_c$ (0.5, 1.0, 2.0 K) and assess the model performance. The accuracy metrics used in our evaluation are presented in Table~\ref{tab:ML_classification}, and they are defined as follows:\\
\begin{equation}
    \text{Accuracy} = \frac{\text{T$_P$} + \text{T$_N$}}{N}
\end{equation}
\begin{equation}
     \text{Precision} = \frac{\text{T$_P$}}{\text{T$_P$} + \text{F$_P$}}
\end{equation}
\begin{equation}
     \text{Recall} = \frac{\text{T$_P$}}{\text{T$_P$} + \text{F$_N$}}
\end{equation}
\begin{equation}
     \text{F1-score} = \frac{2 (\text{Precision $\times$ Recall)}} {\text{precision + Recall}}
\end{equation}
Accuracy, Precision, Recall, and F1-score are performance metrics that evaluate different aspects of a classifier's effectiveness. Accuracy measures the overall correctness of the classifier's predictions. Precision quantifies the correctness of positive predictions, while Recall (sensitivity) assesses the successful identification of positive instances. The F1-score is the harmonic mean of Precision and Recall, providing a balance between the two.
T$_P$, T$_N$, F$_P$, and F$_N$ represent true positive (correctly identifying true superconductors), true negative (correctly identifying true nonsuperconductors), false positive (misidentifying nonsuperconductors as superconductors), and false negative (misidentifying superconductors as nonsuperconductors), respectively. The accuracy metrics of classification models for both Run 1 and Run 3 are displayed in Table.~\ref{tab:ML_classification}.
Across all these metrics, the overall accuracies are reasonable, indicating a useful classification model despite the small dataset. By employing a classification threshold of T$_c$ = 1 K, both ML models exhibit comparable performance. 
Notably, one can observe the enhanced accuracy of Run 3 in contrast to Run 1. This improvement might be attributed to the expanded training dataset size and the balanced representation of both SC and NSC instances after adding stabilized high T$_c$ results. The ML model, when using a T$_c$ threshold of 1 K, achieves an overall accuracy ranging from 60\% to 70\%, precision ranging from 70\% to 90\%, and an F1-score ranging between 50\% and 60\%. The ML models demonstrate lower recall values, ranging from 50\% to 55\%. Considering the data imbalance between the number of SC instances (approximately 40\%) and NSC instances (approximately 60\%), these recall values are reasonable.}

\begin{table}[h!]
\renewcommand{\thetable}{S4}
    \centering
        \caption{Classification accuracy: Results from ML model are presented in the form CGCNN/ALIGNN.}
       \scalebox{1.2}{%
    \begin{tabular}{|c|c|c|c|c|c|c|c|c|}
    \hline
 &   \multicolumn{4}{|c|}{Run 1} & \multicolumn{4}{|c|}{Run 3}\\
 \hline
  T$^{Threshold}_c$ (K)  &Accuracy& Precision& Recall &F1-score &Accuracy& Precision& Recall &F1-score\\
\hline
0.5 & 0.67/0.59 & 0.83/0.58 & 0.57/0.50 & 0.68/0.54 & 0.71/0.67 & 0.74/0.67 & 0.68/0.64  & 0.71/0.66 \\
 \hline
1.0  & 0.66/0.62 & 0.83/0.72 & 0.47/ 0.43 & 0.60/0.54 & 0.64/0.67 & 0.84/0.72 & 0.51/0.55 & 0.64/0.62 \\
 \hline
2.0  & 0.72/0.62 & 0.77/0.77 & 0.43/0.34 & 0.56/0.48 & 0.69/0.61 & 0.9/0.7  & 0.5/0.41 & 0.64/0.52  \\
\hline

\end{tabular}
    }
\label{tab:ML_classification}
\end{table}

\section{Theory of superconductivity: Isotropic Approximation}
We have adopted the DFPT calculation with isotropic Eliashberg approximation to compute SC properties, which provides a harmony between accuracy and efficiency. The critical temperature can be calculated using the Allen-Dynes formula \cite{AD72},
\begin{equation}\label{eq1}
    T_c = \frac{\omega_{log}}{1.2} exp \left[- \frac{1.04 (1 + \lambda)}{\lambda - \mu^*_c(1 + 0.62\lambda)}  \right],
\end{equation}
where, $\lambda = 2\int_0^\infty \frac{d\omega}{\omega} \alpha^2F(\omega)$  is the EPC strength constant, $\omega_{log} = exp \left[\frac{2}{\lambda} \int_0^\infty \frac{d\omega}{\omega} \alpha^2F(\omega) log\omega\right]$, $\alpha^2F(\omega)$ is frequency ($\omega$) resolved Eliashberg spectral function, and $\mu^*_c$ is the Coulomb potential. The spectral function $\alpha^2F(\omega)$ is defined as,
\begin{equation}\label{eq2}
    \alpha^2F(\omega) = \frac{1}{2} \sum_{\nu} \int_{BZ} \frac{d\textbf{q}}{\Omega_{BZ}} \omega_{\textbf{q}\nu} \lambda_{\textbf{q}\nu} \delta(\omega - \omega_{\textbf{q}\nu}).
\end{equation}
Here, $\Omega_{BZ}$ is the volume over the Brillouin zone $\int_{BZ} \frac{d\textbf{q}}{\Omega_{BZ}} \rightarrow \frac{1}{N_\textbf{q}} \sum_{\textbf{q}}$, $\omega_{\textbf{q}\nu}$ is the mode ($\nu$) resolved phonon frequency, and $\lambda_{\textbf{q}\nu}$ is the mode resolved EPC strength constant, 
\begin{equation}\label{eq3}
    \lambda_{\textbf{q}\nu} = \frac{1}{N(\epsilon_F) \omega_{\textbf{q}\nu}} \sum_{mn} \int_{BZ} \frac{d\textbf{k}}{\Omega_{BZ}} |g_{mn,\nu}(\textbf{k},\textbf{q})|^2 \delta(\epsilon_{n\textbf{k}} - \epsilon_F) \delta(\epsilon_{m\textbf{k+q}} - \epsilon_F).
\end{equation}
$N(\epsilon_F)$ is the density of states at the Fermi level $\epsilon_F$, and $g_{mn,\nu}(\textbf{k},\textbf{q})$ is the EPC matrix element which quantifies the scattering process between Kohn-Sham states m\textbf{k+q} and n\textbf{k}. In Ref.~\cite{KA17}, an effective approach to approximate double delta integration is explored, specifically tailored for situations where it is permissible to disregard the dependence on the momentum vector (\textbf{q}). This method finds utility in scenarios such as the study of extensive molecular systems like alkali fullerides, where momentum dependence can be safely overlooked. The net EPC strength constant is computed as
\begin{equation}\label{eq4}
    \lambda = \sum_{q\nu} \lambda_{q\nu}
\end{equation}

For computational feasibility, these Dirac deltas can be approximated by Gaussian functions with a broadening parameter $\sigma$ \cite{WdG05}, and EPC strength is given by Eqs.\ref{eq3} and \ref{eq4} can be redefined as
\begin{equation}\label{eq5}
    \lambda  \approx \frac{1}{N(\epsilon_F ) N_\textbf{q} N_\textbf{k}} \sum_{nm} \sum_\textbf{q} \sum_\textbf{k} \frac{|g_{mn,\nu}(\textbf{k},\textbf{q})|^2}{\omega_{\textbf{q}\nu}}
    \frac{1}{2\pi \sigma^2 } exp \left [ -\frac{(\epsilon_{n\textbf{k}} - \epsilon_F)^2 + ( \epsilon_{m\textbf{k}+\textbf{q}} - \epsilon_F)^2}{\sigma^2} \right ].
\end{equation}

Here, $N_\textbf{q}$ and $N_\textbf{k}$ respectively are the total number of \textbf{q} and \textbf{k} grid points, and $\sigma$ is the smearing used to broaden states at the Fermi-level $\epsilon_F$. With infinitely large \textbf{k}- grids, and $\sigma \rightarrow 0$, the double summation changes back to double delta integration.  
Moreover, $\lambda$ can also obtained from frequency $\omega$ resolved Eliashberg spectral function as \cite{MW68}
\begin{equation}\label{eq3_2}
    \lambda = 2 \int \frac{d\omega \alpha^2F(\omega)}{\omega} = \frac{N(\epsilon_F) <g^2>}{M<\omega^2>},
\end{equation}
where, $<\omega^2>$ average of the square of the phonon frequency $\left (\frac{\int d\omega \omega \alpha^2F(\omega)}{\int \frac{d\omega}{\omega} \alpha^2F(\omega)} \right )$, $<g^2>$ is average over the Fermi surface of the square of electronic phonon coupling matrix element \cite{MW68}.\\
\section{Convergence tests for MgB$_2$ and AlB$_2$}
The ground-state total energy and phonon frequencies at $\Gamma$-point are well converged for MgB$_2$ with \textbf{k}-grids of 8$\times$8$\times$6, compared to the dense grid of 24$\times$24$\times$24. K-mesh grid in materials project database for MgB$_2$ is 8$\times$8$\times$7. Since the superconducting properties of MgB$_2$ with \textbf{k}-grid of 8$\times$8$\times$8 doesn't follow the trend of converged result of denser k-mesh with $\sigma \rightarrow 0$, we chose 8$\times$8$\times$6 so that we can use sufficient \textbf{q}-grids of 4$\times$4$\times$3 instead of using 9$\times$9$\times$9 \textbf{k}-grid and 3$\times$3$\times$3 \textbf{q}-grid. We utilized 8$\times$8$\times$6 \textbf{k}-grid and its multiple to compute the decay parameter. In this work, we are taking $\lambda = $ 0.75 obtained from solving anisotropic Migdal-Eliashberg equation using $\mu^* = $ 0.16 as a reference [Comput. Phys. Commun. \textbf{209}, 116 (2016)]. One can obtain this converged $\lambda$ with denser \textbf{k}- and \textbf{q}- grids and $\sigma \rightarrow 0$, as shown in Figs.~\ref{fig:S2} $-$ \ref{fig:S4}. At last, we check whether the result is transferable to systems with larger unit cell where essentially smaller \textbf{k}-grids provide converged ground-state properties and q-grid, taken half of \textbf{k}-grid, sometime shrinks only to include the $\Gamma$-point. To do that, we performed EPC calculations with 2$\times$2$\times$2 supercell of MgB$_2$ which has 24 atoms per cell, and present the results in Fig.~\ref{fig:S5} and \ref{fig:S6}. Results confirm that \textbf{q}-grid as half as that of the \textbf{k}- grid can provide converged results even for larger systems. Instead of increasing coarse \textbf{k}- and \textbf{q}- grids, one can also utilize extremely large \textbf{k}- grid for interpolating EPC matrix element to achieve convergence, as shown in Figs.~\ref{fig:S4} and \ref{fig:S7} respectively for MgB$_2$ and AlB$_2$. However, it also increases computational complexity for larger systems. 
\begin{table}[H]
\renewcommand{\thetable}{S5}
    \centering
        \caption{Convergence of the ground-state total energy with respect to K-point mesh}
            \scalebox{1.3}{%
    \begin{tabular}{|c|c|}
    \hline
       K-mesh  & Total Energy (eV/atom) \\
       \hline
          6$\times$6$\times$4 & -615.178 \\ 
         \textbf{8$\times$8$\times$6} & -615.179\\
         8$\times$8$\times$8 & -615.180 \\
         9$\times$9$\times$9 & -615.180 \\
          12$\times$12$\times$12 & -615.180 \\
         16$\times$16$\times$16 & -615.180 \\
         \textbf{24$\times$24$\times$24} & -615.180 \\
         \hline
    \end{tabular}
    }
    \label{tab:S1}
\end{table}

\begin{table}[H]
\renewcommand{\thetable}{S6}
    \centering
        \caption{Convergence of phonon frequency (THz) at $\Gamma$-point with respect to K-point mesh}
            \scalebox{1.3}{%
    \begin{tabular}{|c|c|c|c|c|c|c|c|c|c|}
    \hline
       K-mesh  & $\omega_1$ & $\omega_2$ & $\omega_3$ & $\omega_4$ & $\omega_5$ & $\omega_6$ & $\omega_7$ & $\omega_8$ & $\omega_9$ \\
       \hline
      6$\times$6$\times$4 & -0.654 &	-0.654 &	0.343 &	10.18 &	10.18 &	12.07& 20.96 &	24.38	& 24.38 \\
\textbf{8$\times$8$\times$6} &	-0.49 &	-0.49	& 0.48 &	10.11 &	10.11 &	12.11 &	16.45 &	16.45 &	20.75 \\
8$\times$8$\times$8 &	-0.63 &	-0.63 &	0.28 &	10.1 &	10.1 &	12.11 &	17.43 &	17.43 &	20.8 \\
9$\times$9$\times$9	& -0.62	& -0.62	& -0.33 &	9.99 &	9.99 &	12.08 &	14.36 &	14.36 &	20.65 \\
12$\times$12$\times$12 &	-0.64 &	-0.64 &	-0.23 &	10.06 &	10.06 &	12.12 &	17.13 &	17.13 &	20.69 \\
16$\times$16$\times$16 &	-0.64 &	-0.64 &	-0.25 &	10.05 &	10.05 &	12.1 &	17.2 &	17.2 &	20.69 \\
\textbf{24$\times$24$\times$24} &	-0.64 &	-0.64 &	-0.22 &	10.05 &	10.05 &	12.1 &	16.99 &	16.99 &	20.71 \\
         \hline
    \end{tabular}
    }
    \label{tab:S2}
\end{table}
\begin{figure}[H]
\renewcommand{\thefigure}{S5}
    \centering
     \includegraphics[scale = 0.55]{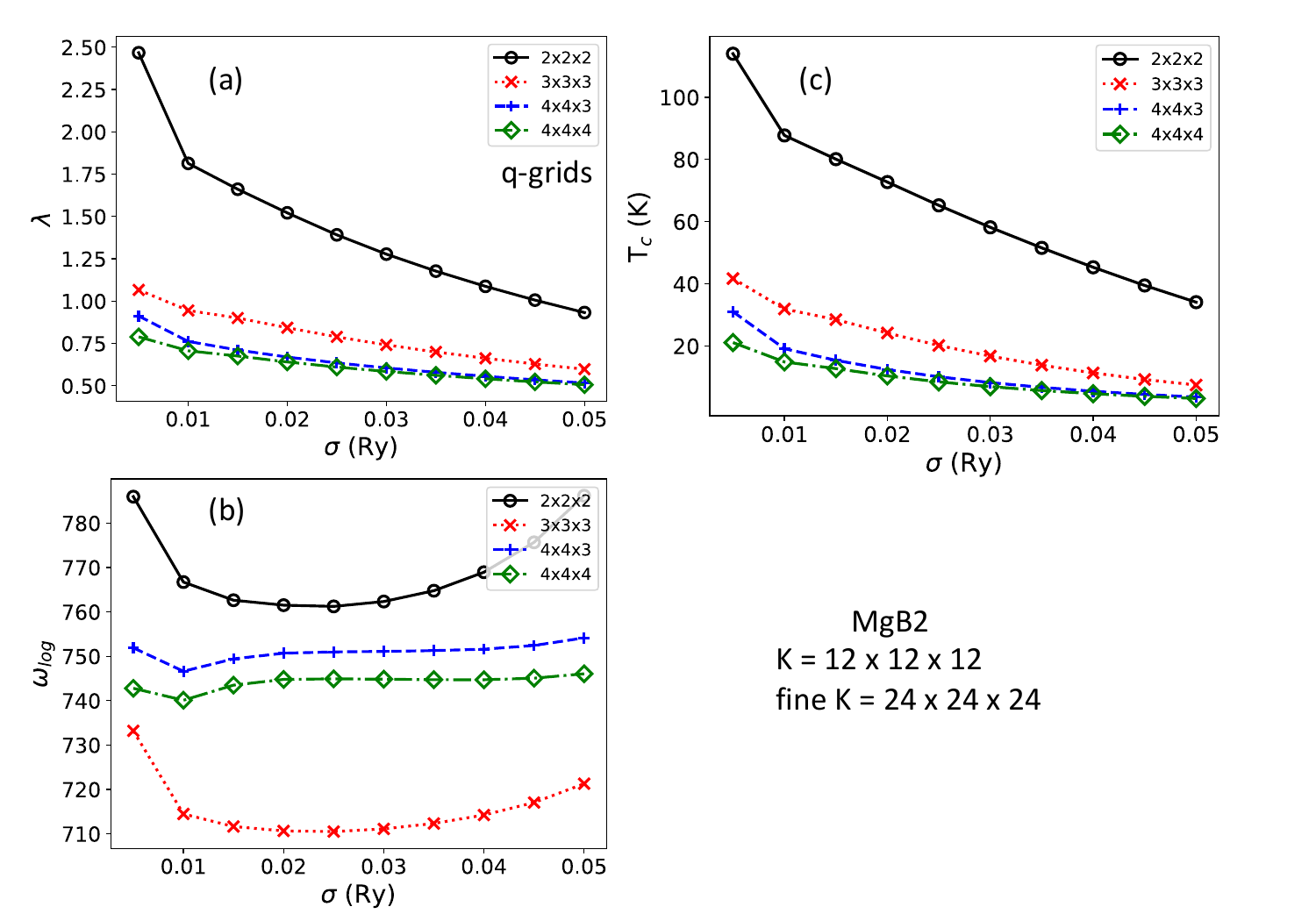}
    \caption{Convergence test for MgB$_2$ results with respect to \textbf{q}-mesh. Unit of T$_c$ and $\omega_{log}$ is Kelvin (K); As \textbf{q} becomes denser, convergence can be achieved across the $\sigma \rightarrow$ 0, with less exponential decay. Fine \textbf{k-} grid is utilized for interpolation.}
    \label{fig:S2}
\end{figure}
\begin{figure}[H]
\renewcommand{\thefigure}{S6}
    \centering
     \includegraphics[scale = 0.55]{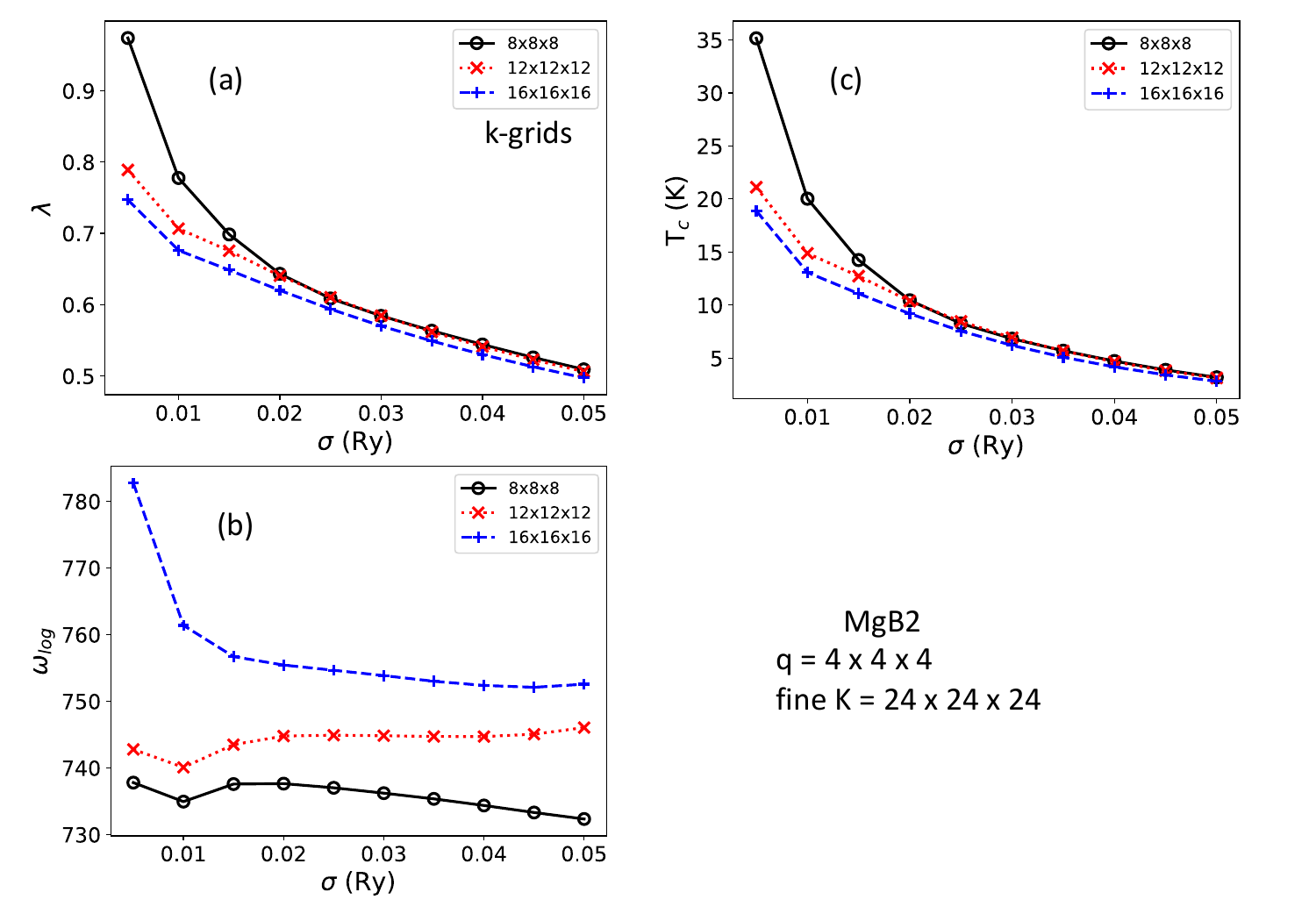}
    \caption{Convergence test for MgB$_2$ results with respect to \textbf{k}-mesh. Unit of T$_c$ and $\omega_{log}$ is Kelvin (K); As \textbf{k} becomes denser, convergence can be achieved across the $\sigma \rightarrow$ 0, with less exponential decay. Fine \textbf{k-} grid is utilized for interpolation.}
    \label{fig:S3}
\end{figure}
\begin{figure}[H]
\renewcommand{\thefigure}{S7}
    \centering
     \includegraphics[scale = 0.55]{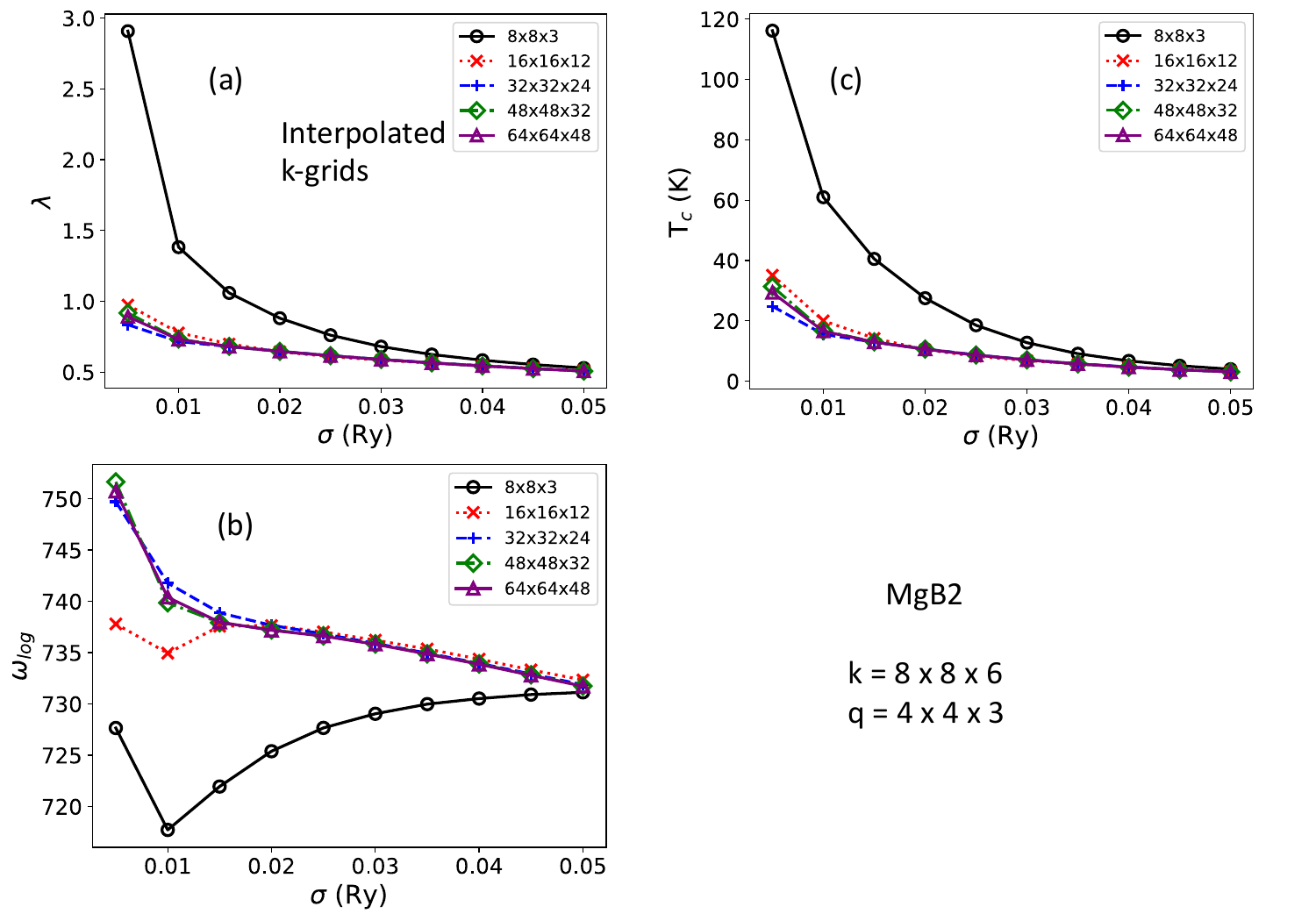}
    \caption{Convergence test for MgB$_2$ results with respect to fine \textbf{k}-mesh used for interpolation. Unit of T$_c$ and $\omega_{log}$ is Kelvin (K); As fine \textbf{k} becomes denser, convergence can be achieved across the $\sigma \rightarrow$ 0, with less exponential decay.}
    \label{fig:S4}
\end{figure}
\begin{figure}[H]
\renewcommand{\thefigure}{S8}
    \centering
     \includegraphics[scale = 0.55]{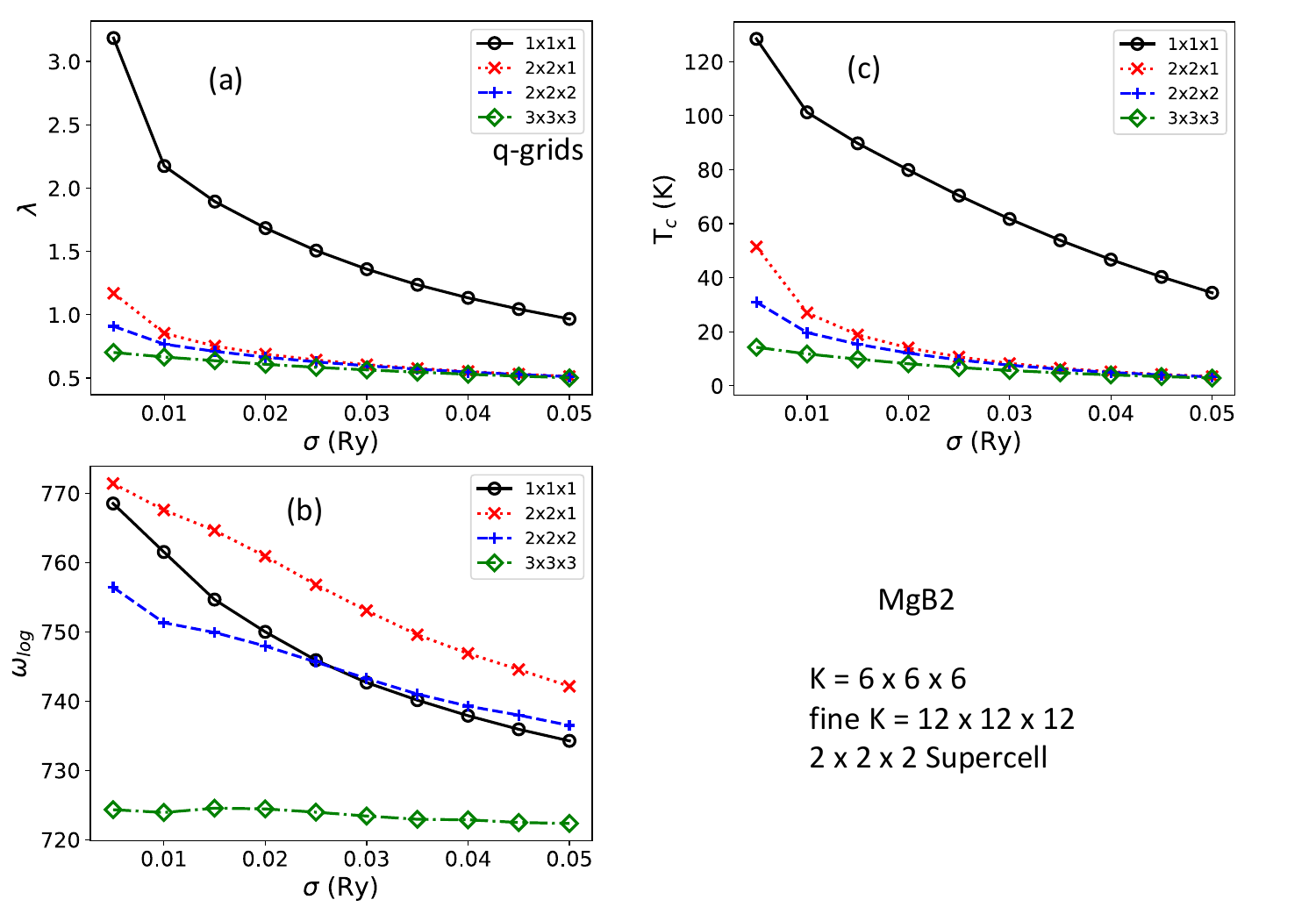}
    \caption{Convergence test for 2$\times$2$\times$2 supercell of MgB$_2$ results with respect to \textbf{q}-mesh. Unit of T$_c$ and $\omega_{log}$ is Kelvin (K); As \textbf{q} becomes denser, convergence can be achieved across the $\sigma \rightarrow$ 0, with less exponential decay. Fine \textbf{k-} grid is utilized for interpolation.}
    \label{fig:S5}
\end{figure}
\begin{figure}[H]
\renewcommand{\thefigure}{S9}
    \centering
     \includegraphics[scale = 0.55]{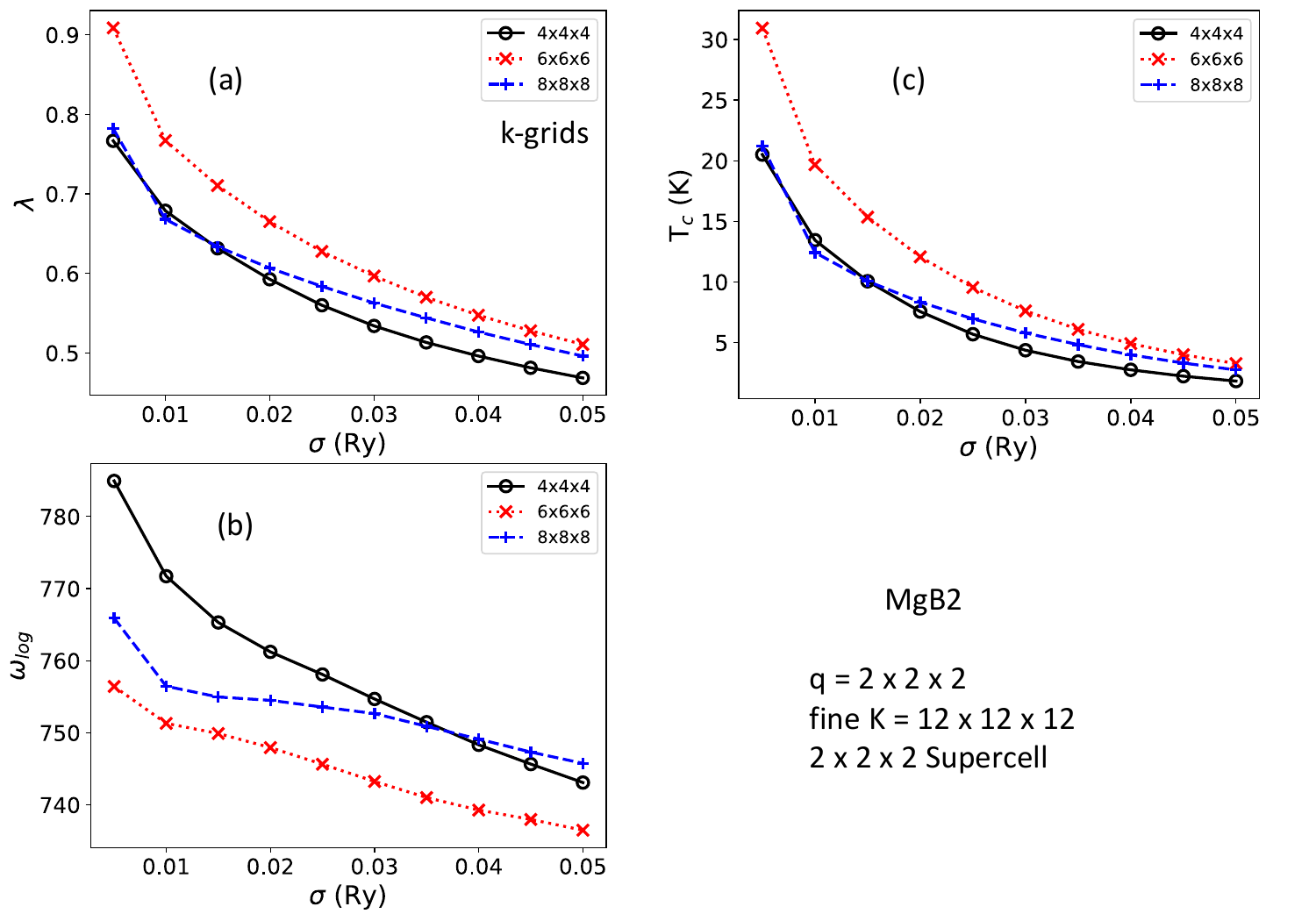}
    \caption{Convergence test for 2$\times$2$\times$2 supercell of MgB$_2$ results with respect to \textbf{k}-mesh. Unit of T$_c$ and $\omega_{log}$ is Kelvin (K); As \textbf{k} becomes denser, convergence can be achieved across the $\sigma \rightarrow$ 0, with less exponential decay. Fine \textbf{k-} grid is utilized for interpolation.}
    \label{fig:S6}
\end{figure}
\begin{figure}[H]
\renewcommand{\thefigure}{S10}
    \centering
     \includegraphics[scale = 0.55]{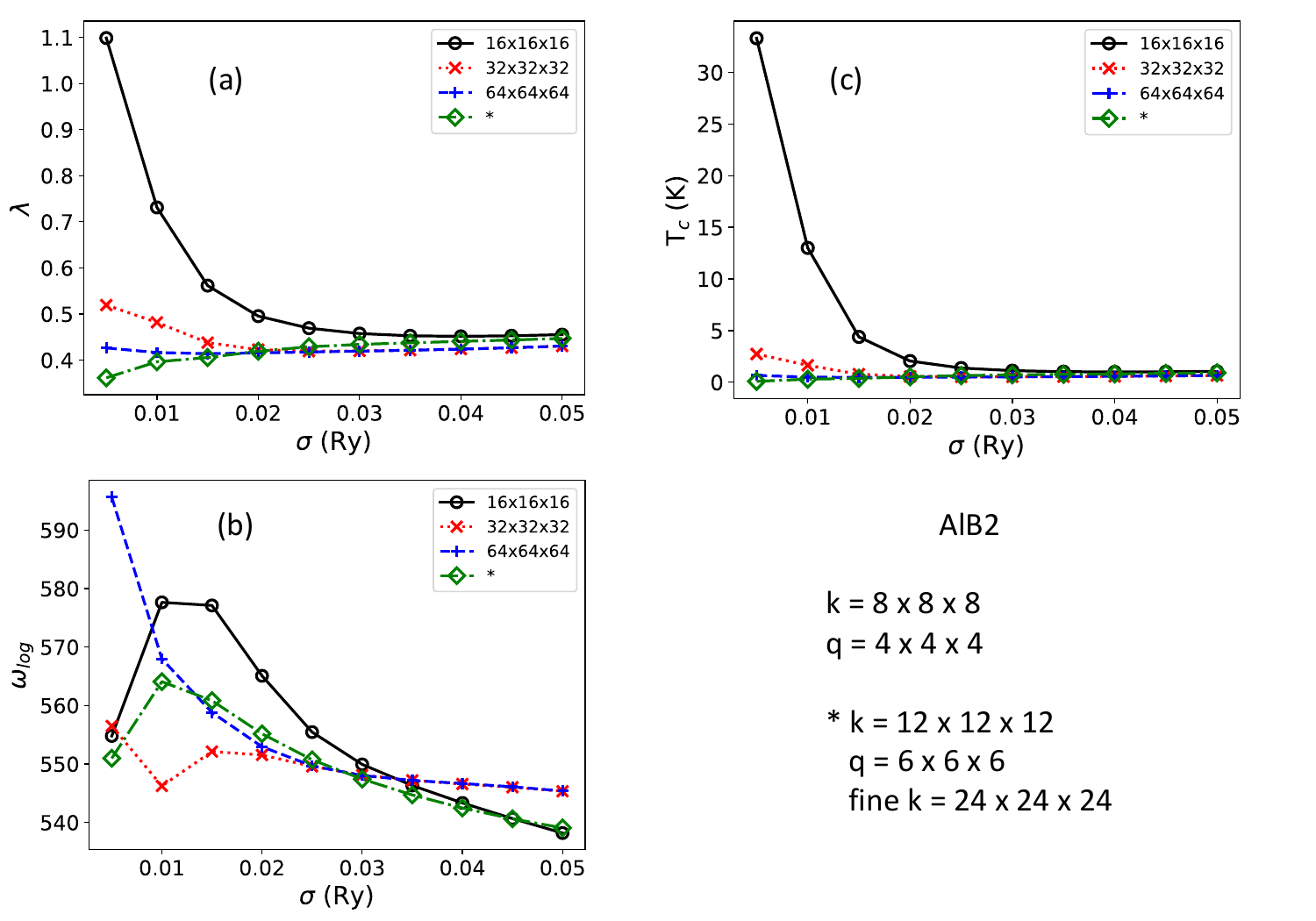}
    \caption{Convergence test for AlB$_2$ results with respect to various \textbf{k}- and \textbf{q}-grids. Unit of T$_c$ and $\omega_{log}$ is Kelvin (K); As grids becomes denser, convergence can be achieved across the $\sigma \rightarrow$ 0, with less exponential decay. Separate results corresponding to \textbf{k}- and \textbf{q-} grids respectively of 12 $\times$ 12 $\times$ 12 and 6 $\times$ 6 $\times$ 6 is represented by ``star (*)" symbol.}
    \label{fig:S7}
\end{figure}
\section{Convergence ansatz}
 Estimation of SC properties requires the computation of double-delta integration, which also defines the nesting function, around the Fermi level E$_F$ over the entire Brillouin zone [Eq. 3]. However, it has a slow convergence with respect to \textbf{k-} and \textbf{q-} grids. In principle, it requires infinitely dense grids, which makes the computation exorbitantly expensive. Therefore, Gaussian broadening technique [Eq. 5]  is employed to compute $\lambda$ with finite \textbf{k}-mesh. 
To compute T$_c$, we employed the DFPT calculation with the isotropic Eliashberg approximation. We used a coarse \textbf{k}-grid obtained from the MP database, which provides reasonably accurate ground-state properties. The \textbf{q}-grid was set to half the size of the \textbf{k}-grid, and a broadening parameter ($\sigma$) of 0.01 Ry was utilized with the coarse MP grid, which accurately captures a converged electron-phonon coupling constant ($\lambda$) of MgB$_2$ computed using the anisotropic Migdal-Eliashberg equations \cite{SMVG16}. However, in practice, $|g_{mn,\nu}(\textbf{k},\textbf{q})|^2$ is computed for a reasonable coarse \textbf{k}- and \textbf{q}- grids and interpolated the matrix elements to fine \textbf{k}- for any \textbf{q}-point, from 32 $\times$ 32 $\times$ 32 to as high as 60 $\times$ 60 $\times$ 60, to achieve numerical convergence, as implemented in Quantum Espresso (QE) code\cite{WdG05}. Furthermore, an efficient interpolation scheme utilizing both dense \textbf{k-} as well as \textbf{q}-grids has been implemented in EPW package that uses wannier orbitals \cite{SMVG16}. However, employing extremely fine grids for interpolation can increase the computational complexity, which is not suitable for highthroughput calculations. Therefore, we have restricted ourselves for choosing fine \textbf{k}-grid only twice of corresponding coarse grid for highthroughput calculations.

 Our investigation revealed that a reasonable number of calculations did not converge, leading to inaccurate predictions, while utilizing a coarse \textbf{k}-grid from MP database and fine grid for interpolation only twice that of the \textbf{k}-grid [Table S7]. For example, AlB$_2$ was predicted to be a superconductor (T$_c$ $\sim$ 11 K) with the coarse grid [\textbf{k}-grid: 8$\times$8$\times$8, \textbf{q}-grid: 4$\times$4$\times$4, fine-\textbf{k}-mesh: 16$\times$16$\times$16] from MP database, whereas a denser grid twice the size of the coarse grid corrected this inaccuracy and predicted AlB$_2$ to be a nonsuperconductor, consistent with experimental observations [Fig. S7].
To identify such cases of convergence failure, we examined the variation of T$_c$ with respect to $\sigma$ and developed a simple ansatz based on the converged results of MgB$_2$, as depicted in Fig. S13. The ansatz involves extracting T$_c$ with a fixed broadening value (as in the case of MgB$_2$) and estimating the decay parameter (A) in the exponential variation of T$_c$ with $\sigma$ as T$_c \sim \exp(-A\sigma^{1/3} + B)$. Unconverged results exhibit larger values of A, which decrease with a denser \textbf{k}-mesh. 

The detail theory of EPC calculations within DFPT formalism in presented in the section ``Theory of superconductivity: isotropic approximation". The variation of density of states at the Fermi level ($N(\epsilon_F)$) with respect to smearing $\sigma$, depends on Gaussian, 
\begin{equation}
G(\Delta,\sigma) = \frac{1}{\sigma^2} exp \left [ -\frac{(\epsilon_{n\textbf{k}} - \epsilon_F)^2 + ( \epsilon_{m\textbf{k}+\textbf{q}} - \epsilon_F)^2}{\sigma^2} \right] \sim \frac{1}{\sigma^2} exp \left[- \frac{\Delta^2}{\sigma^2} \right]
\end{equation}
with $\Delta^2 \sim (\epsilon_{n\textbf{k}} - \epsilon_F)^2 + ( \epsilon_{m\textbf{k}+\textbf{q}} - \epsilon_F)^2$ also depends on $\sigma$. Despite of dependency of $\lambda$ on EPC matrix ($<g^2>$) and phonon frequency ($<\omega^2>$) terms, the variation in $\lambda$ is largely guided by the variation in $N(\epsilon_F)$ or the Gaussian term G. Fig.~\ref{fig:mgb2_conv}(a) shows the variation of G with respect to smearing $\sigma$ for different values of $\Delta$, assuming $\Delta$ independent of $\sigma$. The effect of $\sigma$ on G dictates the variation of $\lambda$ and hence T$_c$, with $\Delta$ depends on the materials as well as on size of the grid to compute double delta integration. A larger $\Delta$ represents more states contributing towards the double delta summation within space around the Fermi-level spanned by $\sigma$. In other word, $\Delta$ will increase with $\sigma$ if the density of states has local maximum close to the Fermi-level, and the variation of EPC properties will be similar to the Gaussian plots corresponding to $\Delta$ $\geq$ 0.05. For Niobium, the variation of EPC properties with respect to $\sigma$ follows similar to blue-dashed curve corresponding to $\Delta$ = 0.01 \cite{WdG05}, whereas for MgB$_2$, the variation follows the Gaussian corresponding to $\Delta$ $<$ 0.01. For $\sigma \rightarrow 0$ and $\Delta \rightarrow 0$, if $\frac{\Delta}{\sigma} \leq 1$ the Gaussian function recovers the delta function, while for $\frac{\Delta}{\sigma} \geq 1$ the Gaussian drops to zero similar to $\Delta \geq$  0.01 cases. Our primary focus is more on exponentially decaying cases.

The effect of smearing $\sigma$ on $\omega_{log}$ is insignificant compared to $N(\epsilon_F)$ or $\lambda$ \cite{KA17}. According to the Bardeen$–$Cooper$–$Schrieffer (BCS) theory \cite{BCS57}, 
\begin{equation}
    T_c \sim \Theta_D exp(-1/\lambda) \sim \Theta_D exp(-\frac{1}{N(\epsilon_F) U})
\end{equation}
where $\Theta_D$ is the Debye cutoff energy and U is electron-phonon coupling potential. One can established the relation between $\lambda$ or $T_c$ on $\sigma$ as $\lambda \sim 1/\sigma^\alpha$ and $T_c \sim \Theta_D exp(-1/\lambda) = exp(-A\sigma^\alpha + B)$ with $\Theta_D \sim exp(B) \times$ constant. Here parameter $\alpha$ is a constant with a positive value if $T_c$ decreases with increasing $\sigma$ and becomes negative otherwise. Here A and B are constants to be determined with A denoting the coefficient of exponential increase (for negative A) or decrease (for positive A) of $T_c$ with respect to $\sigma$. To determine the value of these constants, we perform linear-fit of $logT_c$ vs $\sigma^{\alpha}$ for different \textbf{k}- grids for MgB$_2$, which is frequently studied both theoretically as well as experimentally and often challenging to obtain the converged SC properties \cite{MgB201,KK14,CG22}. Furthermore, one can capture the behavior of T$_c$ vs $\sigma$ for a wide range of T$_c$ using MgB$_2$, as shown in Fig.~\ref{fig:mgb2_conv}(b). We utilize a smaller k-mesh of 6$\times$6$\times$4 and a slightly larger k$^\prime$-grid of 8$\times$8$\times$6 for comparison [Fig.~\ref{fig:mgb2_conv}(b) and (c)].

\begin{figure*}[ht!]
\renewcommand{\thefigure}{S11}
\includegraphics[scale=0.55]{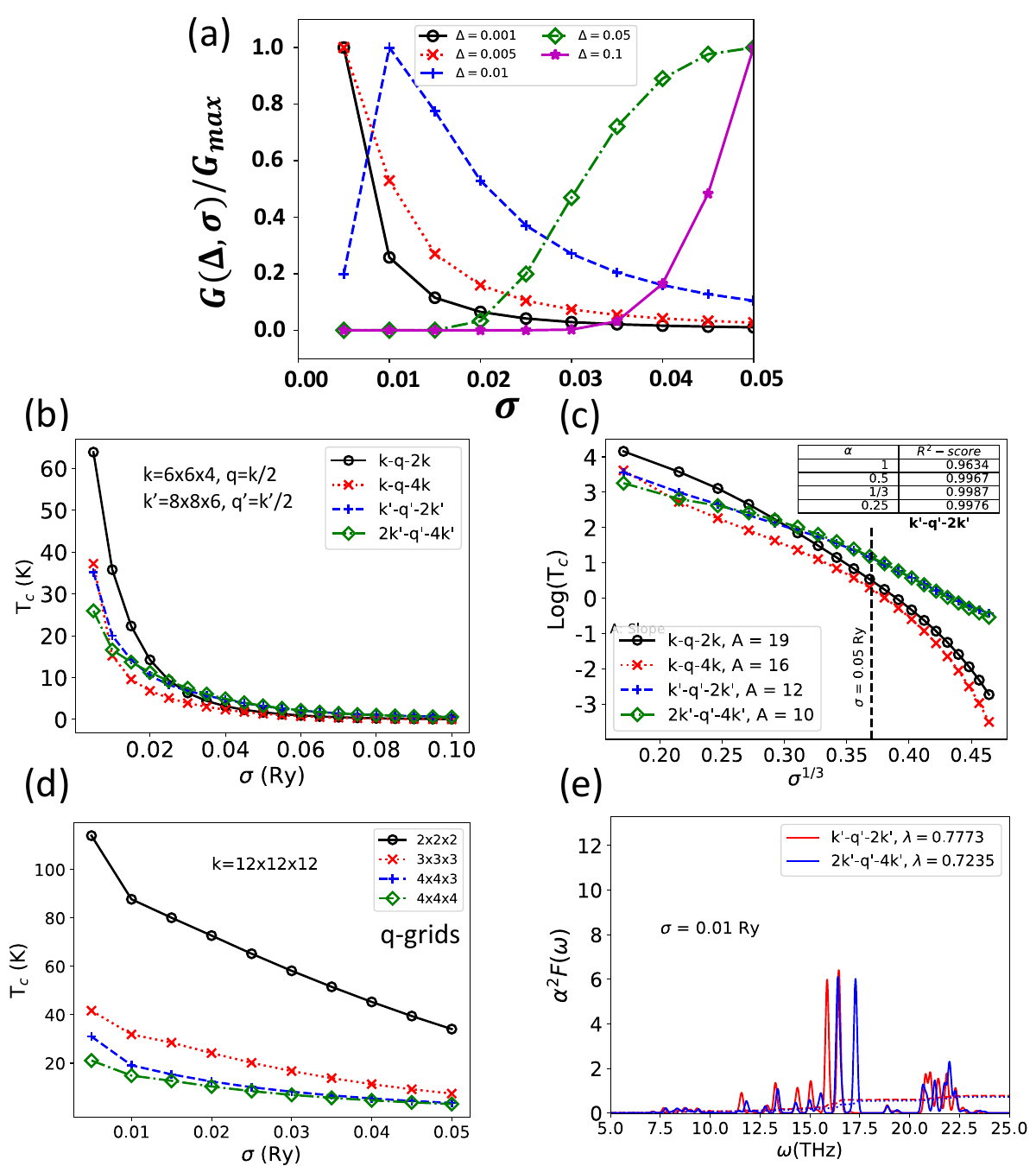}
\caption{Decay Constant Determination from Converged MgB$_2$ data. (a) Variation of $\frac{G(\Delta,\sigma)}{G_{max}}$ with respect to $\sigma$ for different values of $\Delta$; $G_{max}$ being the maximum value of Gaussian for corresponding $\Delta$ within the value of $\sigma$ from 0.005 Ry to 0.05 Ry. (b) T$_c$ vs $\sigma$ for various \textbf{k}- grids for MgB$_2$. \textbf{k}-\textbf{q}-2\textbf{k} represents \textbf{k}-mesh for self-consistent calculation for charge density and EPC, \textbf{q}-mesh for phonon and EPC, and 2\textbf{k}-mesh represents fine grid used for interpolating EPC matrix obtained from \textbf{k} and \textbf{q}-grids; \textbf{k} and \textbf{k}$^\prime$ respectively are 6$\times$6$\times$4 and 8$\times$8$\times$6.  (c) $logT_c$ vs $\sigma^{1/3}$ plot; Linear-fit is performed for data up to $\sigma = 0.05 $ Rydberg (denoted by vertical dashed line), after which lines change slope; $\alpha = 1/3$ fits with an optimal coefficient of determination (R$^2$-score) (Table inside plot); Fitting parameter ``A" decreases from 19 to 10 as \textbf{k-} grid change from coarse to dense; (c) T$_c$ vs $\sigma$ for various \textbf{q}- grids keeping \textbf{k}-grid fixed at 12$\times$12$\times$12. (d) Spectral function $\alpha^2F(\omega)$ vs $\omega$ plots with $\sigma = 0.01$ Ry with k-grid of 8$\times$8$\times$6. }
\label{fig:mgb2_conv}
\end{figure*}

 Fig.~\ref{fig:mgb2_conv} (b) represents the variation of $T_c$ with respect to $\sigma$, while Fig.~\ref{fig:mgb2_conv} (c) presents $logT_c$ vs $\sigma^\alpha$ plot. It exhibits nearly linear behavior up to \(\sigma = 0.05 \, \text{Ry}\), i.e. $\sigma^{1/3}$ = 0.37, after which it deviates from linearity. The linear-fit has optimal coefficient of determination score (R$^2$-score) for $\alpha < 0.5$ on $logT_c$ vs $\sigma^\alpha$ data for \textbf{k}(charge density and EPC)-\textbf{q}(Phonon and EPC)-2\textbf{k}(interpolating EPC matrix) grids (Fig.~\ref{fig:mgb2_conv} (c)) [Table within the Fig.~\ref{fig:mgb2_conv}(c)]. Therefore, we chose $\alpha = 1/3$ in this work with R$^2$-score of 0.9987. Besides different choice of $\alpha$ leads to different values of A and B, it doesn't have a significant role. For coarse grid, the T$_c$ has a larger dependency on $\sigma$ and shows larger exponential decay, compared to more converged calculations on denser grids. The exponential decay parameter ``A" decreases from 19 to 10 with grids changing from coarse to dense one. This analysis suggests that A$_{MgB_2}$ = 12$-$13 can be used as a cutoff for this work, the calculations can be considered unconverged for A $>$ A$_{MgB_2}$, while the calculations can be considered converged for A $<$ A$_{MgB_2}$ for accurate predictions. Fig.~\ref{fig:mgb2_conv}(d) shows the convergence of the T$_c$ with respect to \textbf{q}-point mesh, keeping k-grid fixed at 12$\times$12$\times$12. A \textbf{q}-point grid as half as that of the ground-state \textbf{k}-point grid is sufficient to provide the converged results for MgB$_2$. Fig.~\ref{fig:mgb2_conv} (e) represents the spectral functions for various grids with k=8$\times$8$\times$6. Denser grid (2\textbf{k}) slightly blue shifts the spectral function peaks at lower frequency range (15.5-16.5 THz), while peaks at higher frequency ranges (20-22.5 THz) remain unaffected. This results slight change in $\lambda$ from 0.78 to 0.72 and T$_c$ changing from 20 K to 15-16 K [Figs. S5-S7]. A $\lambda$ of 0.78 agrees with previous theoretical value of $\sim$ 0.75 from anisotropic Migdal-Eliashberg calculation using $\mu^* = $ 0.16 \cite{SMVG16}, $\lambda = $ 0.71 from fully anisotropic SCDFT \cite{SCDFT07}, and $\lambda = $ 0.73 from previous isotropic Eliashberg approximation \cite{BHR01} at slightly larger $\sigma = $ 0.015 Ry. This indicates a \textbf{k}-grid obtained from MP database with $\sigma = 0.01$ Ry already provide converged results in the case of MgB$_2$. However, it is not always the case for other materials. Based on these results, parameters $\sigma$ = 0.01 Ry with $\mu^*_c = 0.16$ seems to be reasonable choice for smearing with 8 $\times$ 8 $\times$ 6 to 16 $\times$ 16 $\times$ 12 \textbf{k}-grids for MgB$_2$, and for other systems for the sake of comparison. These parameters could also depend on pseudopotentials. In order to achieve converged results, it is necessary to use denser \textbf{k}- and \textbf{q}-grids. Subsequently, a double-delta integration should be performed, selecting the results that correspond to the limit of $\sigma \rightarrow 0$. Other details convergence tests of MgB$_2$ with respect to Brillouin-zone sampling are presented in Figs. S5-S10.
\begin{figure}[H]
 \renewcommand{\thefigure}{S12}
    \centering
    \includegraphics[scale=0.3]{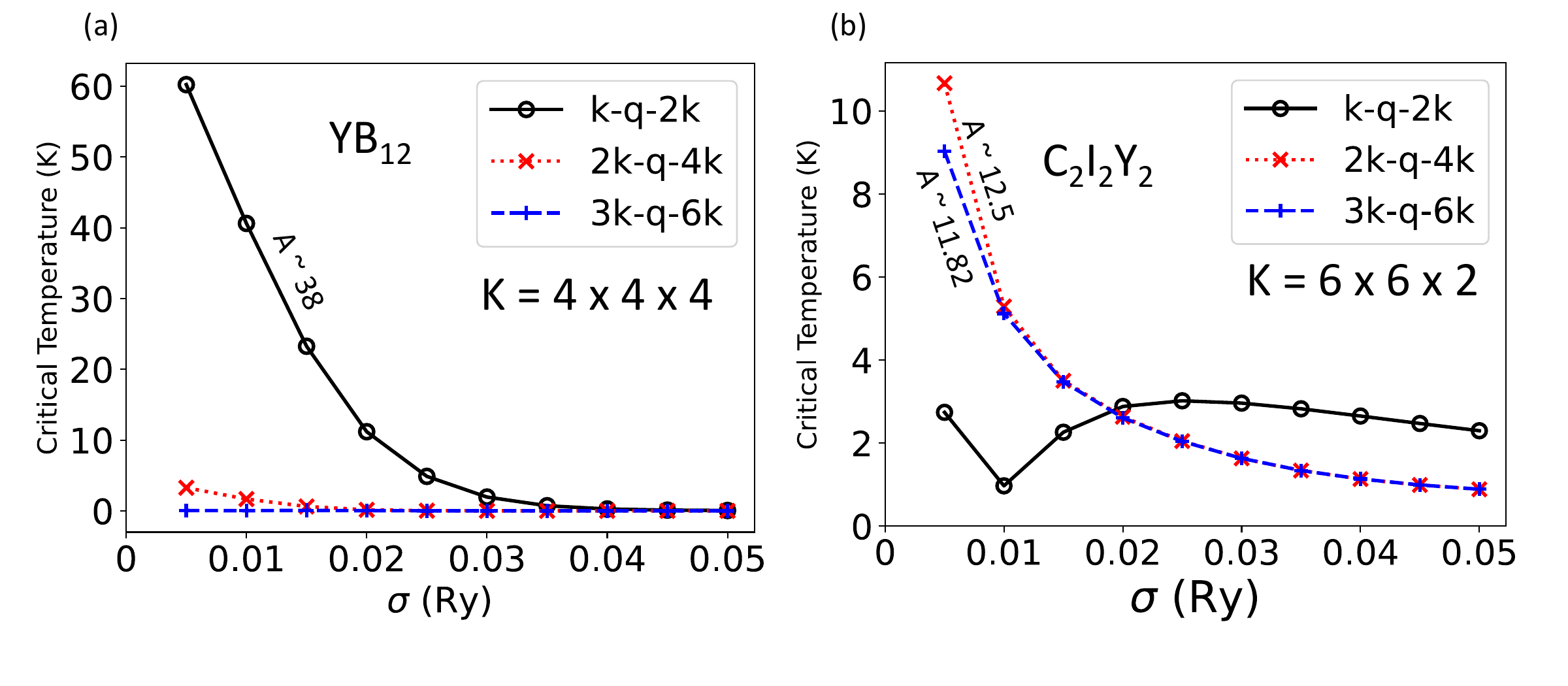}
    \caption{T$_c$ vs $\sigma$ for various k- grids; (a) for YB$_{12}$ and (b) for C$_2$I$_2$Y$_2$.}
    \label{fig:s8}
\end{figure}
 In Fig.~\ref{fig:s8}, we show two cases of YB$_{12}$ and C$_2$I$_2$Y$_2$ in which the coarse \textbf{k}- grids from Materials Project and fine \textbf{k}- grid only twice of that result to qualitatively inaccurate results, unusually high T$_c$ for the former while predicting negligible T$_c$ for the latter. With coarse grids, the exponential decay parameter of YB$_{12}$ is around 38, much larger than the critical value estimated for MgB$_2$ (A$_{MgB_2}$ $\sim$ 12). Increasing k-mesh grids two times in each direction, T$_c$ drops from 40.6 K to 1.67 K at $\sigma =$ 0.01 Ry and drops to almost zero for $\sigma >$ 0.01 Ry indicating non-superconductor, which is in good agreement with the experiment \cite{YB1205}. We found a couple of cases such as C$_2$I$_2$Y$_2$  when $T_c$ first decreases and increases later with $\sigma$ (we set A = 0 in Fig.~\ref{fig:s10}). Such behavior hasn't been observed in G($\Delta$,$\sigma$) vs $\sigma$ plots for a wide range of $\Delta$ (Fig. 8(a) of the main text). The coarse grids result to a very low  T$_c$ of 0.97 K for C$_2$I$_2$Y$_2$. However, the denser mesh corrects the behavior with an exponential decay parameter of 12.5 much closer to the reference value of MgB$_2$. Also, the critical temperature T$_c$ improves to 5.29 K agreeing much more strongly to the experiment \cite{C2I2Y2}. Note that, this analysis only works for compounds with non-zero $T_c$ for a wide range of $\sigma$. For $T_c <\sim $ 1 K at $\sigma = 0.01 $ Ry (computational details for choosing $\sigma = 0.01$ Ry), $T_c <$  0.1 K for $\sigma > 0.01 $ Ry, and rapidly drops to zero for larger $\sigma$, we identify the compound to be non-superconducting with the value of A as zero (A = 0 $<$ A$_{MgB_2}$, already converged). To summarize the ansatz, we present a simple schematics of the process shown as in Fig.~\ref{fig:s9}.
\begin{figure}[H]
 \renewcommand{\thefigure}{S13}\centering
\includegraphics[scale=0.45]{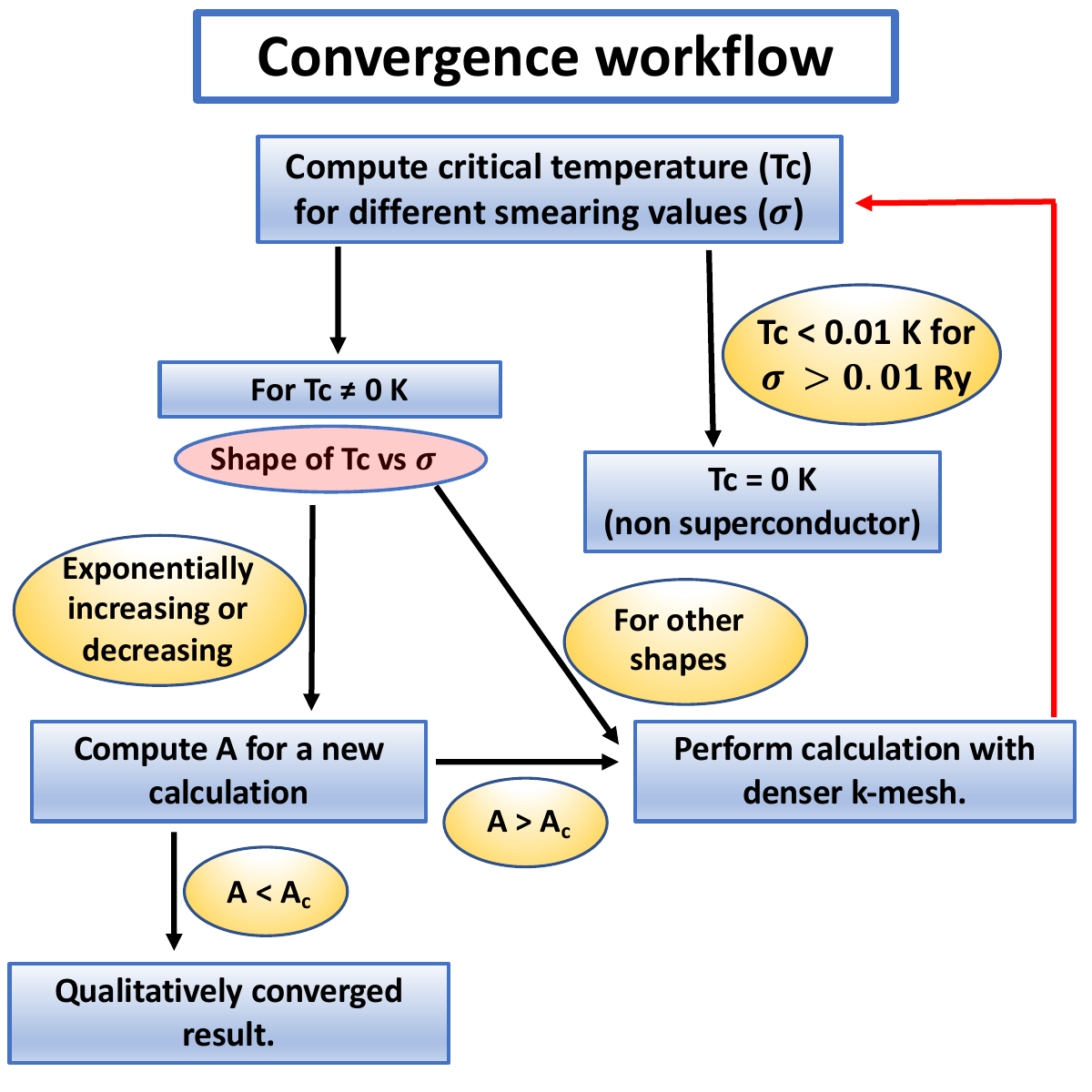}
\caption{Schematics of the convergence testing procedure explained in the main text. A$_c$ is the value of the decay parameter computed for MgB$_2$. First, we compute \(T_c\) for different values of \(\sigma\). If the variation of \(T_c\) with \(\sigma\) decreases exponentially, we compute \(A\). If \(A < A_c\), then the results have converged. If the variation increases exponentially and \(A < 0\), satisfying \(A < A_c\), the results are also considered converged. If \(A > A_c\) or if the variation follows a different pattern, the calculations are repeated with a denser k-mesh. If \(T_c\) decreases extremely rapidly with increasing \(\sigma\), then the material is considered a non-superconductor.}
\label{fig:s9}
\end{figure}

\newpage

\begin{table}[H]
 \renewcommand{\thetable}{S7}
    \centering
     \caption{Comparing results between coarse and fine grids for compounds with A $>$ A$_{MgB_2}$; $T^{eff}_c$ is the critical temperature calculated from efficient EPC calculations, using \textbf{k}-point mesh grid from MP database; $T^{Corr}_c$ is the critical temperature calculated using denser \textbf{k}-point mesh and fine \textbf{k}- mesh grids (2 times of MP database K-point mesh), while \textbf{q}-grid is kept fixed; }
         \scalebox{1.3}{%
    \begin{tabular}{|c|c|c|c|}
    \hline
       Compound  & $T^{eff}_c$ (K) & $T^{Corr}_c$ (K) & $T^{Expt}_c$ (K)\\
       \hline
         YB$_{12}$(Fm-3m)& 40.6 & 1.67  & 0 \cite{YB1205}\footnote{No transition observed above 2.5 K} \\
         CaNiBN (P4/nmm) & 0.96 & 2.43 & 2.2 \cite{CaNiBN12} \\ 
         BaC6 (P6$_3$/mmc) & 6.50 & 1.70 & 0 \cite{BaC6}\\
         La3PbC (Pm-3m) & 3.20 & 0.50 & 0 \cite{La3XC}\footnote{No transition observed above 1 K} \\
         La3SnC (Pm-3m) & 5.09 & 0.81 & 0 \cite{La3XC} \footnote{No transition observed above 1 K}\\
         CsC8 (P6/mmm) & 5.97 & 0.4 & 0.13 \cite{CsC8} \\
         AlB2 (P6/mmm) & 11.3 & 0.3 & 0 \cite{AlB2}\\
         C2I2Y2 (C2/m) & 0.97 & 5.29 & 9.97 \cite{C2I2Y2} \\
         C2Cl2Y2 (C2/m) & 0.508 & 2.73 & 2.3 \cite{C2Cl2Y2}\\
         \hline
    \end{tabular}
   }
    \label{tab1}
\end{table}

\begin{table}[H]
 \renewcommand{\thetable}{S8}
    \centering
        \caption{Comparing results between coarse and fine grids for compounds with A $<$ A$_{MgB_2}$; $T^{eff}_c$ is the critical temperature calculated from efficient EPC calculations, using \textbf{k}-point mesh grid from MP database; $T^{Corr}_c$ is the critical temperature calculated using denser \textbf{k}-point mesh and fine \textbf{k}- mesh grids (2 times of MP database K-point mesh), while \textbf{q}-grid is kept fixed; Here results do not change much qualitatively from coarse to denser mesh. Vertical line in the table after LaB6 separates data from accurate to inaccurate qualitative predictions, compared to experimental results.}
            \scalebox{1.3}{%
    \begin{tabular}{|c|c|c|c|}
    \hline
       Compound  & $T^{eff}_c$ (K) & $T^{Corr}_c$ (K) & $T^{Expt}_c$ (K)\\
       \hline
         WB (Cmcm)& 5.69 & 4.9 & 2.8 \cite{WB12}\\ 
         ReB2 (P6$_3$/mmc) & 0 & 0 & 0 \cite{ReB2}\\
         LaB6 (Pm-3m)& 0.42 & 0 & 0.005 \cite{LaB6} \\
         Ta2C (P-3m1) & 0.0 & 0.0 & 0 \cite{Ta2C} \\
          YIr3B2 (P6/mmm) & 5.5 & 5.54 & N/A \footnote{Not reported} \\
         \hline
         RuB2 (Pmmn)& 0.59 & 0.32  & 1.5 \cite{RuB2}\\
         VC (Fm-3m) & 31.67 & 19.9 & 3.2 \cite{Lang77}\\
         ZnNi3C (Pm-3m) & 15.37 & 15.4 & 0 \cite{ZnNi3C}\footnote{No transition observed above 2.0 K} \\
         YRh3B2 (P6/mmm) & 5.8 & 4.2 & 0 \cite{YRh3B2} \footnote{No transition observed above 1.5 K}\\
         YRu3B2 (P6/mmm)& 3.7 & 2.12 & 0 \cite{YRu3B2}\footnote{No transition observed above 1.2 K} \\
         \hline
    \end{tabular}
}
    \label{tab2}
\end{table}

Next, we present results obtained from EPC calculations for 113 known boron and carbon superconducting (N = 53) and non-superconducting (N = 60) compounds from SuperCon database \cite{supercon}. Fig.~\ref{fig:s10} represents the distribution of $T_c$ with respect to exponential decay parameter A for efficient calculations for $\sigma = 0.01$ Ry with $\mu^*_c = 0.16$. Based on the convergence check ansatz, we found that around 15 \% (NoConv) of the results are not fully converged, while 70 \% of them are qualitatively inaccurate (NoConv-False), compared to available experimental results. Similarly, we have 12 \% of cases that have A $<$ A$_{MgB_2}$ with qualitatively inaccurate predictions (Conv-False), which could be attributed from either inaccuracy of approximations to compute $T_c$ or inaccuracies within available experimental results. This work not only addresses the 15 \% (NoConv) cases but also assists in validating true (Conv-True) results.\\
\begin{figure}[H]
    \centering
    \renewcommand{\thefigure}{S14}
    \includegraphics[scale=0.6]{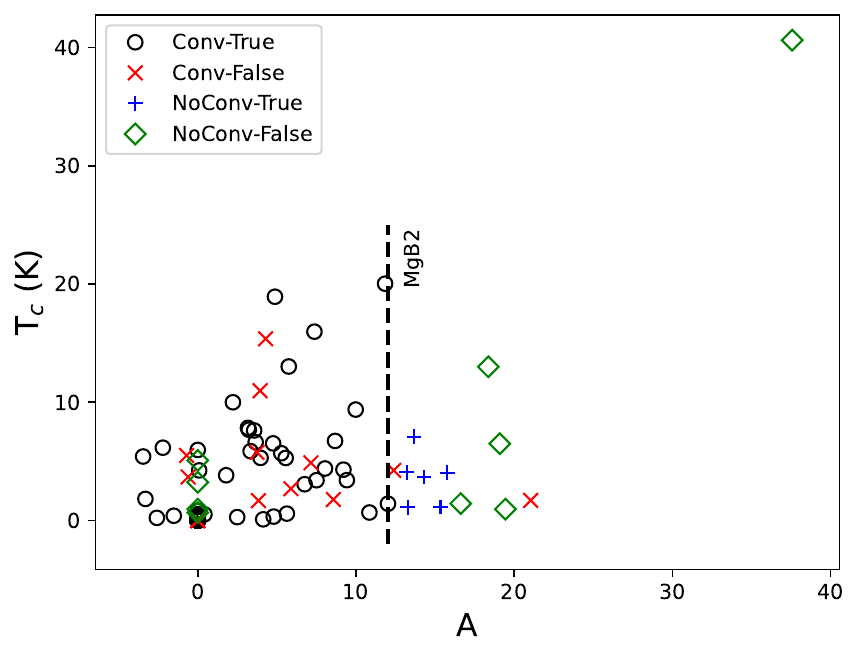}
    \caption{Critical temperature $T_c$ vs exponential decay parameter (A) plot; Black open circle and red cross symbols represent converged (A $<$ A$_{MgB_2}$) true and false predictions respectively; Blue plus and green diamond symbols represent unconverged (A $>$ A$_{MgB_2}$) true and false predictions respectively; A vertical dashed line at $A$ = 12 represents the MgB$_2$ result; When $T_c$ doesn't exponentially decay or increase with $\sigma$ (parabolic in nature, first decrease and increase later), those results are simply identified as unconverged, and presented in the plot with $A$ = 0; With denser grids, these results of parabolic $T_c$ vs $\sigma$ nature changes to exponential decay. {Here, a true prediction indicates agreement with the available experimental results, while a false prediction indicates disagreement.}}
    \label{fig:s10}
\end{figure}

We present a few qualitatively inaccurate predictions with A $>$ A$_{MgB_2}$ and improved results with a denser grid in Table~\ref{tab1}, while Table~\ref{tab2} shows unaffected results (qualitatively) with denser grids for the cases with A $<$ A$_{MgB_2}$. For example, AlB$_2$ is predicted to be superconductors with a coarse grid, while it changes to a non-superconductor with a denser grid, consistent with the experimental prediction. Similarly, $T_c$ of superconductors such as CaNiBN, C2I2Y2, and C2Cl2Y2 change from below 1 K (non-superconductor) to a larger value, and show a better agreement with the experiment. On the contrary, the qualitative results remain unchanged for converged calculations represented by A $<$ A$_{MgB_2}$, regardless of their agreement with the experimental references. 

\section{Application of convergence ansatz on known compounds of SuperCon}
In this section, we compare the computed $T_c$ with available experimental results, as shown in Fig.~\ref{fig:s11} for efficient EPC calculations with and without improved data for A $>$ A$_{MgB_2}$ (left panel) and A $<$ A$_{MgB_2}$ (right panel) cases respectively. We also highlight some significantly deviated results with red rectangles for efficient calculations. With improved data, computed critical temperatures $T_c$ have much better agreement compared to the references, as indicated by the red rectangle enclosing more data points (right panel of Fig.~\ref{fig:s11}), compared to the plot on the left. The results obtained from the \textit{ab-initio} calculations demonstrate a notable level of accuracy with the mean absolute error (MAE) of 2.21 K for the critical temperature, compared to experiments, highlighting the reliability and robustness of this ansatz test. This is in comparison to the MAE of 3.18 K observed in efficient calculations. Please note that we have exclusively incorporated DFPT T$_c$ results for systems that are dynamically stable.
Despite the general agreement between DFPT and experimental findings, there exist notable discrepancies. {Instances like ZnNi3C (Pm-3m), with a calculated T$_c$ of 15 K, deviate enormously from the experimental values of 0 K \cite{PGPLLC03}. A comparable investigation of ZnNi3C by Hoffmann \textit{et al.}\cite{HCSM22}, utilizing different pseudopotentials from the PSEUDODOJO project\cite{VGBVHGR18}, revealed phonon instabilities in these compounds. Notably, these unstable phonon modes were identified to possess significant mode-resolved EPC strength\cite{HCSMdata22}, a finding deserving attention, as we address these concerns in the ``Imaginary Phonon Modes and Superconductivity" section. Similarly, Nb2InC (P6$_3$/mmc) with a calculated T$_c$ of 0.34 K,  VC (Fm-3m) at 20 K, and WB (I4$_1$/amd) at 18.5 K, present a problematic cases. Experimental T$_c$ measurements stand at 3.2 K \cite{Lang77}, 4.3 K \cite{KHFSHAA12}, and 7.5 K \cite{BFSDM09} for VC, WB, and Nb2InC, respectively.} These discrepancies can be ascribed to the constraints of both theoretical approaches and experimental procedures. For instance, older experimental investigations initially reported critical temperatures that were later rectified by more recent experiments or vice versa. For instance, in Ref~\cite{MGACHM68}, superconductivity below 4.7 K was claimed in certain YB$_{12}$ samples, yet subsequent experiments, such as those in \cite{CSPKPO05}, failed to corroborate these findings.  Furthermore, employing more sophisticated theoretical methodologies like solving the many-body anisotropic Migdal-Eliashberg equations or utilizing SCDFT to estimate T$_c$ can significantly enhance the accuracy of calculations compared to isotropic methods. As an illustration, when examining MgB$_2$, T$_c$ calculated using isotropic Eliashberg theory falls within the 15-20 K range. However, in Ref.~\cite{KTMKM22}, T$_c$ values of 34-42 K were reported over a range of $\mu^*$ values, demonstrating the improvement achievable with these advanced techniques.

In Table~\ref{tab5}, we show some of the inaccurate results predicting nonsuperconductors with 2D \textbf{q-} grids (4$\times$4$\times$1). Emergence of strong el-ph coupling and T$_c$ with denser 3D \textbf{q-} grid (3$\times$3$\times$3) is unlikely for such nonsuperconducting predictions with T$_c$ $<$ 0.01 K for $\sigma$ $>$ 0.01 Ry [Fig.~\ref{fig:s9}].

\begin{table}[H]
 \renewcommand{\thetable}{S9}
    \centering
        \caption{Comparison of superconducting properties predicting nonsuperconductors that do not agree with the experiments. Denser \textbf{k-} and \textbf{q-} grids do not improve the results.}
            \scalebox{1.3}{%
    \begin{tabular}{|c|c|c|c|c|c|c|}
    \hline
       Compound  & Grids  & $\lambda$ & $\omega_{log}$ (K) & $T_c$ (K) & $T^{Expt}_c$ (K)\\
       \hline
        Nb2InC & 12$\times$12$\times$12, 3$\times$3$\times$3 & 0.39   & 285  & 0.1 & 7.5 \cite{KBHP22}\\
         & 8$\times$8$\times$2, 4$\times$4$\times$1 &  0.43  & 297 & 0.34 &\\
         \hline
        Ti2InC & 12$\times$12$\times$12, 3$\times$3$\times$3 & 0.18   & 355 & 0.0 & 3.1 \cite{BSDM07}\\
        & 16$\times$16$\times$4, 4$\times$4$\times$1 & 0.13   & 352 & 0.0 &\\
        
         \hline
    \end{tabular}
}
    \label{tab5}
\end{table}

\section{DFPT results for dynamically stable compounds}
\begin{table}[htbp]
\renewcommand{\thetable}{S10}
\centering
\small
\caption{EPC properties of dynamically stable compounds}

\begin{tabular}{|c|c|c|c|c|c|c|}
\hline
ID & Compound & SG & FE (eV/atom) & $\lambda$ & $\omega_{log}$ (K) & T$_c$ (K) \\
\hline
mp-2252&B2Sc1&191&-0.84&0.34&606&0.032 \\
mp-10020&C1Sc1&225&-0.14&0.61&442&5.086 \\
mp-29941&C1Sc2&164&-0.42&0.24&351&0.0 \\
mp-1232372&B2C2Sc2&194&-0.6&0.21&565&0.0 \\
mp-27693&B8C8Sc4&55&-0.46&0.45&560&1.092 \\
mp-10343&B1C2Sc2&139&-0.57&0.41&467&0.383 \\
mp-10139&B1Sc3Sn1&221&-0.61&0.1&331&0.0 \\
mp-12062&Ag1B2&191&0.52&0.88&568&21.563 \\
mp-7817&B12Y1&225&-0.24&0.48&605&1.69 \\
mp-1084&B12Zr1&225&-0.21&0.61&306&3.409 \\
mp-345&B1Hf1&225&-0.41&0.51&281&1.225 \\
mp-1890&B4Mo4&141&-0.5&0.4&364&0.228 \\
mp-999198&B2Mo2&63&-0.49&0.63&338&4.398 \\
mp-451&B1Zr1&225&-0.37&0.47&383&0.946 \\
mp-763&B2Mg1&191&-0.13&0.78&734&20.027 \\
mp-450&B2Nb1&191&-0.69&0.72&365&7.68 \\
mp-1773&B4Re2&194&-0.43&0.22&409&0.0 \\
mp-1077&B4Ru2&59&-0.29&0.42&364&0.356 \\
mp-1472&B2Zr1&191&-0.99&0.14&581&0.0 \\
mp-2680&B6La1&221&-0.56&0.26&343&0.0 \\
mp-2203&B6Y1&221&-0.4&1.97&81&9.999 \\
mp-1079500&B2Ca2N2Ni2&129&-0.97&0.53&464&2.474 \\
mp-1106180&B6Re14&186&-0.22&0.72&180&3.83 \\
mp-15671&B2Re6&63&-0.2&0.72&209&4.4 \\
mp-2850&B4Os2&59&-0.21&0.61&263&3.07 \\
mp-21502&B4Rh8&62&-0.21&0.65&183&2.692 \\
mp-7857&B4Ti4&62&-0.83&0.27&443&0.0 \\
mp-20881&B2La2N2Ni2&129&-1.05&0.64&301&4.241 \\
mp-9219&B2C1La1Pt2&139&-0.64&0.76&262&6.732 \\
mp-3465&B2La1Rh3&191&-0.67&0.84&174&5.859 \\
mp-6114&B2La3N3Ni2&139&-1.15&0.61&362&4.056 \\
mp-20234&B4Li8Pt12&212&-0.57&1.31&96&7.615 \\
mp-4984&B4Mo10Si2&140&-0.4&0.81&257&7.838 \\
mp-1189073&B8Os4Sc4&62&-0.68&0.63&330&4.305 \\
mp-1105309&B4Ge2Ta10&140&-0.59&0.29&234&0.0 \\
mp-1078866&B4C4Y2&127&-0.44&0.52&455&2.323 \\
mp-15955&B4C4Y2&131&-0.3&0.5&392&1.531 \\
mp-1024989&B2C1Pd2Y1&139&-0.5&1.14&249&15.971 \\
mp-5984&B8Rh8Y2&137&-0.59&0.77&223&5.866 \\
mp-1190832&B16Ru4Y4&55&-0.62&0.28&501&0.0 \\
mp-980205&B16Os4Y4&55&-0.57&0.28&404&0.0 \\
mp-2091&B4V6&127&-0.72&0.23&473&0.0 \\
mp-260&B2Cr2&63&-0.52&0.45&412&0.68 \\
mp-9973&B2V2&63&-0.85&0.23&508&0.0 \\
mp-1183252&B1Ir1&187&-0.2&0.47&214&0.539 \\
mp-567164&B2Rh2&194&-0.39&0.15&313&0.0 \\
mp-1063752&B2Rh2&194&-0.18&0.91&173&7.127 \\
mp-4472&B2C2Mo4&63&-0.25&0.78&199&5.414 \\
mp-10112&B2Ir3La1&191&-0.65&0.62&151&1.828 \\
mp-2536&B2Ni4&140&-0.29&0.22&316&0.0 \\
\hline
\end{tabular}
\label{tab:compound-properties}
\end{table}

\begin{table}[htbp]
\renewcommand{\thetable}{S11}
\centering
\caption{EPC properties of dynamically stable compounds}
\begin{tabular}{|c|c|c|c|c|c|c|}
\hline
ID & Compound & SG & FE (eV/atom)& $\lambda$ & $\omega_{log}$ (K) & T$_c$ (K) \\
\hline
mp-1491&B2V1&191&-0.74&0.29&494&0.001 \\
mp-8431&B4Ca2Rh4&70&-0.61&0.26&325&0.0 \\
mp-1008487&B2W2&63&-0.36&0.68&274&4.902 \\
mp-10142&B4Ta3&71&-0.77&0.36&357&0.046 \\
mp-865&Ca1B6&221&-0.41&0.16&538&0.0 \\
mp-14445&B4Ca2Ir4&70&-0.68&0.26&338&0.0 \\
mp-944&Al1B2&191&-0.04&0.4&564&0.295 \\
mp-10852&B4C4La2&127&-0.44&0.38&420&0.113 \\
mp-7783&B4C4La2&131&-0.31&0.5&459&1.862 \\
mp-2967&B2Co2La1&139&-0.48&0.31&304&0.002 \\
mp-5992&B2C1Ir2La1&139&-0.57&0.36&250&0.035 \\
mp-568083&B2C1La1Ni2&139&-0.44&0.33&299&0.009 \\
mp-6794&B2C1La1Rh2&139&-0.6&0.35&284&0.022 \\
mp-571428&B6Mg3Ni9&181&-0.38&0.2&353&0.0 \\
mp-4938&B2Co3Sc1&191&-0.55&0.37&275&0.07 \\
mp-6939&B4Ir4Sr2&70&-0.63&0.28&351&0.0 \\
mp-7348&B4Rh4Sr2&70&-0.56&0.23&335&0.0 \\
mp-3515&B2Co2Y1&139&-0.56&0.24&383&0.0 \\
mp-1024941&B2Rh3Y1&191&-0.71&0.76&162&4.181 \\
mp-4382&B2Ru3Y1&191&-0.47&0.62&172&2.121 \\
mp-9956&Al2C2Cr4&194&-0.17&0.24&403&0.0 \\
mp-3271&Al1C1Ti3&221&-0.59&0.28&271&0.0 \\
mp-1025497&Al2C2V4&194&-0.52&0.23&425&0.0 \\
mp-4448&Y3Al1C1&221&-0.42&0.1&254&0.0 \\
mp-1190604&Al8C4Nb12&213&-0.46&0.35&256&0.022 \\
mp-1190760&Al8C4Ta12&213&-0.45&0.4&191&0.111 \\
mp-1214417&Ba2C12&194&-0.09&0.53&300&1.709 \\
mp-9961&C2Cd2Ti4&194&-0.54&0.28&233&0.0 \\
mp-21075&C1Hf1&225&-0.94&0.13&457&0.0 \\
mp-1096993&C2Hf2&194&-0.74&0.46&416&0.865 \\
mp-1002124&Hf1C1&216&-0.3&0.05&424&0.023 \\
mp-10611&C1La3Pb1&221&-0.43&0.49&151&0.516 \\
mp-1206443&C1La3Sn1&221&-0.49&0.53&146&0.806 \\
mp-5443&C2Nb4Sn2&194&-0.42&0.45&271&0.527 \\
mp-20661&C2Pb2Ti4&194&-0.57&0.31&321&0.004 \\
mp-631&C1Ti1&225&-0.81&0.16&601&0.0 \\
mp-1282&C1V1&225&-0.41&1.14&312&19.989 \\
mp-20648&C4V8&60&-0.47&0.3&348&0.001 \\
mp-1008632&C1V2&164&-0.45&0.24&319&0.0 \\
mp-2795&C1Zr1&225&-0.8&0.15&484&0.0 \\
mp-1014307&C2Zr2&194&-0.64&0.41&480&0.321 \\
mp-570112&C4Cr6&63&-0.07&0.55&343&2.407 \\
mp-28861&C8Cs1&191&-0.05&0.4&766&0.438 \\
mp-568643&C16Rb2&70&-0.04&0.5&755&2.902 \\
mp-16290&C1Ni3Zn1&221&-0.06&1.94&126&15.406 \\
mp-1066566&C2Ni1Y1&38&-0.33&0.47&395&1.074 \\
mp-1079635&C2Ga2Mo4&194&-0.12&0.77&249&6.627 \\
mp-2305&C1Mo1&187&-0.08&0.18&512&0.0 \\
mp-1552&C4Mo8&60&-0.11&0.7&272&5.298 \\
mp-1221498&C1Mo2&164&-0.05&0.6&287&3.119 \\
\hline
\end{tabular}
\label{tab:compound-properties}
\end{table}

\begin{table}[htbp]
\renewcommand{\thetable}{S12}
\centering
\caption{EPC properties of dynamically stable compounds}
\begin{tabular}{|c|c|c|c|c|c|c|}
\hline
ID & Compound & SG & FE (eV/atom)& $\lambda$ & $\omega_{log}$ (K) & T$_c$ (K) \\
\hline
mp-1009832&C1Ta1&216&-0.06&0.69&213&4.034 \\
mp-7088&C1Ta2&164&-0.6&0.22&250&0.0 \\
mp-2034&C4W8&60&-0.02&0.77&212&5.512 \\
mp-33065&C2W4&58&-0.01&0.79&213&6.119 \\
mp-2367&C2La1&139&-0.16&0.5&258&0.978 \\
mp-313&C2Y1&139&-0.19&0.83&282&9.383 \\
mp-1208630&C12Sr2&194&-0.04&0.67&310&5.291 \\
mp-1018048&C2La1Ni1&38&-0.25&0.71&346&6.943 \\
mp-1206284&Br2C2La2&12&-1.12&0.7&174&3.416 \\
mp-20315&C2In2Ti4&194&-0.67&0.13&352&0.0 \\
mp-643367&Br2C2Y2&12&-1.08&0.77&251&6.614 \\
mp-37919&Br2C2Y2&59&-0.08&1.21&109&7.789 \\
mp-23062&C2I2Y2&12&-0.86&0.77&196&5.177 \\
mp-1206889&C2Cl2Y2&12&-1.44&0.6&258&2.725 \\
mp-6576&B2C1Ni2Y1&139&-0.51&0.59&315&3.159 \\
mp-2192&B2Pt2&194&0.07&1.24&188&13.786 \\
mp-1009218&C1Mo1&216&0.5&1.71&103&11.153 \\
mp-632442&Al4C3&1&0.31&0.6&256&2.704 \\
mp-1076&KB6&221&-0.03&0.51&823&3.715 \\
mp-1078278&CrB4&58&-0.31&0.25&622&0.0 \\
mp-1079437&FeB4&58&-0.17&0.9&468&18.723 \\
mp-1080111&B3Mo&166&-0.31&0.29&574&0.001 \\
mp-1106184&MnB4&14&-0.29&0.14&559&0.0 \\
mp-1213975&CaB4&127&-0.39&0.25&551&0.0 \\
mp-1228730&B24Mo7&187&-0.14&1.33&359&28.901 \\
mp-2315&NaB15&74&-0.05&0.1&763&0.0 \\
mp-262&Na3B20&65&-0.06&0.4&782&0.513 \\
mp-27710&CrB4&71&-0.3&0.23&755&0.0 \\
mp-576&B13C2&166&-0.07&0.97&835&39.304 \\
mp-637&YB4&127&-0.59&0.43&388&0.457 \\
mp-7283&LaB4&127&-0.56&0.28&465&0.0 \\
mp-1218188&SrLaB12&123&-0.51&0.36&306&0.041 \\
mp-1232339&LiC12&63&-0.0&0.43&1055&1.24 \\
mp-1227841&BaLaB12&123&-0.49&0.36&304&0.052 \\
mp-1001581&LiC6&191&-0.0&0.39&1043&0.397 \\
mp-1021323&LiC12&191&-0.01&0.33&949&0.035 \\
mp-1001835&LiB&194&-0.17&0.21&328&0.0 \\
mp-1002188&TcB&187&-0.28&0.4&445&0.267 \\
mp-1009695&CaB2&191&-0.14&0.68&444&7.604 \\
mp-1224328&HfNbB4&191&-0.86&0.33&462&0.013 \\
mp-1019317&TcB2&194&-0.44&0.27&497&0.0 \\
mp-1025170&Ti3B4&71&-0.93&0.25&517&0.0 \\
mp-1217974&TaB4W3&25&-0.48&0.6&277&2.921 \\
mp-1080021&Nb2B3&63&-0.75&0.45&409&0.777 \\
mp-1217965&TaB2Mo&38&-0.66&0.57&348&2.788 \\
mp-1102394&Nb5B6&65&-0.77&0.4&398&0.209 \\
mp-1108&TaB2&191&-0.65&1.22&253&18.116 \\
mp-1217924&TaNbC2&166&-0.52&1.41&326&28.387 \\
mp-1217818&TaVB4&191&-0.69&0.6&387&4.047 \\
mp-1206441&V5B6&65&-0.83&0.26&488&0.0 \\

\hline
\end{tabular}
\label{tab:compound-properties}
\end{table}

\begin{table}[htbp]
\renewcommand{\thetable}{S13}
\centering
\caption{EPC properties of dynamically stable compounds}
\begin{tabular}{|c|c|c|c|c|c|c|}
\hline
ID & Compound & SG & FE (eV/atom)& $\lambda$ & $\omega_{log}$ (K) & T$_c$ (K) \\
\hline
mp-1216709&TiNbB2&38&-0.81&0.31&472&0.005 \\
mp-20857&CoB&62&-0.4&0.11&357&0.0 \\
mp-2580&NbB&63&-0.77&0.31&415&0.005 \\
mp-1215250&ZrB4Mo&191&-0.62&1.3&193&15.047 \\
mp-569803&B2W&194&-0.33&0.48&323&0.999 \\
mp-999120&TcB&194&-0.22&0.42&378&0.335 \\
mp-1215178&ZrTiB4&191&-1.0&0.09&617&0.0 \\
mp-9208&V2B3&63&-0.8&0.29&488&0.001 \\
mp-999118&TcB&194&-0.12&1.32&332&26.488 \\
mp-12054&Cr2B3&63&-0.43&0.56&358&2.753 \\
mp-10114&ScB4Ir3&176&-0.59&0.18&258&0.0 \\
mp-9880&YB2C&135&-0.52&0.32&506&0.012 \\
mp-995282&LiAlB4&55&-0.05&0.5&693&2.659 \\
mp-1008527&B2CN&115&-0.44&0.8&644&18.759 \\
mp-1207086&MgAlB4&191&-0.15&0.34&769&0.059 \\
mp-1215209&ZrTaB4&191&-0.84&0.39&436&0.177 \\
mp-1215223&ZrNbB2&38&-0.71&0.35&404&0.04 \\
mp-1216966&TiCrB4&191&-0.68&0.59&339&3.385 \\
mp-1217028&TiB2Mo&38&-0.82&0.25&461&0.0 \\
mp-1220697&Nb2AlB6&191&-0.47&0.54&431&2.781 \\
mp-1224283&HfTaB4&191&-0.85&0.41&423&0.354 \\
mp-1224291&HfTaB2&38&-0.77&0.4&354&0.199 \\
mp-1224347&HfNbB2&38&-0.76&0.37&387&0.073 \\
mp-1226228&CrB2Mo&38&-0.48&0.62&331&3.919 \\
mp-13341&YB4Rh&55&-0.61&0.62&483&6.008 \\
mp-568985&La2B3Br&187&-0.84&0.35&212&0.018 \\
mp-569002&Y3ReB7&63&-0.33&0.81&330&9.953 \\
mp-1008526&B2CN&156&-0.42&1.11&566&34.9 \\
mp-569121&La2B3Cl&174&-1.02&0.41&209&0.158 \\
mp-1191641&YVB4&55&-0.7&0.3&473&0.001 \\
mp-1216398&VCrB2&38&-0.69&0.39&460&0.19 \\
mp-1216475&V3ReB4&25&-0.72&0.39&391&0.169 \\
mp-1216667&TiVB4&191&-0.9&0.27&516&0.0 \\
mp-1217026&TiB4Mo&191&-0.68&0.69&348&6.374 \\
mp-1217095&Ti3B4Mo&6&-0.83&0.28&454&0.0 \\
mp-1217898&TaTiB4&191&-0.88&0.29&441&0.001 \\
mp-1217997&Ta3AlB8&191&-0.5&0.54&316&1.858 \\
mp-1220349&NbB2W&38&-0.58&0.57&341&2.717 \\
mp-1220351&NbVB4&191&-0.71&0.51&421&1.948 \\
mp-1220383&NbB4Mo3&38&-0.56&0.58&354&3.101 \\
mp-1008525&B2CN&160&-0.43&1.66&577&60.733 \\
mp-10057&VCoB3&63&-0.62&0.26&438&0.0 \\
mp-1226242&CrB2W&38&-0.43&0.74&290&6.877 \\
mp-1228634&B4MoIr&187&-0.25&0.5&280&1.054 \\
mp-1196778&Y2B6Ru&55&-0.66&0.22&511&0.0 \\
mp-22709&TaNiB2&62&-0.63&0.3&324&0.001 \\
mp-569116&Sc2B6Rh&55&-0.73&0.52&414&2.079 \\
mp-1224274&HfTiC2&166&-0.83&0.42&464&0.462 \\
mp-1215222&ZrNbC2&166&-0.66&0.73&397&8.92 \\
mp-1224334&HfNbC2&166&-0.73&0.65&385&5.753 \\
\hline
\end{tabular}
\label{tab:compound-properties}
\end{table}

\begin{table}[htbp]
\renewcommand{\thetable}{S14}
\centering
\caption{EPC properties of dynamically stable compounds}
\begin{tabular}{|c|c|c|c|c|c|c|}
\hline
ID & Compound & SG & FE (eV/atom)& $\lambda$ & $\omega_{log}$ (K) & T$_c$ (K) \\
\hline
mp-38818&HfNbB4&71&-0.86&0.4&420&0.254 \\
mp-1086667&ScNiC2&129&-0.28&0.52&368&1.864 \\
mp-1217975&Ta3TiB4&25&-0.84&0.26&394&0.0 \\
mp-9530&Y4C7&14&-0.28&0.13&323&0.0 \\
mp-9459&Y4C5&55&-0.33&0.84&225&7.678 \\
mp-1188534&ScCrC2&59&-0.34&0.62&389&4.881 \\
mp-1224170&HfZrC2&123&-0.87&0.25&483&0.0 \\
mp-1334&Y2C&166&-0.29&0.29&201&0.0 \\
mp-1232379&YBC&194&-0.29&0.28&427&0.0 \\
mp-29896&Y2B3C2&65&-0.5&0.6&375&3.915 \\
mp-612670&YNiBC&129&-0.52&0.2&421&0.0 \\
mp-567692&B4C3Ni4Y3&139&-0.51&0.31&393&0.003 \\
mp-12737&B2C1Rh2Y1&139&-0.6&0.38&285&0.098 \\
mp-1087495&Tc3B&63&-0.27&0.8&254&7.66 \\
mp-978989&Tc7B3&186&-0.32&0.68&227&4.037 \\
mp-1068296&Fe(BW)2&71&-0.43&0.45&279&0.465 \\
mp-11750&Ti6Si2B&189&-0.66&0.22&333&0.0 \\
mp-1238800&CaBC&194&-0.07&0.42&501&0.494 \\
mp-27261&Ba7(BIr)12&166&-0.48&0.47&233&0.626 \\
mp-29980&Nb4B3C2&63&-0.59&1.01&219&11.258 \\
mp-29979&Nb3B3C&63&-0.66&1.22&229&16.438 \\
mp-541849&Al3(BRu2)2&123&-0.59&0.26&308&0.0 \\
mp-1188408&Zr5Sn3B&193&-0.68&0.45&202&0.343 \\
mp-1206909&CaBPd3&221&-0.56&1.1&103&6.187 \\
mp-9985&NbNiB&63&-0.56&0.34&256&0.018 \\
mp-1080829&Ti6Ge2B&189&-0.62&0.21&305&0.0 \\
mp-1025192&Ta4C3&221&-0.45&1.46&138&12.644 \\
mp-1215211&ZrNbB4&191&-0.85&0.31&479&0.005 \\
mp-1220641&Nb3B4W3&38&-0.43&0.46&299&0.66 \\
mp-10721&Ti2C&227&-0.64&0.14&386&0.0 \\
mp-27919&Ti8C5&166&-0.72&0.22&455&0.0 \\
mp-1218000&Ta4C3&160&-0.55&0.52&183&0.919 \\
mp-1220752&Nb10(SiB)3&42&-0.65&0.24&310&0.0 \\
mp-1189539&Hf2Al3C5&194&-0.13&0.47&1454&3.645 \\
mp-9958&Ti2GeC&194&-0.81&0.48&276&0.847 \\
mp-1025524&Zr2TlC&194&-0.63&0.21&188&0.0 \\
mp-1216707&TiNbC2&166&-0.67&0.73&422&9.623 \\
mp-1079992&Zr2PbC&194&-0.66&0.4&230&0.146 \\
mp-1207413&Zr5Sn3C&193&-0.68&0.53&180&1.013 \\
mp-1217106&Ti2C&166&-0.63&0.35&357&0.037 \\
mp-1217822&TaVC2&166&-0.46&0.8&343&10.296 \\
mp-12990&Ti2AlC&194&-0.7&0.25&371&0.0 \\
mp-3871&Ti2SnC&194&-0.73&0.24&353&0.0 \\
mp-1025427&Ta2GaC&194&-0.52&0.45&231&0.385 \\
mp-1078712&Hf2TlC&194&-0.63&0.21&171&0.0 \\
mp-1079076&Hf2PbC&194&-0.63&0.41&194&0.15 \\
mp-1079908&Ti2SiC&194&-0.79&0.4&334&0.205 \\
mp-1220365&NbVC2&166&-0.39&0.83&365&11.893 \\
mp-21023&Ti3SnC2&194&-0.79&0.18&421&0.0 \\
mp-22144&Ta2InC&194&-0.38&0.37&233&0.062 \\
\hline
\end{tabular}
\label{tab:compound-properties}
\end{table}

\begin{table}[htbp]
\renewcommand{\thetable}{S15}
\centering
\caption{EPC properties of dynamically stable compounds}
\begin{tabular}{|c|c|c|c|c|c|c|}
\hline
ID & Compound & SG & FE (eV/atom)& $\lambda$ & $\omega_{log}$ (K) & T$_c$ (K) \\
\hline
mp-13137&Hf2CS&194&-1.51&0.19&303&0.0 \\
mp-1220725&Nb2CN&166&-0.8&0.88&310&11.685 \\
mp-1216616&V2CN&166&-0.8&1.0&322&16.224 \\
mp-1214755&BPd6&15&-0.2&0.16&169&0.0 \\
mp-7424&BPd2&58&-0.27&0.42&159&0.16 \\
mp-1078540&Ni6Ge2B&189&-0.31&0.34&190&0.011 \\
mp-1078623&Zr2BIr6&225&-0.81&0.36&184&0.03 \\
mp-12073&Ba(BIr)2&139&-0.5&0.37&269&0.051 \\
mp-7349&Ba(BRh)2&139&-0.48&0.29&295&0.001 \\
mp-7705&NbFeB&187&-0.12&1.76&162&18.063 \\
mp-1215258&ZrBeB&187&-0.58&0.45&436&0.751 \\
mp-1208348&Ta5Ga3B&193&-0.42&0.31&190&0.002 \\
mp-605839&Li2B2Rh3&55&-0.5&0.27&302&0.0 \\
mp-8308&Ca3Ni7B2&166&-0.34&0.13&305&0.0 \\
mp-1206490&Nb2B2Mo&127&-0.64&0.28&367&0.0 \\
mp-31052&LaBPt2&180&-0.92&0.28&142&0.0 \\
mp-1223681&La2(Ni2B)3&44&-0.33&0.46&227&0.517 \\
mp-1097&B2Ta2&63&-0.81&0.41&366&0.297 \\
mp-28930&C16K2&70&-0.03&0.57&1102&8.897 \\
mp-7832&B4W4&141&-0.37&1.58&186&18.463 \\
mp-28613&B3Li3Pt9&189&-0.51&0.4&155&0.088 \\
mp-569759&B4Rh8Zn5&65&-0.42&0.26&222&0.0 \\
mp-571419&Al4C5Zr2&166&-0.35&0.36&387&0.06 \\
mp-1207385&Al8C8Zr2&164&-0.25&0.26&482&0.0 \\
mp-1189895&B2Ge6Ta10&193&-0.45&0.61&181&2.053 \\
mp-10140&B1Sc3Tl1&221&-0.36&0.54&216&1.324 \\
mp-1216165&B1Si6Y10&162&-0.65&0.4&166&0.103 \\
mp-20175&C2In2Sc4&194&-0.55&0.22&170&0.0 \\
mp-20983&C2In2V4&194&-0.34&0.35&349&0.03 \\
mp-1224263&B4Hf1Ti1&191&-1.02&0.1&592&0.0 \\
mp-1224184&B4Hf1Zr1&47&-1.0&0.15&507&0.0 \\
mp-8307&B2Ca2Ni8&191&-0.3&0.33&129&0.004 \\
mp-4079&Al1C1Sc3&221&-0.59&0.04&303&0.048 \\
mp-1103814&C3K5N6&229&-0.52&0.08&190&0.0 \\
mp-1224285&C2Hf1Ta1&166&-0.81&0.62&359&4.486 \\
mp-1215219&C2Ta1Zr1&166&-0.74&0.65&370&5.558 \\
mp-570499&B2La5N6&12&-1.51&0.25&351&0.0 \\
mp-569935&B2La3N4&71&-1.5&0.26&378&0.0 \\
mp-1223086&C6La2Y1&12&-0.16&0.56&271&2.049 \\
mp-1221519&C1Mo2N1&25&-0.35&0.37&473&0.105 \\
mp-1222150&Al1B10Mg4&191&-0.14&0.59&768&7.488 \\
mp-1189984&C8Mo4Y4&62&-0.24&0.61&409&4.562 \\
mp-3380&C8La4Rh4&76&-0.33&0.27&263&0.0 \\
mp-4262&BeAlB&216&-0.05&0.16&578&0.0 \\
mp-5971&YBPt2&180&-1.0&0.18&191&0.0 \\
mp-9596&La(BIr)4&86&-0.58&0.47&201&0.503 \\
mp-1105186&Cu3B5Pt9&189&-0.21&0.4&164&0.099 \\
mp-1106165&Nb5Si3B&193&-0.69&0.46&282&0.638 \\
mp-1106398&V5Ge3B&193&-0.46&0.38&303&0.11 \\
mp-1188194&Ta3B2Ru5&127&-0.53&0.33&146&0.006 \\
\hline
\end{tabular}
\label{tab:compound-properties}
\end{table}

\begin{table}[htbp]
\renewcommand{\thetable}{S16}
\centering
\caption{EPC properties of dynamically stable compounds}
\begin{tabular}{|c|c|c|c|c|c|c|}
\hline
ID & Compound & SG & FE (eV/atom) & $\lambda$ & $\omega_{log}$ (K) & T$_c$ (K) \\
\hline
mp-29723&LaB2Ru3&191&-0.41&0.67&148&2.477 \\
mp-22759&CoBW&62&-0.43&0.37&342&0.071 \\
mp-28786&Zn(BIr)2&139&-0.32&0.08&319&0.0 \\
mp-9999&Ni(BMo)2&71&-0.48&0.3&352&0.001 \\
mp-1076987&TaNiB&63&-0.62&0.3&253&0.001 \\
mp-3348&LiBIr&70&-0.51&0.16&325&0.0 \\
mp-2760&Nb6C5&12&-0.55&0.51&393&1.632 \\
mp-32679&Nb10C7&12&-0.48&0.45&301&0.537 \\
mp-1226378&Cr2C&164&-0.04&0.38&376&0.118 \\
mp-2318&Nb2C&164&-0.45&0.31&285&0.002 \\
mp-974437&Re2C&194&-0.03&0.32&348&0.007 \\
mp-10037&AlCo3C&221&-0.22&0.1&212&0.0 \\
mp-9987&Nb2PC&194&-0.75&0.44&345&0.493 \\
mp-21003&Y2ReC2&62&-0.42&0.29&282&0.0 \\
mp-28767&Sc5Re2C7&65&-0.48&0.21&384&0.0 \\
mp-567462&Sc3RhC4&12&-0.5&0.68&393&6.91 \\
mp-7130&ScRu3C&221&-0.28&0.4&227&0.12 \\
mp-996161&Nb3AlC2&194&-0.52&0.36&384&0.048 \\
mp-996162&Nb2AlC&194&-0.51&0.46&337&0.768 \\
mp-1078811&Nb2GeC&194&-0.51&0.45&307&0.531 \\
mp-1080835&V2GaC&194&-0.52&0.27&375&0.0 \\
mp-1189574&YWC2&62&-0.27&0.54&389&2.392 \\
mp-4992&ScCrC2&194&-0.34&0.62&415&4.966 \\
mp-8044&V2PC&194&-0.69&0.4&409&0.227 \\
mp-10046&V2AsC&194&-0.53&0.47&354&0.878 \\
mp-1212439&Hf5Al3C&193&-0.47&0.2&2988&0.0 \\
mp-1217764&Ta2CN&123&-0.82&2.7&127&19.495 \\
mp-37179&Ta2CN&141&-0.84&1.94&162&19.757 \\
mp-4384&Nb2CS2&166&-1.12&1.03&222&11.827 \\
mp-559976&Ta2CS2&164&-1.19&0.55&251&1.71 \\
mp-995201&Ti5Si3C&193&-0.82&0.55&253&1.784 \\
mp-1025441&Ta2AlC&194&-0.52&0.44&272&0.374 \\
mp-1079546&Nb2GaC&194&-0.52&0.44&263&0.424 \\
mp-1220371&NbAlVC&164&-0.47&0.44&362&0.491 \\
mp-3732&Ti2CS&194&-1.42&0.2&473&0.0 \\
mp-4563&Ti3TlC&221&-0.44&0.29&229&0.0 \\
mp-1216139&Y4C4I3Br&8&-0.91&0.84&174&5.825 \\
mp-1220491&Nb6V2(CS2)3&12&-1.09&0.54&278&1.768 \\
mp-1220693&Nb2CuCS2&156&-0.9&0.85&207&7.255 \\
mp-1215225&ZrTaCN&160&-1.15&0.78&301&8.323 \\
mp-1025205&Y2Re2Si2C&12&-0.62&0.73&172&3.887 \\
mp-1215184&ZrTiCN&160&-1.33&0.43&437&0.5 \\
mp-1224279&HfTiCN&160&-1.41&0.4&420&0.267 \\
mp-1009894&Zr1C1&216&-0.19&0.01&444&0.405 \\
mp-1068661&ZrBRh3&221&-0.75&0.01&197&0.176 \\
mp-1145&B2Ti1&191&-1.06&0.11&638&0.0 \\
\hline
\end{tabular}

\label{tab:compound-properties}
\end{table}

\begin{table}[htbp]
\renewcommand{\thetable}{S17}
\centering
\caption{EPC properties of dynamically stable compounds}
\begin{tabular}{|c|c|c|c|c|c|c|}
\hline
ID & Compound & SG & FE (eV/atom) & $\lambda$ & $\omega_{log}$ (K) & T$_c$ (K) \\
\hline
mp-1080664&B4Cr4&141&-0.53&0.27&449&0.0 \\
mp-1994&B2Hf1&191&-1.02&0.16&518&0.0 \\
mp-2331&B4Mo2&166&-0.43&0.4&437&0.239 \\
mp-20689&B4Nb6&127&-0.63&0.26&365&0.0 \\
mp-13415&B4Ta6&127&-0.67&0.24&306&0.0 \\
mp-910&C1Nb1&225&-0.46&1.11&309&18.924 \\
mp-1094093&C2Nb2&194&-0.35&1.14&300&19.284 \\
mp-999388&C2Nb2&194&-0.33&0.84&455&15.269 \\
mp-999377&C2Nb2&194&-0.09&0.71&320&6.7 \\
mp-1086&C1Ta1&225&-0.58&0.68&236&4.239 \\
mp-1207750&Y5(SiB4)2&127&-0.68&0.33&401&0.014 \\
mp-1217023&TiB2W&38&-0.78&0.3&419&0.001 \\
mp-1079333&B2CN&51&-0.53&0.46&1045&2.381 \\
mp-13854&B3Ru2&194&-0.33&0.14&445&0.0 \\
mp-14019&NiB&63&-0.24&0.24&376&0.0 \\
mp-1018050&CrC&187&-0.0&0.23&610&0.0 \\
mp-1215480&Zr3NbC4&166&-0.73&0.77&346&9.324 \\
mp-1216691&TiVC2&166&-0.62&0.53&446&2.47 \\
mp-1215174&ZrTiC2&123&-0.71&0.25&461&0.0 \\
mp-1215218&ZrMoC2&166&-0.35&1.79&120&13.656 \\
mp-4893&Hf2SnC&194&-0.78&0.31&205&0.003 \\
mp-5659&Ti3SiC2&194&-0.82&0.37&345&0.081 \\
mp-1092281&Ti2TlC&194&-0.57&0.14&295&0.0 \\
mp-3747&Ti3AlC2&194&-0.76&0.26&342&0.0 \\
mp-3886&Zr2AlC&194&-0.64&0.56&163&1.181 \\
mp-1188856&V5Si2B&140&-0.39&0.37&302&0.067 \\
mp-1216445&V9Cr3B8&10&-0.7&0.25&459&0.0 \\
mp-1216643&V10Si6B&162&-0.63&0.33&323&0.014 \\
mp-1220688&Nb3Re3B4&38&-0.35&0.53&265&1.47 \\
mp-1226327&Cr3B4Mo3&38&-0.29&0.78&289&8.091 \\
mp-1009817&C1Ta1&187&-0.16&1.54&235&22.8 \\
mp-1542&YB2&191&-0.56&0.42&473&0.453 \\
mp-1216692&TiNbB4&191&-0.89&0.29&481&0.001 \\
mp-4613&Zr2SnC&194&-0.79&0.32&259&0.004 \\
mp-1232384&ZrBC&194&-0.32&1.12&320&19.998 \\
\hline
\end{tabular}

\label{tab:compound-properties}
\end{table}

\end{document}